\documentclass[aps,pre,onecolumn,showpacs,showkeys,a4paper]{revtex4-1}

\usepackage{stmaryrd}

\usepackage{graphicx} 
\usepackage{amsmath}
\usepackage{amsfonts}
\usepackage{amscd}
\usepackage{amssymb}
\usepackage{fullpage}
\usepackage{color}

\usepackage{float}
\usepackage{empheq} 
\usepackage{fancybox}
\usepackage{pdfpages}

\usepackage{epsf}
\usepackage{epsfig}
\usepackage{epstopdf}

\begin{document}

\title{The Physicist's Companion to Current Fluctuations: \\ One-Dimensional Bulk-Driven Lattice Gases}
\author{Alexandre Lazarescu$^{(1)}$}
\affiliation{(1) Instituut voor Theoretische Fysica, K. U. Leuven, Belgium}
\begin{abstract}

One of the main features of statistical systems out of equilibrium is the currents they exhibit in their stationary state: microscopic currents of probability between configurations, which translate into macroscopic currents of mass, charge, etc. Understanding the general behaviour of these currents is an important step towards building a universal framework for non-equilibrium steady states akin to the Gibbs-Boltzmann distribution for equilibrium systems. In this review, we consider one-dimensional bulk-driven particle gases, and in particular the asymmetric simple exclusion process (ASEP) with open boundaries, which is one of the most popular models of one-dimensional transport. We focus, in particular, on the current of particles flowing through the system in its steady state, and on its fluctuations. We show how one can obtain the complete statistics of that current, through its large deviation function, by combining results from various methods: exact calculation of the cumulants of the current, using the integrability of the model ; direct diagonalisation of a biased process in the limits of very high or low current ; hydrodynamic description of the model in the continuous limit using the macroscopic fluctuation theory (MFT). We give a pedagogical account of these techniques, starting with a quick introduction to the necessary mathematical tools, as well as a short overview of the existing works relating to the ASEP. We conclude by drawing the complete dynamical phase diagram of the current. We also remark on a few possible generalisations of these results.

\end{abstract}

\maketitle

\tableofcontents

\section{Introduction}

Our world is a complex and chaotic place. In order to understand it better, we look for the fundamental laws that govern it,  hoping that they are simple enough to be found, and universal enough to be useful. This is how, for instance, by observing the behaviour of various massive objects in different situations, Newton deduced his laws of motion, or how, by studying how gases react to changes in their environment, Clapeyron discovered the ideal gas law. Both of these laws are strikingly simple, and, although they are both approximations, they apply to an extremely wide range of systems.

In fact, since a gas is a collection of massive objects, it must obey both laws, but at different levels of description: at the microscopic level, each individual atom follows Newton's laws ; at the macroscopic level, the whole gas follows Clapeyron's law. Moreover, Newton's equation of motion, although conceptually simple, has $6N$ variables for a gas with $N$ atoms (the position and velocity of each atom), and the ideal gas law has only $3$ (density, temperature, pressure). How these two descriptions are compatible is not entirely obvious, although this example is one of the simplest (but there are more complex ones, such as how neurons make a brain, to give an example from the other extreme). Understanding how one goes from one law to the other is just as essential as knowing the two laws themselves. It is the goal of statistical physics to bridge that gap, and to find out how simplicity can emerge from large numbers.

~

Let us consider a system made of a large number of components, each obeying simple laws. There are several formalisms that we can use to that effect. If the system is isolated and stationary, with a well-defined interaction potential between its constituents, then it is reasonable to assume that its states are distributed according to the Gibbs-Boltzmann law: the system is at equilibrium. Given that assumption, one can then compute averages of macroscopic observables with respect to that distribution, and recover all the thermodynamic properties of the system, usually by obtaining the free energy, which gives the probability distribution of macroscopic variables. Non-analyticities in that free energy are the sign of phase transitions, which are the most interesting features of equilibrium systems, and have been studied extensively. However, in that framework, one is limited to static observables, as the Gibbs-Boltzmann distribution gives no information on dynamics.

Should one be interested in dynamical observables, or if the system is driven by some external field and cannot be described by an interaction potential, one may take a step back and assume, not the form of the stationary distribution, but of the rates of transition between microstates and of the distribution of the time elapsed between successive transitions. This is equivalent to assume a certain probability density for dynamical trajectories in space and time, instead of a probability for a static configuration. The simplest and most common choice here is to assume Markovian dynamics: the evolution of a trajectory depend only on its state, and not on its history, which implies Poisson-distributed waiting times. Under reasonable assumptions, and if the dynamics do not depend explicitly on time, the system can be shown to relax to a steady state, the distribution of which is not a given, in contrast to equilibrium: it is a contraction of the distribution on trajectories, and is in general difficult to obtain.

We may then distinguish two classes of Markovian dynamics: those that have detailed balance, and those that do not. In the first case, the steady state will have an equilibrium distribution, and the microscopic currents of probability between microstates will all be exactly zero: the state is not merely stationary, but entirely inert. Without detailed balance, some of those currents will be non-zero, and their magnitude is an indication of the work done by the environment in order to maintain that flow. That probability current, which translates into macroscopic currents of particles, charges, heat, etc. is what defines the system as being out of equilibrium. It is that type of systems which will be the focus of our attention.

~

\begin{figure}[ht]
\begin{center}
 \includegraphics[width=0.8\textwidth]{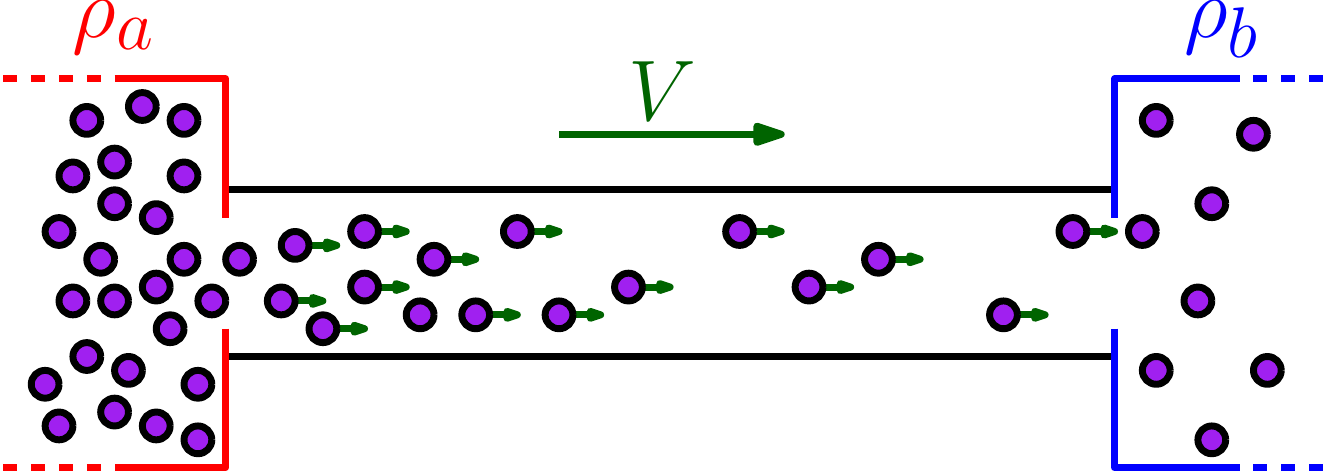}
  \caption{Sketch of a one-dimensional non-equilibrium system. Particles move in a one-dimensional channel, between two reservoirs at fixed densities $\rho_a$ and $\rho_b$. The particles interact with each-other, and are subject to a driving field $V$. Because of the field, and the possible imbalance between the reservoirs, there is a net flux of particles from one side of the system to the other.}
\label{I-fig-neq}
 \end{center}
 \end{figure}

Non-equilibrium systems are quite ubiquitous in nature, and in particular in biology: any system which is meant to transport objects, cells, energy, etc. from one point to another is, by definition, out of equilibrium. In many cases, the system is one-dimensional (think of cells in a blood vessel, for instance, or molecular motors on actin), and the main quantity of interest is the flux that goes through it (see fig.\ref{I-fig-neq}). The question is then to deduce the macroscopic behaviour of the system, and in particular the statistics of that current, from the microscopic definition of the model, and identify what, in that behaviour, may be generic among similar systems.

There are many approaches to that problem. One of them, which has yielded major results in many other sub-fields of statistical physics, is to find a toy model which is simple enough to be mathematically tractable, yet complex enough to be physically relevant. In the case of equilibrium statistical physics, the Ising model has played such a role, and has had a central part in our understanding equilibrium phase transitions. Its counterpart for non-equilibrium systems is the simple exclusion process: particles jump stochastically from site to site on a finite lattice, and may not jump to a site which is already occupied. One-dimensional versions of this model, with uniform jumping rates and either periodic or open boundary conditions, are exactly solvable, and allow for very precise calculations, such as relatively simple expressions for the full steady state distribution and current fluctuations. However, these results and calculations do not extend easily to other models, precisely because resolvability is not generic.

Another approach is to start from an effective mesoscopic description, which requires more assumptions but applies to a broader range of systems. For equilibrium models, one may think, for instance, of the mean-field approach, where it is assumed that spatial correlations can be neglected. A similar framework for one-dimensional non-equilibrium particle systems is the macroscopic fluctuation theory, where the probability of a trajectory is assumed to depend only on the local averages of the density and current of particles.

It is then quite natural to combine these two approaches: the first can be used to verify the assumptions made for the second, and the second can help in generalising the results obtained through the first. In this review, we propose to give an overview of the methods and results pertaining to these two approaches in the case of the one-dimensional asymmetric simple exclusion process with open boundaries.

~

The layout of this review is as follows.

In section \ref{I}, we give the main definitions and results related to the mathematical framework that is relevant to our interests: that of large deviations. In particular, we look at large deviations of time-additive observables for Markov processes in continuous time, and we define the so-called ‘s-ensemble’, which is a statistical ensemble for Markov processes where the current is seen as a free parameter.

In section \ref{II}, we get acquainted with the asymmetric simple exclusion process. In the first part of the section, we give the definition of the model and briefly review existing variants and results. In the second part, we look at the steady state of the system, first through the mean-field approach, and then through the exact expression for the stationary distribution (the so-called `matrix Ansatz').

In section \ref{III}, we present an exact expression for the complete generating function of the cumulants of the current in the open ASEP, and we take the limit of large sizes to extract the asymptotic behaviour of that expression, obtaining a different result for each phase in the ASEP's diagram. We then look at what the corresponding behaviours are for the large deviation function of the current in each phase, which are valid for small fluctuations of the current.

In section \ref{IV}, we take the limit of infinitely high or low currents, and calculate the corresponding limits for the large deviation function through direct diagonalisation of the conditioned process. In the low current limit, we obtain a perturbative expansion around a diagonal matrix. In the high current limit, we obtain a system equivalent to an open XX spin chain, diagonalisable through free fermion techniques.

In section \ref{V}, we use the macroscopic fluctuation theory to obtain the full dynamical phase diagram of the current for the open ASEP. We compare these results with those obtained from the exact calculations of the previous sections. We also remark that the method used in that section is in principle applicable to any one-dimensional bulk-driven particle gas.

~

The present review is largely based on the author's PhD manuscript \cite{Lazarescu2013}. It is intended to give a self-contained and reasonably detailed account of the tools involved in determining the large deviations of the current in the asymmetric simple exclusion process, as much as of the results that they yield. For the sake of brevity and legibility, some of the finer details of those calculations, as well as most technical aspects of everything related to the integrability of the model, have been omitted, but should the readers be in need of clarifications, they may refer to \cite{Lazarescu2013} or to the references given in section \ref{II-1}. That section contains most of the bibliographical references of this review, and although it is far from being exhaustive, it should provide an adequate starting point for the curious reader.

\newpage

\section{A crash course in large deviations}
\label{I}

This first section contains a brief introduction to the mathematical objects that we will be manipulating in the rest of the review, namely: large deviation functions. We first define the concept of large deviations in general, and give a few useful theorems. We then apply the concept to time-additive observables for Markov processes in continuous time, and look at two specific examples with interesting properties: the time-integrated empirical vector, and the entropy production. For a thorough review of this topic, one may refer to \cite{Touchette20091} and \cite{Touchette2011}.

\subsection{Definition and a few useful results}
\label{I-1}

Consider a system defined by a size $N$, and an observable $a$ intensive in $N$, which has a probability distribution ${\rm P}_N(a)$ for each $N$. It is said that $a$ obeys a large deviations principle with rate $g(a)$ if the limit
\begin{equation}\label{I-1-g}
g(a)=\lim\limits_{N\rightarrow\infty}\Bigl[-\frac{\log({\rm P}_N(a))}{N}\Bigr]
\end{equation}
is defined and finite for every $a$. In other terms, $g(a)$ is the rate of exponential decay of ${\rm P}_N(a)$ with respect to $N$. Its minimum is the most probable value of $a$. We then write:
\begin{equation}\label{I-1-Pgtossa}\boxed{
{\rm P}_N(a)\approx {\rm e}^{-N g(a)}
}\end{equation}
where the $\approx$ signifies precisely what is written in eq.(\ref{I-1-g}).

Note that $N$ is not necessarily an actual size or a number of elements: it can be a time span, a number of events, or any variable that can be taken to infinity. Also note that $a$ doesn't have to be a scalar observable. It can be a function (in which case $g[a]$ is a large deviation functional), or any mathematical object for which a probability can be defined in the system under consideration.

~~

From the large deviation function of $a$, one can obtain that of an observable $b=f(a)$, where the function $f$ is not necessarily injective, by writing
\begin{equation}
{\rm P}_N(b)=\int {\rm P}_N(a)\delta\bigl(b-f(a)\bigr){\rm d}a=\int  {\rm e}^{-N g(a)}\delta(b-f(a)){\rm d}a\sim {\rm e}^{-N \min\limits_{f(a)=b}[g(a)]},
\end{equation}
the last expression being obtained from a saddle-point approximation for large $N$. If $f$ is injective, one simply obtains a change of variables from $a$ to $b$. If, on the contrary, $f$ is not injective, which is to say that $b$ is a contraction of $a$, we get the {\it contraction principle}: the large deviation function $\tilde{g}(b)$ is given by
\begin{equation}\label{I-1-contraction}\boxed{
\tilde{g}(b)=\min\limits_{f(a)=b}[g(a)].
}\end{equation}
In principle, $a$ may be any mathematical object and $f$ any function, but we will mostly consider linear transformations, where $a$ is a vector and $f$ a matrix with a non-maximal rank.

~

An alternative way to treat a probability distribution which decays fast enough is through its rescaled cumulants $E_k$, or their exponential generating function $E(\mu)=\sum\limits_{k=1}^{\infty}E_k\frac{\mu^k}{k!}$, defined through
\begin{equation}\label{I-1-ma}
{\rm e}^{N E(\mu)}\equiv\langle{\rm e}^{\mu N a}\rangle=\int {\rm P}_N(a){\rm e}^{\mu N a} {\rm d}a
\end{equation}
which is to say that the generating function of cumulants is the logarithm of the exponential generating function of moments.

If we replace ${\rm P}_N(a)$ by its limit under the large deviation principle: ${\rm P}_N(a)\rightarrow {\rm e}^{-N g(a)}$, we get, in the large $N$ limit:
\begin{equation}\label{I-1-Eg}
{\rm e}^{N E(\mu)}\rightarrow \int {\rm e}^{-N(g(a)-\mu a)}da.
\end{equation}
which yields, through a saddle-point approximation,
\begin{equation}\label{I-1-Eg2}
E(\mu)=\max_a[\mu a-g(a)]
\end{equation}
or equivalently
\begin{equation}\label{I-1-Eg3}\boxed{
E(\mu)=\mu a^\star-g(a^\star)~~,~~\frac{{\rm d}}{{\rm d}a} g(a^\star)=\mu
}\end{equation}
where $a^\star$ is the value of $a$ at which the maximum in eq.(\ref{I-1-Eg2}) is attained. That is to say that $E$ and $g$ are Legendre transforms of one another and contain essentially the same information (unless $g$ has a non-convex part). The inverse transformation formula is then
\begin{equation}\label{I-1-Eg3.5}
g(a)=\min_\mu[\mu a-E(\mu)]
\end{equation}
or
\begin{equation}\label{I-1-Eg4}\boxed{
g(a)=\mu^\star a-E(\mu^\star)~~,~~\frac{{\rm d}}{{\rm d}\mu} E(\mu^\star)=a
}\end{equation}
where $\mu^\star$ is the value of $\mu$ at which the maximum in eq.(\ref{I-1-Eg4}) is attained. This last equation is part of the G\"artner-Ellis theorem, which states that if $E(\mu)$, defined through eq.(\ref{I-1-ma}), is well-behaved, then $a$ obeys a large deviation principle with a rate $g(a)$ obtained through eq.(\ref{I-1-Eg3.5}) \cite{Touchette20091}.

~

One of the most useful features of the cumulant approach is how it combines with the contraction principle. Consider a vector $a$ and a non-injective matrix $f$. The generating function of the cumulants of a contracted observable $b=f\cdot a$ is given by
\begin{equation}
\tilde{E}(\tilde{\mu})=\max_b[\tilde{\mu}\cdot b-\tilde{g}(b)]=\max_b[\tilde{\mu} \cdot b-\min\limits_{f\cdot a=b}[g(a)]]=\max_a[\tilde{\mu} \cdot f\cdot a-g(a)]=E(\tilde{\mu}\cdot f)
\end{equation}
which is to say that the function $\tilde{E}$ is in fact the function $E$ applied to a variable $\tilde{\mu}\cdot f$ which has fewer degrees of freedom than $\mu$ (because $\tilde{\mu}$ is conjugate to $b$ which has fewer degrees of freedom than $a$). In other words, contracting at the level of cumulants reduces to taking special values of $\mu$, which is often much easier than finding the minimum in eq.(\ref{I-1-contraction}).

~

One last remark we should make is that ${\rm P}_N(a)$ has a sub-exponential pre-factor which we may (and did) neglect entirely, as long as it has no poles in $a$. In the case that it does, all the saddle-point approximations that we performed have to be modified to take into account contour integrals around these poles, which may dominate the large $N$ limit \cite{Touchette2010}.

\subsection{Dynamical large deviations for Markov processes}
\label{I-2}

We will now see what can be said of large deviations in the context of continuous time Markov processes on a finite state space.

\subsubsection{Definitions}
\label{I-2-1}

Consider a Markov matrix $M$ acting on states $\{\cal C\}$, with rates $w({\cal C}',{\cal C})$ from $\cal C$ to $\cal C'$, and escape rates $r({\cal C})=\sum\limits_{{\cal C}'} w({\cal C}',{\cal C})$:
\begin{equation}\label{I-2-M}
M=\sum\limits_{\mathcal{C}\neq\mathcal{C}'}w(\mathcal{C},\mathcal{C}')|\mathcal{C}\rangle\langle\mathcal{C}'|-\sum\limits_{\mathcal{C}}r(\mathcal{C})|\mathcal{C}\rangle\langle\mathcal{C}|.
\end{equation}

This defines the time evolution of a probability vector $|P_{t}\rangle$ through the master equation
\begin{equation}\label{I-2-MP}
\frac{d}{dt}|P_{t}\rangle=M|P_{t}\rangle
\end{equation}
of which the solution is, formally:
\begin{equation}\label{I-2-eMP}
|P_{t}\rangle={\rm e}^{t M}|P_{0}\rangle
\end{equation}
where $|P_{0}\rangle$ is the initial probability distribution. In the limit of long times, if $M$ is not reducible, this vector converges to a steady state
\begin{equation}\label{I-2-Pstar}
\lim\limits_{t\rightarrow\infty} |P_t\rangle=|P^{\star}\rangle~~~~{\rm with}~~~~M|P^{\star}\rangle=0.
\end{equation}

Equivalently, we can define the probability density ${\rm P}\bigl({\cal C}(t)  \bigr)$ of a history ${\cal C}(t)$ going through configurations ${\cal C}_i$ with waiting times $t_i$: the Markovianity of the process tells us that the waiting times are Poisson-distributed with respect to the escape rates, which gives us
\begin{equation}
{\rm P}[\mathcal{C}(t)]={\rm e}^{-t_N r({\cal C}_N)}w({\cal C}_N,{\cal C}_{N-1})~{\rm e}^{-t_{N-1} r({\cal C}_{N-1})}\dots{\rm e}^{-t_2 r({\cal C}_2)}w({\cal C}_2,{\cal C}_1){\rm e}^{-t_1r({\cal C}_1)}
\end{equation}
where time increases from right to left. This can then be recast in a more compact form:
\begin{equation}\label{I-2-Phist}\boxed{
\log\Bigl({\rm P}[\mathcal{C}(t)]\Bigr)=-\int\limits_{t=0}^{t_f} r\bigl(C(t)\bigr){\rm d}t+\sum\limits_{i=1}^{N-1}\log{w({\cal C}_{i+1},{\cal C}_i)}.
}
\end{equation}
The entries of ${\rm e}^{t_f M}$ are then given by the sum of the probabilities of all paths that start and end at the corresponding microstates.

\subsubsection{Time-additive observables}
\label{I-2-2}

We can now define time-dependent observables on those histories and look at their large deviations in the long time limit. Let us consider an observable $A_t$ defined as a functional of a history ${\cal C}(t)$:
\begin{equation}\label{I-2-At}
A_t={\rm F}[\mathcal{C}(t)].
\end{equation}

For the sake of simplicity, we are only interested in observables that are additive in time, meaning that if a history ${\cal C}(t)$ is the concatenation of two shorter ones ${\cal C}_1(t)$ and ${\cal C}_2(t)$, which we will write as $\mathcal{C}_1(t)\oplus\mathcal{C}_2(t)$, the functional $F$ distributes over them:
\begin{equation}\label{I-2-Fadd}
{\rm F}[\mathcal{C}_1(t)\oplus\mathcal{C}_2(t)]={\rm F}[\mathcal{C}_1(t)]+{\rm F}[\mathcal{C}_2(t)].
\end{equation}
This forces $F$ to be local in time (independent of time correlations). We also assume $F$ to be time-invariant, i.e. independent on the value of the initial time $t_0$ in ${\cal C}(t)$.

These constraints allow us to find the general form of such functionals. Consider first a history without any transitions: ${\cal C}(t)={\cal C}_1$. In this case, time additivity can be used to show that $F[{\cal C}(t)]$ is proportional to the duration $t_1$ of the process. The proportionality coefficient may depend on ${\cal C}_1$, and we will call it $V({\cal C}_1)$. Consider now a history with one transition: the system is in ${\cal C}_1$ for a duration $t_1$, and in ${\cal C}_2$ for a duration $t_2$. By cutting out, using additivity, the portion of history before $t_1-\varepsilon$ and that after $t_1+\varepsilon$, with $\varepsilon$ going to $0$, one is left with just the transition, which has a contribution to $F$ that depends only on ${\cal C}_1$ and ${\cal C}_2$. We will call it $U({\cal C}_2,{\cal C}_1)$.

\begin{figure}[ht]
\begin{center}
 \includegraphics[width=0.7\textwidth]{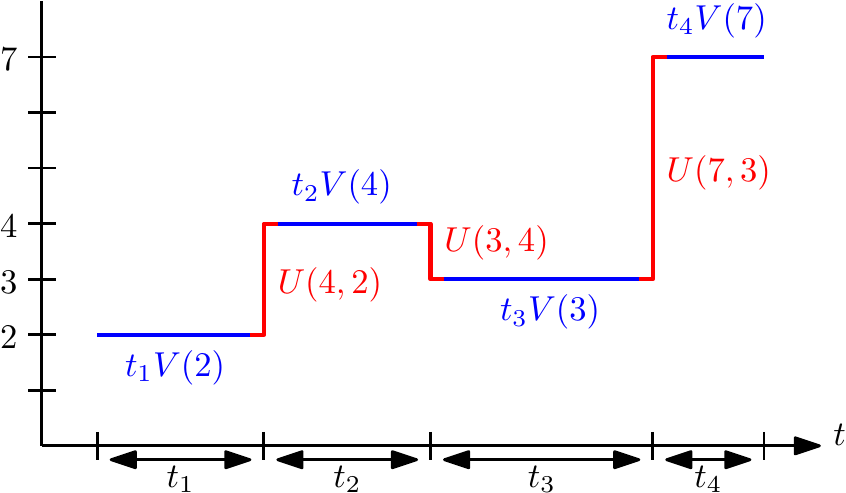}
  \caption{Functional $F$ over a schematised history. Each straight portion contributes a simple term to the whole function. Waiting periods (in blue) give a contribution that is extensive in time, and depends only on one configuration. Transitions (in red) give a term that depends on the two configurations involved.}
\label{I-fig-add}
 \end{center}
 \end{figure}

Putting those pieces together, and considering that any history can be decomposed into portions containing at most one transition, we can finally write:
\begin{equation}\label{I-2-F}\boxed{
{\rm F}[\mathcal{C}(t)]=\int_{t_0}^{t_f} V\bigl(\mathcal{C}(t)\bigr)dt+\sum\limits_{i=1}^{N}U(\mathcal{C}_{i},\mathcal{C}_{i-1})
}\end{equation}
which is expressed schematically on fig.-\ref{I-fig-add}.

The first part of that expression, containing $V$, is a {\it state observable}, depending only on the empirical vector, which is to say the relative time spent in each microstate $\cal C$. It contains no direct information on the currents between microstates, although it may depend indirectly on the system being in or out of equilibrium, through time-correlations. The second part, containing $U$, is a {\it jump observable}, and is a direct measure of those currents. Each $U(\mathcal{C}_{i},\mathcal{C}_{i-1})$ can be seen as a counter for the transition between $\mathcal{C}_{i-1}$ and $\mathcal{C}_{i}$: every time it is used in the evolution of the system, the value of $A_t$ increases by one quantum of $U$. Whether each value of $U$ is taken as an independent variable, or given a precise value, determines what information is monitored regarding the way those transitions are used, but in many cases, one thing that makes a crucial difference on the resulting behaviour of $A_t$ is whether the system is in equilibrium or not, as we will see shortly.

~~

But first, let us have a look at the cumulants of this observable. The generating function $E_t(\nu)$ of the cumulants of $a_t=\frac{1}{t}A_t$ (the intensive version of $A_t$) can be expressed as:
\begin{equation}\label{I-2-Amean}
{\rm e}^{t E(\nu)}=\langle {\rm e}^{t\nu a_t}\rangle=\int {\rm e}^{\nu{\rm F}[\mathcal{C}(t)]}{\rm P}[\mathcal{C}(t)]{\cal D}[\mathcal{C}(t)]
\end{equation}
where ${\cal D}[\mathcal{C}(t)]$ is the measure associated with histories (this can be simply defined in the discrete time case, and then taken as a formal limit for small $\delta t$).

Replacing ${\rm P}[\mathcal{C}(t)]$ by the expression in eq.(\ref{I-2-Phist}), we see that
\begin{align}\label{I-2-Phist2}
\log\bigl({\rm e}^{\nu{\rm F}[\mathcal{C}(t)]}{\rm P}[\mathcal{C}(t)]\bigr)=-\int\limits_{t=0}^{t_f} \Bigl(r\bigl(C(t)\bigr)-\nu V[C(t)]\Bigr){\rm d}t+\sum\limits_{i=1}^{N-1}\big(\log{w({\cal C}_{i+1},{\cal C}_i)}+\nu U({\cal C}_{i+1},{\cal C}_i)\big)
\end{align}
which is to say that it is the (un-normalised) weight produced by a modified Markov matrix $M_\nu$ defined as
\begin{equation}\label{I-2-MF}\boxed{
M_\nu=\sum\limits_{\mathcal{C}\neq\mathcal{C}'}{\rm e}^{\nu U(\mathcal{C},\mathcal{C}')}w(\mathcal{C},\mathcal{C}')|\mathcal{C}\rangle\langle\mathcal{C}'|-\sum\limits_{\mathcal{C}}\bigl(r(\mathcal{C})-\nu V(\mathcal{C})\bigr)|\mathcal{C}\rangle\langle\mathcal{C}|
}\end{equation}
(where we take $U(\mathcal{C},\mathcal{C})=0$). By replacing $M$ by $M_\nu$ in (\ref{I-2-eMP}), we get
\begin{equation}\label{I-2-PtF}
|P_\nu(t)\rangle={\rm e}^{t M_\nu}|P_{0}\rangle=\sum\limits_{\mathcal{C}_N}\int {\rm e}^{\nu{\rm F}[\mathcal{C}(t)]}{\rm P}[\mathcal{C}(t)]{\cal D}[\mathcal{C}(t)]~|\mathcal{C}_N\rangle
\end{equation}
where, as intended, the probabilities of histories have received an extra ${\rm e}^{\nu{\rm F}[\mathcal{C}(t)]}$ factor. We can finally sum over the final configuration ${\mathcal{C}_N}$ by projecting to the left on the uniform vector $\langle 1|$ (of which all entries are $1$) and write
\begin{equation}\label{I-2-Amean2}
{\rm e}^{t E(\nu)}=\langle 1|P_\nu(t)\rangle=\langle 1|{\rm e}^{t M_\nu}|P_{0}\rangle.
\end{equation}
Note that we use $\nu$ as a generic parameter, which can in fact be a function of the configuration (as in section \ref{I-2-3}) or of the transition.

~~

Moreover, as long as $\nu$, $V$ and $U$ are real, the Perron-Frobenius theorem applies: the largest eigenvalue $\Lambda_\nu$ of $M_\nu$ is non-degenerate. If we write the corresponding eigenvectors as $|P_\nu\rangle$ and $\langle \tilde{P}_\nu|$, we get, for large times,
\begin{equation}\label{I-2-MPF}
{\rm e}^{t M_\nu} \approx {\rm e}^{t \Lambda_\nu}|P_\nu\rangle\langle \tilde{P}_\nu|.
\end{equation}
By combining equations (\ref{I-2-Amean2}) and (\ref{I-2-MPF}) for $t$ large, we get ${\rm e}^{t E(\nu)}\approx{\rm e}^{t\Lambda_\nu } \langle 1|P_\nu\rangle\langle \tilde{P}_\nu|P_0\rangle$, which is to say that
\begin{equation}\label{I-2-MPF2}\boxed{
E(\nu)=\Lambda_\nu.
}\end{equation}

The {\it generating function of the cumulants} of any additive observable is therefore equal to the {\it largest eigenvalue of the associated deformed Markov matrix} $M_\nu$. This is a classic result from the Donsker-Varadhan theory of temporal large deviations \cite{Donsker2010,Donsker1975,Donsker1976,Donsker1983,deuschel1989large}. Notice that this is a property of the deformed matrix, and not of the initial or final configurations: regardless of those, the long time behaviour of the generating function of the cumulants will be the same. One can easily make sense of this: if the duration of the process is large enough, then the system only takes a small time at first to reach its steady state, and gets out of it very near the end. The rest of the evolution can be considered to be around the steady state, whatever the initial and final distributions are, and the part of $A_t$ which is extensive in time comes only from there. The initial distribution only gives a time-independent term $\langle 1|P_\nu\rangle\langle \tilde{P}_\nu|P_0\rangle$, which is negligible (unless, as we mentioned earlier, it has poles).

~~

The eigenvectors $|P_\nu\rangle$ and $\langle\tilde{P}_\nu|$ also carry information on current fluctuations. Considering eq.(\ref{I-2-PtF}) for $t$ large, we can regroup all the histories for which ${\rm F}[\mathcal{C}(t)]=t f$ and the final configuration is $\mathcal{C}_t$. Writing that probability ${\rm P}(f~\&~\mathcal{C}_t)$ as ${\rm P}(\mathcal{C}_t|f){\rm P}(f)$ (where ${\rm P}(A|B)$ is the probability of $A$, conditioned on $B$), and invoking the large deviations principle ${\rm P}(f)\approx {\rm e}^{-t g(f)}$, we get
\begin{equation}\label{I-2-PtF3}
{\rm e}^{t E(\nu)}|P_\nu\rangle\approx\sum\limits_{\mathcal{C}_t}|\mathcal{C}_t\rangle\int {\rm P}(\mathcal{C}_t|f){\rm e}^{t(\nu  f-g(f))}{\rm d}f.
\end{equation}
Finally, just as in eq.(\ref{I-1-Eg}), a saddle-point approximation on ${\rm e}^{t(\nu  f-g(f))}$ yields ${\rm e}^{t E(\nu)}$ and fixes the value of $f$ to $\frac{{\rm d}}{{\rm d}\nu} E(\nu)$. Injecting this in (\ref{I-2-PtF3}), we get
\begin{equation}\label{I-2-PtF4}
P_\nu({\mathcal C})={\rm P}\Bigl({\mathcal C}_t={\mathcal C}~\Big|~f\!=\!\frac{{\rm d}}{{\rm d}\nu} E(\nu)\Bigr).
\end{equation}
This tells us that the vector $|P_\nu\rangle$ is in fact the probability vector of the final configuration, knowing that the value of $f$ through the evolution of the system was $\frac{{\rm d}}{{\rm d}\nu} E(\nu)$ \cite{Nemoto2014a}. A similar calculation on the left eigenvector $\langle\tilde{P}_\nu|$ shows it to be the probability vector of what the initial configuration was, knowing that the value of $f$ is $\frac{{\rm d}}{{\rm d}\nu} E(\nu)$:
\begin{equation}\label{I-2-PtF4}
\tilde{P}_\nu({\mathcal C})={\rm P}\Bigl({\mathcal C}_0={\mathcal C}~\Big|~f\!=\!\frac{{\rm d}}{{\rm d}\nu} E(\nu)\Bigr)
\end{equation}
Finally, the product of the two gives the probability of observing a configuration at any point during the evolution of the system (but far enough from the initial and final time), conditioned on the value of $f$:
\begin{equation}\label{I-2-sensemble}\boxed{
P_\nu({\mathcal C})\tilde{P}_\nu({\mathcal C})={\rm P}\Bigl({\mathcal C}~\Big|~f\!=\!\frac{{\rm d}}{{\rm d}\nu} E(\nu)\Bigr).
}\end{equation}
Note that these vectors are implicitly normalised appropriately.

Those three distributions are quite different from one another, and in particular their most probable states are in general not the same. We will be mostly interested in eq.(\ref{I-2-sensemble}), for reasons that will become apparent later, in section \ref{III}. When the observable of interest is the entropy production (which we will examine in section \ref{I-2-4}), the statistical ensemble defined by those probabilities, where $\nu$ is considered as a free parameter (similar to a temperature, as it is conjugate to the entropy), is sometimes called the `s-ensemble' \cite{jack2010large}. It is quite natural to consider this ensemble: since entropy production plays an important part in the system being out of equilibrium, being able to control its value and look at how the system responds provides us with useful information on its behaviour. Moreover, much like in equilibrium, where one can think of a Lagrange multiplier as an extra interaction, one may think of $M_\mu$ as equivalent to a process with modified dynamics, which has typical values for observables given by the fluctuation values in the original process \cite{Chetrite2014,Chetrite2015}. The Markov matrix for that new process can be obtained through a so-called `Doob transform', which consists in shifting its eigenvalues by $-E(\nu)$ and conjugating the matrix through $\tilde{P}_\nu$ so that its largest eigenvalue be $0$ and its dominant left eigenvector be uniform. Its steady state is then given by $P_\nu({\mathcal C})\tilde{P}_\nu({\mathcal C})$, which is another argument for that distribution being the most natural one.

In the following sub-sections, we will consider two special cases of time-additive observables: the empirical vector, which is a state observable, and the entropy production, which is a jump observable.

\subsubsection{Large deviations of the empirical vector}
\label{I-2-3}

The first special case that we will consider is that of the time-integrated empirical vector, which is the vector of the total time spent in each configuration. We will write its intensive counterpart as $\rho$, defined by
\begin{equation}
\rho(c)=\frac{1}{t}\int_{0}^{t} \delta_{c,\mathcal{C}(\tau)}d\tau
\end{equation}
which contains the fractions of time spent in each configuration $c$. We recognise a special case of eq.(\ref{I-2-F}) where $V(\mathcal{C})=\delta_{c,\mathcal{C}}$ and $U(\mathcal{C}',\mathcal{C})=0$. This is the most general state observable, which can give any other through contraction.

The deformed Markov matrix corresponding to $\rho$ is given by
\begin{equation}\label{I-2-Mh}
M_h=\sum\limits_{\mathcal{C}\neq\mathcal{C}'}w(\mathcal{C},\mathcal{C}')|\mathcal{C}\rangle\langle\mathcal{C}'|-\sum\limits_{\mathcal{C}}\bigl(r(\mathcal{C})-h_{\mathcal{C}}\bigr)|\mathcal{C}\rangle\langle\mathcal{C}|=M+H
\end{equation}
where $H$ is a diagonal matrix with entries $h_{\mathcal{C}}$. The largest eigenvalue $E(h)$ of $M_h$ contains all the cumulants of the empirical vector, including n-point functions. For instance,
\begin{equation}
\frac{\rm d^2}{{\rm d}h_{\mathcal{C}_1}{\rm d}h_{\mathcal{C}_2}}E(0)=\langle \rho(\mathcal{C}_1)\rho(\mathcal{C}_2)\rangle_t-\langle \rho(\mathcal{C}_1)\rangle_t\langle \rho(\mathcal{C}_2)\rangle_t
\end{equation}
where $\langle\cdot\rangle_t$ refers to the average in time (which we will soon need to distinguish from an ensemble average).

~

There is another empirical vector $\pi$ which we may define, by looking at the probability of being at $\mathcal{C}$ at a given time $t$, averaged over a large number $N$ of copies of the process. If the probability vector at time $t$ is $|P_t\rangle$, the exponential generating function of the cumulants of $\pi$ is given by
\begin{equation}
\mathcal{E}(h)=\log\bigl(\langle 1|{\rm e}^H|P_t\rangle\bigr)
\end{equation}
which involves ensemble averages $\langle\cdot\rangle_e$.

It is important to note that these two generating functions, as well as the corresponding large deviation functions, are in principle different: fluctuations over time are not identical to fluctuations among copies of the system. They do however have one common feature. We may write the first cumulant from $E(h)$ as
\begin{equation}
\langle\rho(\mathcal{C})\rangle_t=\frac{\rm d}{{\rm d}h_{\mathcal{C}}}E(0)=\lim\limits_{t\rightarrow\infty}\frac{1}{t}\frac{\rm d}{{\rm d}h_{\mathcal{C}}}\log\bigl(\langle 1|{\rm e}^{t(M+H)}|P_0\rangle\bigr)\Big|_{h=0}=\lim\limits_{t\rightarrow\infty}\frac{1}{t}\int_{0}^{t}{\rm P}_\tau(\mathcal{C}){\rm d}\tau={\rm P}^{\star}(\mathcal{C})
\end{equation}
where we recall that ${\rm P}^{\star}$ is the distribution of the steady state of $M$. That same first cumulant from $\mathcal{E}(h)$ gives, in the limit of long times:
\begin{equation}
\langle\pi(\mathcal{C})\rangle_e=\frac{\rm d}{{\rm d}h_{\mathcal{C}}}\mathcal{E}(0)=\lim\limits_{t\rightarrow\infty}\frac{\rm d}{{\rm d}h_{\mathcal{C}}}\log\bigl(\langle 1|{\rm e}^H|P_t\rangle\bigr)\Big|_{h=0}=\frac{\rm d}{{\rm d}h_{\mathcal{C}}}\log\bigl(\langle 1|{\rm e}^H|P^\star\rangle\bigr)\Big|_{h=0}={\rm P}^{\star}(\mathcal{C}).
\end{equation}
Combining this identity with any contraction of $\rho$ or $\pi$, we conclude that {\it time averages and ensemble averages of state observables are identical}, which is to say that the Markov process we are considering is {\it ergodic}. All the other cumulants are in principle different, which can be understood by the fact that time-correlations (i.e. the dynamics of the system) play a role in $E$ which they don't in $\mathcal{E}$. This implies that the absolute minima of both large deviation functions have the same locus, but that their shape (stiffness, skewness, etc.) around those minima is different.

~

It is also interesting to note that this remains true for averages in a conditioned process. Let us for instance take a generic deformed Markov matrix $M_\mu$, with dominant eigenvectors $|P_\mu\rangle$ and $\langle\tilde{P}_\mu|$, and consider the two generating functions
\begin{equation}
E(\mu,h)=\lim\limits_{t\rightarrow\infty}\frac{1}{t}\log\bigl(\langle 1|{\rm e}^{t(M_\mu+H)}|P_0\rangle\bigr)
\end{equation}
and
\begin{equation}
\mathcal{E}(\mu,h)=\log\bigl(\langle \tilde{P}_\mu|{\rm e}^H|P_\mu\rangle\bigr)
\end{equation}
(where the product between the two eigenvectors $|P_\mu\rangle$ and $\langle\tilde{P}_\mu|$ means we are looking at a time which is somewhere in the middle of a long evolution). A calculation similar to what we did earlier yields:
\begin{equation}\boxed{
\langle\rho_\mu(\mathcal{C})\rangle_t=\frac{\rm d}{{\rm d}h_{\mathcal{C}}}E(\mu,0)={\rm P}_\mu(\mathcal{C})\tilde{{\rm P}}_\mu(\mathcal{C})=\frac{\rm d}{{\rm d}h_{\mathcal{C}}}\mathcal{E}(\mu,0)=\langle\pi_\mu(\mathcal{C})\rangle_e
}\end{equation}
where the product ${\rm P}_\mu(\mathcal{C})\tilde{{\rm P}}_\mu(\mathcal{C})$ is normalised.

\subsubsection{Large deviations of the entropy production}
\label{I-2-4}

We now consider a special jump observable: the entropy production $S_t$ for an evolution between times $0$ and $t$, defined as
\begin{equation}\label{I-2-ent1}
S_t[{\cal C}(\tau)]=\log\Biggl( \frac{{\rm P}[{\cal C}(\tau)]}{{\rm P}[{\cal C}^R(\tau)]} \Biggr),
\end{equation}
which measures how probable a history $\mathcal{C}(\tau)$ is compared to its time-reversal $\mathcal{C}^R(\tau)=\mathcal{C}(t-\tau)$. It corresponds to a time-additive observable with
\begin{equation}\label{I-2-ent}
V({\cal C})=0~~,~~U({\cal C},{\cal C}')=\log\bigl( w({\cal C}',{\cal C})\bigr)-\log \bigl(w({\cal C},{\cal C}')\bigr).
\end{equation}
so that the corresponding deformed Markov matrix is given by:
\begin{equation}\label{I-2-Ment}
M_\nu=\sum\limits_{\mathcal{C},\mathcal{C}'}w(\mathcal{C},\mathcal{C}')^{1+\nu}w(\mathcal{C}',\mathcal{C})^{-\nu}|\mathcal{C}\rangle\langle\mathcal{C}'|.
\end{equation}
We notice that it has a peculiar symmetry: if we replace $\nu$ by $-1-\nu$, the exponents in $M_\nu$ are swapped, which has the same effect as transposing the matrix. That is to say,
\begin{equation}\label{I-2-GC}\boxed{
M_\nu=~^t\!M_{-1-\nu}.
}\end{equation}
This has interesting consequences on its eigensystem: all the eigenvalues are symmetric with respect to $\nu\leftrightarrow -1-\nu$, and the associated right and left eigenvectors are exchanged. This is the famous `Gallavotti-Cohen symmetry' \cite{Lebowitz99agallavotti-cohen,Kurchan1998,PhysRevLett.74.2694}, which is one of the only universal features known for non-equilibrium systems.

In particular, the generating function of the cumulants of the intensive entropy production $s$, and the conditional probabilities that we have defined earlier, verify:
\begin{equation}\label{I-2-GC2}
E(\nu)=E(-1-\nu)~~~~,~~~~P_\nu({\cal C})=\tilde{P}_{-1-\nu}({\cal C}).
\end{equation}
which becomes, for the large deviation function,
\begin{equation}\label{I-2-GC2}
g(s)-g(-s)=-s
\end{equation}
or, equivalently,
\begin{equation}\label{I-2-GC4}
{\rm P}(-s)={\rm e}^{-t s}{\rm P}(s)
\end{equation}
for $t\rightarrow\infty$. This last equation is called the `fluctuation theorem'. It was first observed by Evans, Cohen and Morriss in \cite{PhysRevLett.71.2401}, then proven by Evans and Searles in \cite{Evans1994}, and later led to Gallavotti and Cohen's formulation of their symmetry. The theorem means that a negative entropy production rate is much less probable than its positive counterpart, but not impossible. This does not, as it might seem, contradict the second law of thermodynamics, which is expressed only for the average of $s$. It in fact validates it, since it implies that the mean value of $s$ must be positive.

~

We may finally note that, for equilibrium systems, where the detailed balance condition imposes that
\begin{equation}\label{I-2-DB}
P^\star({\cal C})w({\cal C}',{\cal C})=P^\star({\cal C}')w({\cal C},{\cal C}')
\end{equation}
for any two configurations ${\cal C}$ and ${\cal C}'$, the deformed Markov matrix turns out to be similar to the un-deformed one, so that they have the same eigenvalues. Consequently, the generating function of the cumulants of the entropy production rate is identically zero, and its large deviation function is a delta function:
\begin{equation}\label{I-2-Eeq}
E(\nu)=0~~~~{\rm and}~~~~{\rm P}(s)=\delta(s).
\end{equation}
There is therefore no entropy production whatsoever in the case of an equilibrium system. This, as in eq.(\ref{I-2-MPF2}), is a property of the transition rates, and not of the initial or final distributions. There can still be a conservative exchange of entropy between the initial and final configurations of a history (if their equilibrium probabilities are not equal).

\newpage

\section{The asymmetric simple exclusion process}
\label{II}

In this section, we are introduced to the Asymmetric Simple Exclusion Process. After giving its definition, we briefly go through the existing literature related to that model, including its many variants and connections with other problems in physics, mathematics and biology. We then examine the steady state of the model, first through a mean-field approach and then through an exact calculation.

\subsection{Definition of the model and variants}
\label{II-1}

\begin{figure}[ht]
\begin{center}
 \includegraphics[width=0.8\textwidth]{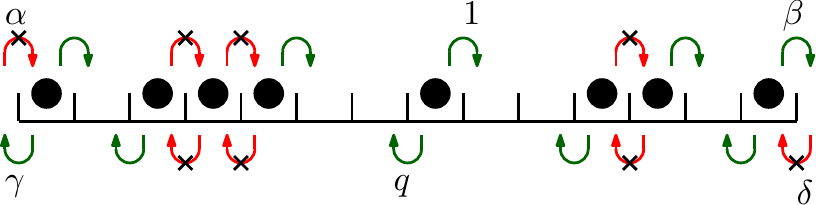}
  \caption{Dynamical rules for the ASEP  with open boundaries. The rate
 of forward jumps has been normalised to 1. Backward jumps occur with rate
 $q < 1$. All other parameters are arbitrary. The jumps shown in green are allowed by the exclusion constraint. Those shown in red and crossed out are forbidden.}
\label{II-fig-PASEP}
 \end{center}
 \end{figure}

Consider a one-dimensional lattice with $L$ sites (or a row of $L$ boxes), numbered from $1$ to $L$. Each site can be empty, or carry one particle. Those particles jump stochastically from site to site, with a rate $p$ if the jump is to the right, from site $i$ to site $i+1$ (and which will be set to $p=1$ by choosing the rate of forward jumps as a time scale), and a rate $q<1$ if the jump is to the left, from site $i$ to site $i-1$. The jumping rate is larger to the right than to the left in order to mimic the action of a field driving the particles in the bulk of the system. Each end of the system is connected to a reservoir of particles, so that they may enter the system at site $1$ with rate $\alpha$ or at site $L$ with rate $\delta$, and leave it from site $1$ with rate $\gamma$ or from site $L$ with rate $\beta$. Those rates allow us to define the effective densities of the two reservoirs. In all of these operations, the only constraint that must be obeyed is that of exclusion, which is to say that there cannot be more than one particle on a given site at a given time, so that a particle cannot jump to a site that is already occupied. These rules are represented schematically on fig.-\ref{II-fig-PASEP}.

Configurations of the system are written as strings of $0$'s and $1$'s, where $0$ indicates an empty site, and $1$ an occupied site. The Markov matrix governing that process is a sum of local jump operators, each carrying the rates of jumps over one of the bonds in the system:
\begin{equation}\label{II-1-M}
M=m_0+\sum\limits_{i=1}^{L-1}M_i +m_L
\end{equation}
with
\begin{equation}\label{II-1-M2}
m_0=\begin{bmatrix} -\alpha & \gamma \\ \alpha & -\gamma  \end{bmatrix}~,~ M_{i}=\begin{bmatrix} 0 & 0 & 0 & 0 \\ 0 & -q & 1 & 0 \\ 0 & q & -1 & 0 \\ 0 & 0 & 0 & 0 \end{bmatrix}~,~m_L=\begin{bmatrix} -\delta & \beta \\  \delta & -\beta  \end{bmatrix}.
\end{equation}
It is implied here that $m_0$ acts as written on site $1$ (and is represented in basis $\{0,1\}$ for the occupancy of the first site), and as the identity on all the other sites. Likewise, $m_L$ acts as written on site $L$, and $M_i$ on sites $i$ and $i+1$ (and is represented in basis $\{00,01,10,11\}$ for the occupancy of those two sites). Each of the non-diagonal entries represents a transition between two configurations that are one particle jump away from each other.

As we mentioned earlier, because of the asymmetry of the jumps, the particles flow to the right, which results in a macroscopic current deeply connected to the non-equilibrium nature of the system. It is a special case of time-additive observable (as presented in section \ref{I-2-2}) where $V=0$ and $U$ is taken as $1$ for the transitions where a particle jumps to the right, and $-1$ for those where one jumps to the left. We will see in section \ref{III} that it is strongly related to the entropy production (they are in fact equal, up to a constant), and we will be referring to probabilities conditioned on the current as the s-ensemble as well. All the results presented in this review are related to the characterisation of that current and its fluctuations.

\subsubsection{Variants of the ASEP}
\label{II-1-a}

There are a few simpler cases that one can consider. The first is to force the particles to jump only to the right, by taking $q=\gamma=\delta=0$. In this case, the model is called the totally asymmetric simple exclusion process (or TASEP), and we will often use it in our calculations, as its behaviour is identical to that of the ASEP for all intents and purposes, but much easier to deal with.

The second is the opposite limit, where the jumps are as probable to the left as they are to the right: $q=1$. This case is called the symmetric simple exclusion process (or SSEP). It is an example of a `boundary-driven diffusive systems' (as opposed to the ASEP, which is bulk-driven by the asymmetry, and is therefore not diffusive). Its behaviour is quite different from that of the ASEP, and we will not consider this limit in the present review.

Sitting somewhere between the SSEP and the ASEP is the weakly asymmetric simple exclusion process (or WASEP), where the asymmetry $1-q$ is taken to scale with the size of the system as $L^{-1}$. This is done in order to make the integral of the field in the bulk, which is of order $L(1-q)$, comparable with the difference of chemical potential between the reservoirs, which is a constant with respect to $L$. The ASEP and the WASEP correspond to two different ways to take the large $L$ limit in the system: in the ASEP, no rescaling is done to the driving field, so that the large size limit corresponds to a system of increasing length, with the lattice spacing remaining constant, which is relevant to model a system which is really discrete (think for instance of ribosomes on a long string of mRNA, or any other example of discrete biological transport). In the WASEP, on the contrary, the field is rescaled as $L^{-1}$, so that the large size limit corresponds to a system of fixed length, with a smaller and smaller lattice spacing, going to a continuous system when $L$ reaches infinity. We will be using the WASEP as a starting point in section \ref{V}.

~~

One can also consider different geometries for the model. Take for instance the ASEP with periodic boundary conditions, i.e. on a ring (fig.-\ref{II-fig-variants}-b). In this case, the system is not connected to any reservoir, and the number of particles is conserved. This makes it somewhat easier to deal with: the steady-state distribution is uniform, and the coordinate version of the Bethe Ansatz can be used to solve it, as we will see in section \ref{III-2-a}.

The ASEP can be defined on an infinite lattice instead (c.f. lower part of fig.-\ref{II-fig-variants}-d). In this case, there is in general no  convergence to a steady state (for generic initial conditions), and the observable of choice is instead the large time behaviour of the transient regime.

Finally, one can put more than one type of particles in the system, and consider the multispecies ASEP (fig.-\ref{II-fig-variants}-c). The exchange rates must then be defined between any two different species of particles. The simplest case to consider (and the most tractable one) is that where the types of particles are numbered, from $0$ (for holes) to $K$ (for the `fastest' particles), and where a particle of type $k$ sees all lower types $k'<k$ as holes, which is to say that the rates of exchange of two particles of types $k_1<k_2$ are $1$ for $k_2 k_1\rightarrow k_1 k_2$ and $q$ for $k_1 k_2\rightarrow k_2 k_1$ (those rates are represented on fig.-\ref{II-fig-variants}-c, where different species of particles bear different colours, and are numbered by their rank).

\subsubsection{Brief overview of the ASEP's family tree}
\label{II-1-b}

\begin{figure}[hp]
\begin{center}
 \includegraphics[width=\textwidth]{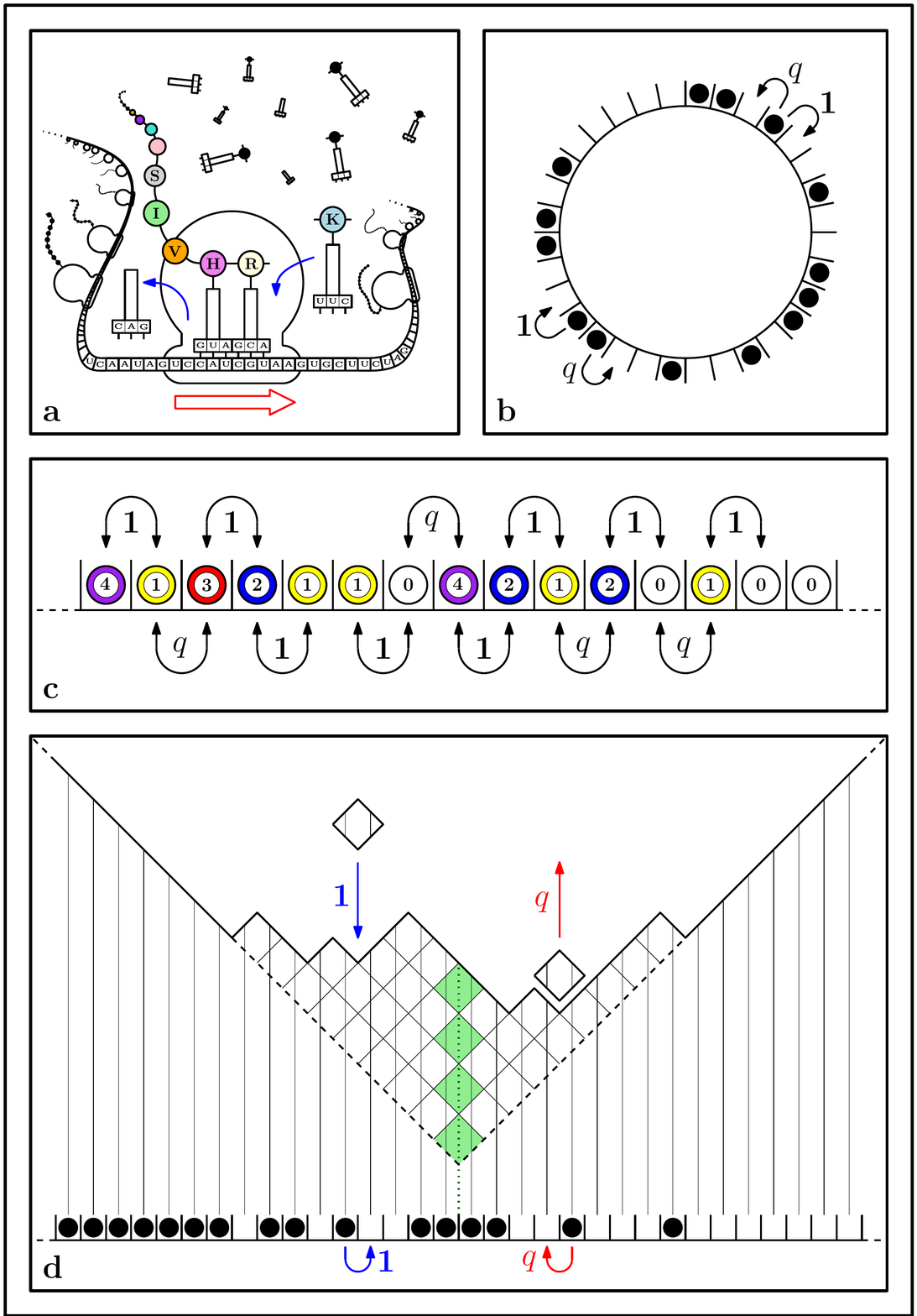}
  \caption{The ASEP's family album: a) Ribosomes on mRNA, b) Periodic ASEP, c) Multispecies ASEP, d) Surface growth.}
\label{II-fig-variants}
 \end{center}
 \end{figure}

We mentioned biological transport earlier for a good reason: the first definition of an ASEP-like model was made in 1968 in \cite{MacDonald1968,MacDonald1969} precisely in order to study the dynamics of ribosomes on mRNA (fig.-\ref{II-fig-variants}-a). It is still used today in that context, often with a few modifications to make it slightly more realistic, such as making the particle reservoirs finite \cite{1742-5468-2008-06-P06009} or even shared between several systems \cite{Greulich2012}, changing the jumping rates from site to site \cite{Greulich2008}, changing the jumping cycle by adding an inactive state for particles \cite{Ciandrini2010}, allowing them to attach or detach in the middle of the chain \cite{Reese2011}, and so on. These are only a few recent examples, but a thorough review can be found in \cite{0034-4885-74-11-116601}.

~~

It has also been noticed that the ASEP is strongly related to the XXZ spin chain with spin $\frac{1}{2}$ \cite{sandow1994partially}: the Markov matrix of the ASEP and the Hamiltonian of the spin chain are related through a matrix similarity. This fact goes deeper than a simple mapping between two systems: the XXZ spin chain is well known and well studied, as it has the mathematical property of being `integrable', meaning that it can be solved exactly, for instance through the Bethe Ansatz \cite{Faddeev1996}, and that we can expect precise analytical results from it \cite{Baxter1982}. For that reason, many results have been obtained for the ASEP by adapting the Bethe Ansatz to its formalism \cite{Prolhac2008,Prolhac2009,prolhac2010tree,Prolhac2008a,Crampe2010,crampe2011matrix,1742-5468-2006-12-P12011,de2005bethe,simon2009construction}, and even results that have been found by other means (such as those presented in \cite{derrida1993exact}) are in fact consequences of that property. The downside of this fact, one might argue, is that those methods are only transposable to other integrable systems, but the undeniable upside is that we might have access to very precise results, which could lead to discovering universal features of non-equilibrium systems.

~~

Another model related to the ASEP is that of random surface growth \cite{kardar1986dynamic} (fig.-\ref{II-fig-variants}-d). In that model, a wall, made of square blocks (with corners pointing up and down), grows by a procedure where blocks can fall in valleys at rate $1$, or lift off of peaks at rate $q$. The relation to the ASEP is rather obvious if one replaces upward slopes (when reading from left to right) by holes, and downward slopes by particles (adding a block means replacing `down up' by `up down', i.e. $10$ by $01$, and removing one is the opposite operation). In this context, the situation that is usually considered is that of the infinite ASEP, with a simple initial condition, such as a given mean density to the right of site $0$ and another one to the left (the simplest one being all $1$s to the left and all $0$s to the right \cite{johansson2000shape,tracy:095204}, as represented by the dashed line in fig.-\ref{II-fig-variants}-d), although more general ones can be considered \cite{PhysRevLett.104.230602}. One of the most interesting quantities, just as for finite size, is the total current of particles that went over the bond at the centre of the system, which is equal to the number of blocks that have been added above that site, i.e. to the height of the surface (represented in green in fig.-\ref{II-fig-variants}-d; each green block corresponds to one of the particles that crossed to the right of the system). After a first breakthrough by Johansson in \cite{johansson2000shape}, the fluctuations of that height were conjectured \cite{Prahofer2002} and then proven \cite{BenArous2011} to be related to the famous Tracy-Widom distributions governing the eigenvalues of random matrices \cite{ferrari2010interacting,sasamoto2007fluctuations}. Some even more complex quantities have been studied, such as the general n-point correlations of the height \cite{Prolhac2011}. Moreover, there is a whole class of systems, the KPZ universality class (named after Kardar, Parisi and Zhang, authors of the seminal article that started it all \cite{kardar1986dynamic}), that are governed by the same laws, such as the directed random polymer \cite{amir2011probability,Calabrese2010}, or the delta-Bose gas \cite{Imamura2011}. One can in particular recover the KPZ equation \cite{Spohn2012} from the WASEP, or an equivalent solid on solid model, through a particular rescaling \cite{Bertini1997}, as well as some of the critical exponents through renormalisation group techniques \cite{Lecomte2007a}.

One of the main features of the KPZ universality class is its dynamical exponent $z=\frac{3}{2}$: in the language of the exclusion process, the time it will take on average for a local perturbation of the density of particles, or a marker particle, to spread over a region of size $L$ scales as $t\sim L^{3/2}$, as opposed to $t\sim L^2$ for a diffusion and $t\sim L$ for a ballistic system. That property relates to the relaxation of the system at long but finite times, which we will not consider in the present work (as we will be looking at steady states, i.e. at infinite times), but we will observe the appearance of that exponent in certain places, such as in equation (\ref{IV-1-EHDMC2}).

Experimental evidence of the relevance of that model has been obtained in a liquid crystal undergoing a phase transition \cite{PhysRevLett.104.230601}. This subject has generated many more works than could be summed up here, and the reader can find more information in reviews such as \cite{corwin2012kardar,halpin1995kinetic,kriecherbauer2010pedestrian,Quastel2010}.

~~

The ASEP can be related to many more models and mathematical objects, such as chains of quantum dots \cite{Karzig2010}, alternating sign matrices \cite{Batchelor2001,DeGier2002} (through its connection to the XXZ chain), continued fractions \cite{Blythe2009}, Brownian excursions \cite{Derrida2004a,Majumdar2005,Majumdar2004}, Askey-Wilson polynomials \cite{corteel2011tableaux,sasamoto1999one,Uchiyama2004}, and a large family of combinatorial objects which all have a connection to Catalan numbers \cite{gerard2009canopy}.

\subsubsection{Earlier results}
\label{II-1-c}

All these interesting connections notwithstanding, the ASEP is a very popular model in itself \cite{Schütz20011,Derrida199865,1742-5468-2007-07-P07023} (it has even been referred to as the Ising model of non-equilibrium systems \cite{Schmittmann19953}), and has been the subject of a tremendous number of works.

The SSEP, for one, has established itself as an archetype of diffusive systems with interactions, for which many universal results have been found, such as the cumulants of the current in a periodic system \cite{appert2008universal} or an open one \cite{PhysRevLett.92.180601}. Those results all have to do with the so-called `macroscopic fluctuation theory' (or MFT) \cite{Bertini2002,Bertini2007,bertini2005current,Bertini2006}, developed to deal with the fluctuations of diffusive systems through a hydrodynamic approach \cite{spohn1991large}. As for results more specific to the SSEP or the WASEP, the large deviation functional of the density profiles was expressed in \cite{Enaud2004}, leading to the joint large deviations functional for the current and the density \cite{Bodineau2006} which we will be using in section \ref{V}. The cumulants of the current for the open SSEP were found in \cite{Derrida2004}, and were observed to depend on a single variable and not on the two boundary densities independently. This lead to the discovery of a surprising symmetry connecting the non-equilibrium SSEP (with different reservoir densities) to a system at equilibrium \cite{Tailleur2007,Tailleur2008,Lecomte2010}. The full cumulants of the current for the periodic WASEP were found in \cite{Prolhac2009a}. In that case, as was found in \cite{PhysRevE.72.066110} and further analysed in \cite{Simon2011,Belitsky2013}, the system undergoes a phase transition in the s-ensemble where, for a low enough current, the optimal density profiles become time-dependent. A similar transition can be found for the activity (the sum of jumps, regardless of direction) in the SSEP \cite{Lecomte2012}, or even for combinations of current and activity \cite{Jack2015}. Recently, the large deviations of the position of a tracer in the one-dimensional single file diffusion (a continuous equivalent to the SSEP) using MFT \cite{Krapivsky,Mallick}. See \cite{1742-5468-2007-07-P07023} for a review of some of these results.

~~

The periodic ASEP, with its fixed number of particles and its trivial steady state (all the configurations are equally probable, as long as they have the correct number of particles), has mostly been studied for the fluctuations of the current. The full generating function of those was found for the TASEP in 1998 \cite{derrida1998exact,derrida1999universal}, and although the second cumulant for the ASEP was found prior to that \cite{Derrida1997}, the complete generating function was only obtained more than 10 years later \cite{Prolhac2008,Prolhac2009,prolhac2010tree,Prolhac2008a}. Some other results were obtained for the periodic TASEP, such as the gap (i.e. the characteristic time of the transient regime) \cite{Golinelli2004,Golinelli2005}, and, very recently, the whole distribution of the spectrum of the Markov matrix and the behaviour of the low excitations \cite{Prolhac2013,Prolhac2014a,Prolhac2015}. The s-ensemble was also investigated, for the limit of very large currents, and the probabilities of the configurations were found to be those of a Dyson-Gaudin gas (the discrete analogue of a Coulomb gas) \cite{Popkov2010}. We will come back to that last observation in section \ref{IV-2}.

~~

The open ASEP is richer than the periodic case, but much harder to handle. The structure of the steady state itself is quite intricate: it was first found in \cite{derrida1992exact} for the TASEP thanks to some surprising recurrence relations between the weights of the configurations for successive sizes. It was then generalised to the ASEP by expressing those relations in algebraic form \cite{derrida1993exact}, giving birth to the `matrix Ansatz', which we will present in section \ref{II-2-b}. Depending on the values of the two reservoir densities, the system can find itself in three different phases, which was discovered for the TASEP in \cite{krug1991boundary} as an interesting feature of non-equilibrium systems (since, for equilibrium systems with short range interactions, transitions cannot be induced by boundaries). This phase diagram was refined in \cite{Schutz1993} where sub-phases were found with different correlation lengths. Those results were extended to the ASEP in \cite{Sasamoto1999,sasamoto1999one} (part of which we present in section \ref{II-2-b}). The 2-point correlation function \cite{Derrida1993a} and then the complete n-point function \cite{Derrida2004a} were calculated for the TASEP, for some values of the boundary densities, and the same was later done for the ASEP in \cite{Uchiyama2004,uchiyama2005correlation}. Most of these results rely on the matrix Ansatz, and a review of those results and methods can be found in \cite{1751-8121-40-46-R01}. See also \cite{Derrida199865} for a review of various results for the steady state of the open ASEP. More recently, a matrix Ansatz was found for the steady state of the two-species open ASEP \cite{Crampe}.

Other properties of the steady state were analysed, such as the static density, current and activity distributions \cite{Queiroz2012,depken2005exact}, the large deviation function of the density profiles \cite{derrida2002exa}, or the reverse bias regime (where the boundaries impose a current floving to the left) \cite{DeGier2011,Blythe2000}. A hydrodynamic description, named `domain wall theory' (or DWT) \cite{Dudzinski2000,Kolomeisky1998,Santen2002,spohn1991large,Varadhan1996} where states of the system are approximated by regions of constant density separated by discontinuities called shocks, was proposed to describe the large scale dynamics of the system, even in the transient regime, and the hydrodynamic quasi-potential was obtained along with the optimal relaxation pathways in \cite{Bahadoran}, but no full equivalent of the MFT has yet been devised.

One of the main reasons why the open ASEP is more difficult to study than its circular sibling is that the Bethe Ansatz cannot be used as easily in this case. The coordinate version of the Ansatz which we presented in section \ref{II-2-a}, where the particles are treated as plane waves, relies on the number of particles being fixed, and breaks down in the open case. Variants of the coordinate Bethe Ansatz have been used successfully to build excited eigenstates of the system for some special cases of the boundary parameters \cite{Crampe2010,crampe2011matrix,simon2009construction} (in particular with triangular boundary matrices), and, in conjunction with numerical analysis, to find the relaxation speed of the system (i.e. the gap of the Markov matrix) \cite{1742-5468-2006-12-P12011,de2005bethe,Proeme2011}, as well as the asymptotic large deviation function of the current inside of the Gaussian phases \cite{PhysRevLett.107.010602}. A generalisation of the matrix Ansatz was used in \cite{derrida1995exact} to calculate the second cumulant of the current in the totally asymmetric case. In \cite{Lazarescu2014}, an alternative to the Bethe Ansatz, named the `Q-operator' method, is used to obtain an expression for the complete generating function of the cumulants of the current of the open ASEP, with the exact same structure as for the periodic case \cite{prolhac2010tree}, proving a conjecture previously emitted in \cite{gorissen2012exact}. We will give a brief account of that method in section \ref{III-2-b}.

~~

Many variants of the ASEP have also been studied. A matrix Ansatz, akin to the one we mentioned before, was found for the steady state of the periodic multispecies ASEP \cite{Evans2009,prolhac2009matrix}. The case of a single defect particle was analysed, for itself \cite{Boutillier2002,derrida1999bethe,Sasamoto2000} or used as a way to mark the position of a shock \cite{Derrida1993,Mallick1996}. Different updates procedures were considered and compared for the discrete time case \cite{evans1999exact,rajewsky1998asymmetric}. A system with two interacting chains was studied in \cite{Evans2011}. The ASEP was also considered with entry and exit of particles in the bulk of the system \cite{Evans2003}, disordered \cite{Harris2004} or smoothly varying \cite{Stinchcombe2011} jumping rates, a single slow bond \cite{Janowsky1994,Janowsky1992}, repulsive nearest-neighbour interactions \cite{Popkov1999}, or on a two-dimensional grid \cite{PhysRevB.28.1655,Carlos2015}.

Finally, on the numerical front, the ASEP has been used to develop and test numerical algorithms aimed at producing and analysing rare events, such as a variant of the `density matrix renormalisation group' (DMRG) algorithm, first in \cite{Hieida1998} and later in \cite{Gorissen2009,1751-8121-44-11-115005}, as well as numerical implementations of the MFT \cite{Bunin2012,Bunin2012a,Bunin2013}, and the so-called `cloning algorithm' \cite{giardina2011simulating,giardina2006direct,Lecomte2007,Tailleur2009}.

\subsection{Steady state of the open ASEP}
\label{II-2}

Before we look into the fluctuations of the current in the ASEP, let us first see what can be said of the average current and of the steady state of the model. We will start with a simple mean field approach, which gives good results in the large size limit, and we will then present the exact expressions of these quantities, in therms of the so-called `matrix Ansatz' \cite{derrida1993exact}.

\subsubsection{Mean field}
\label{II-2-a}

As we recall, the master equation reads:
\begin{equation}\label{II-2-MP}
\frac{d}{dt}|P_{t}\rangle=M|P_{t}\rangle
\end{equation}
where $M$ is the Markov matrix of the open ASEP:
\begin{equation}\label{II-2-M}
M=m_0+\sum\limits_{i=1}^{L-1}M_i +m_L
\end{equation}
with
\begin{equation}\label{II-2-M2}
m_0=\begin{bmatrix} -\alpha & \gamma \\ \alpha & -\gamma  \end{bmatrix}~,~ M_{i}=\begin{bmatrix} 0 & 0 & 0 & 0 \\ 0 & -q & 1 & 0 \\ 0 & q & -1 & 0 \\ 0 & 0 & 0 & 0 \end{bmatrix}~,~m_L=\begin{bmatrix} -\delta & \beta \\  \delta & -\beta  \end{bmatrix}.
\end{equation}
We shall write the configurations of the system as ${\cal C}=\{n_i\}_{i:1..L}$, where $n_i\in\{0,1\}$ is the occupancy of site $i$. If we trace equation (\ref{II-2-MP}) over all $n_j$'s except for one at site $i$ which is taken to be $1$ (which means projecting it onto $\langle 1|\delta_{n_i,1}$), we get an equation for the time evolution of the mean density at that site:
\begin{equation}\label{II-1-nMP}
\frac{d}{dt}\langle n_i\rangle=\langle 1|\delta_{n_i,1}M|P_{t}\rangle.
\end{equation}
The term $\delta_{n_i,1}$ only affects two matrices from the sum in eq.(\ref{II-2-M}), and each of the matrices is stochastic individually, so that only a few terms remain in the right-hand side. A straightforward calculation yields the following expression:
\begin{equation}\label{II-1-dn}\boxed{
\frac{d}{dt}\langle n_i\rangle=J_{i-1}-J_i
}\end{equation}
with
\begin{align}
J_0&=\alpha\langle(1-n_1)\rangle-\gamma\langle n_1\rangle,\label{II-1-Ji1}\\
J_i&=\langle n_{i-1}(1-n_i)\rangle-q\langle(1-n_{i-1})n_i\rangle,\label{II-1-Ji2}\\
J_L&=\beta\langle n_L\rangle-\delta\langle(1-n_L)\rangle.\label{II-1-Ji3}
\end{align}
Equation (\ref{II-1-dn}) shows, as expected, that matter is conserved in the system: the variation of density at a site is equal to the current coming from the left minus the current leaving to the right. No approximation has been made so far, but we cannot solve these equations as they are: each current involves two-point correlations, so that the evolution of the densities $\langle n_i\rangle$ is not autonomous. In principle, one would then have to compute, in a similar way, quantities such as $\frac{d}{dt}\langle n_in_{i+1}\rangle$, which would involve three-point correlations, and so on until one reaches the final $L$-point functions \cite{derrida1992exact,Schutz1993}.

We will not be so ambitious, and will instead do a brutal approximation which will allow us to solve everything directly: we will assume that all the local densities are uncorrelated, which is to say $\langle n_i n_{j}\rangle\sim\langle n_i\rangle\langle n_j\rangle$. The resulting system of equations is autonomous. We are interested in the steady state, for which all time derivatives are zero, so that eq.(\ref{II-1-dn}) imposes that all the currents $J_i$ are equal (in a one-dimensional system, a gradient-free current needs to be constant). Equations (\ref{II-1-Ji1}~-~\ref{II-1-Ji3}) then become
\begin{align}
J&=\alpha(1-\langle n_1\rangle)-\gamma\langle n_1\rangle\label{II-1-Jmf1}\\
&=\langle n_{i-1}\rangle(1-\langle n_i\rangle)-q\langle n_i\rangle(1-\langle n_{i-1}\rangle)\label{II-1-Jmf2}\\
&=\beta\langle n_L\rangle-\delta(1-\langle n_i\rangle)\label{II-1-Jmf3}.
\end{align}
The second of these equations gives a recursion relation between $\langle n_i\rangle$ and $\langle n_{i-1}\rangle$, which can then be used $L-1$ times to express $\langle n_L\rangle$ as a function of $\langle n_1\rangle$. Then, the first and last equations can be used to get a single equation on $J$, which fixes its value as a function of all the parameters of the system ($L$, $q$ and the four boundary rates). These calculations can be found in \cite{derrida1992exact} for the TASEP. We are only interested in the large size limit, so we will instead use a method similar to that of \cite{krug1991boundary}.

Writing $\langle n_i\rangle=\rho(\frac{i-1/2}{L})$, and taking $L$ to infinity, equation (\ref{II-1-Jmf2}) becomes
\begin{equation}\label{II-1-Jx}\boxed{
J=(1-q)\rho(1-\rho)-\frac{1+q}{2L}\nabla\rho
}\end{equation}
with
\begin{equation}
\frac{d}{dt}\rho(x)=-\nabla J=0.
\end{equation}
Note that we have rescaled time in this last equation, by a factor $L$, and that we are keeping the gradient in eq.(\ref{II-1-Jx}) even though its pre-factor goes to zero, as this will give us more information on the shape of the density profile for large sizes, and is in fact necessary to find the correct steady state.

There remains the problem of finding the correct boundary conditions from eqs.(\ref{II-1-Jmf1}) and (\ref{II-1-Jmf3}), assuming that only the boundary rates from each side influence the corresponding boundary condition. This can be done by considering the situation where the steady state is homogeneous: $\rho(x)=\rho$. Equations (\ref{II-1-Jmf1}) and (\ref{II-1-Jmf2}) then give
\begin{equation}
\alpha(1-\rho)-\gamma\rho=(1-q)\rho(1-\rho)
\end{equation}
which is solved by $\rho=\rho_a=\frac{1}{1+a}$, with
\begin{equation}\label{II-1-a}
a=\frac{1}{2\alpha}\Bigl[(1-q-\alpha+\gamma)+\sqrt{(1-q-\alpha+\gamma)^2+4\alpha\gamma}\Bigr].
\end{equation}
Doing the same at the left boundary, we get a density $\rho_b=\frac{b}{1+b}$ with:
\begin{equation}\label{II-1-b}
b=\frac{1}{2\beta}\Bigl[(1-q-\beta+\delta)+\sqrt{(1-q-\beta+\delta)^2+4\beta\delta}\Bigr].
\end{equation}
Those two densities $\rho_a$ and $\rho_b$ can be considered as the effective densities of the reservoirs to which the system is connected. We will take those densities as boundary conditions $\rho$ are $\rho(0)=\rho_a$ and $\rho(1)=\rho_b$ in all future calculations.

~

Now that the equation and the boundary conditions are set, it is time to solve it. In principle, the standard way to do this would be to propagate the left boundary condition $\rho_a$ to the left side of the system through eq.(\ref{II-1-Jx}), keeping $J$ as an unknown, and identifying the result with $\rho_b$, thus obtaining an equation for $J$ in terms of the boundary densities. Plugging this back into eq.(\ref{II-1-Jx}) and solving it then yields $\rho(x)$. It is in fact much simpler to fix $J$ instead, to find all the density profiles compatible with that value, and to determine which boundary conditions are appropriate only at the end.

\begin{figure}[ht]
\begin{center}
 \includegraphics[width=0.6\textwidth]{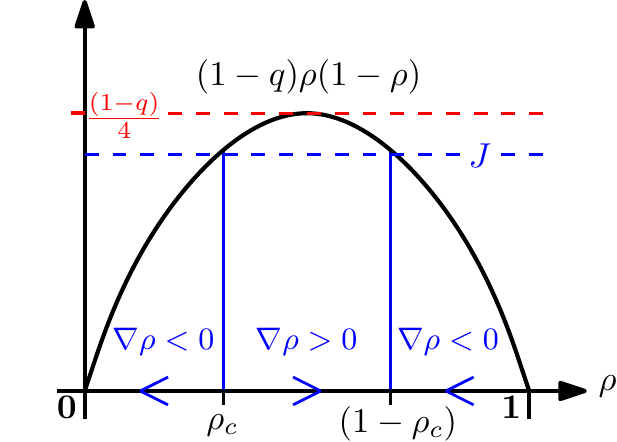}
  \caption{Variations of $\rho$ depending on its position with respect to $\rho_c$ and $1-\rho_c$. For $x$ increasing, $(1-\rho_c)$ is an attractive fixed point, and $\rho_c$ is repulsive.}
\label{fig-DeltaRho}
 \end{center}
 \end{figure}

Looking at equation (\ref{II-1-Jx}), we see that the sign of $\nabla\rho$ depends on the difference between $J$ and $(1-q)\rho(1-\rho)$. Moreover, the gradient of $\rho$ needs to be of order $L$ to compensate for any finite difference between these terms. We can therefore argue that $J$ cannot be larger than $\frac{1-q}{4}$ (which is the maximal value taken by $(1-q)\rho(1-\rho)$) or smaller than $0$, otherwise $|\nabla\rho|$ would be larger than some constant of order $L$, and $\rho$ would diverge. This means that there is a density $\rho_c\leq 1/2$ such that $J=(1-q)\rho_c(1-\rho_c)$. We then have (fig.-\ref{fig-DeltaRho}):
\begin{align}
&\nabla\rho<0~~~~{\rm for}~~~~\rho<\rho_c,\\
&\nabla\rho>0~~~~{\rm for}~~~~\rho_c<\rho<(1-\rho_c),\\
&\nabla\rho<0~~~~{\rm for}~~~~(1-\rho_c)<\rho,
\end{align}
which is to say that $\rho$ gets away from $\rho_c$ and closer to $(1-\rho_c)$.

\begin{figure}[ht]
\begin{center}
 \includegraphics[width=0.78\textwidth]{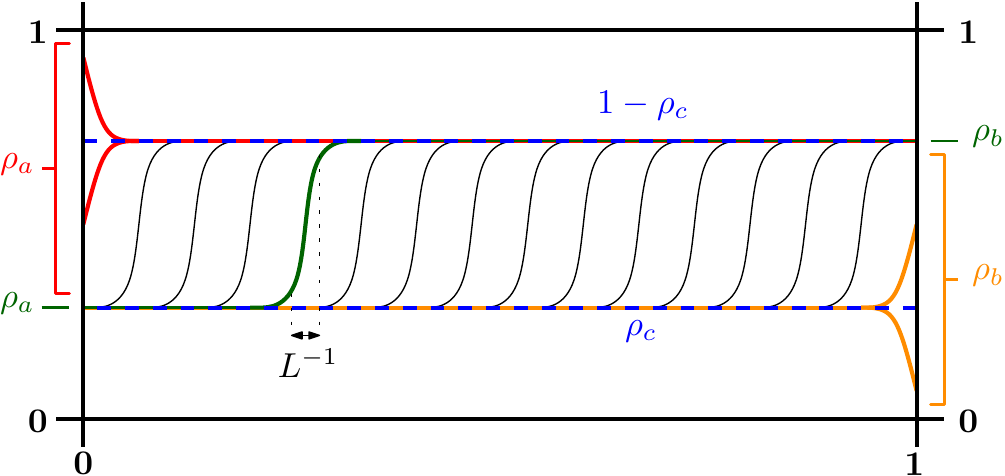}
  \caption{Possible density profiles for a given $\rho_c$. All the profiles with $\rho_a$ in the red region converge to $\rho_b=1-\rho_c$. All the profiles with $\rho_b$ in the orange region come from $\rho_a=\rho_c$.}
\label{fig-Flow}
 \end{center}
 \end{figure}

Moreover, it is straightforward to check that $\rho(x)$ is a hyperbolic tangent between $\rho_c$ and $1-\rho_c$, and approximately an exponential otherwise, with a scale $\frac{1}{L}$. On fig.-\ref{fig-Flow}, we draw some of the possible profiles for a certain value of $J<\frac{1-q}{4}$.

Since $\rho_c$ is repulsive and $1-\rho_c$ attractive, there are in fact only a few possibilities:

\begin{itemize}

\item $\rho_a=\rho_c$ and $\rho_b<1-\rho_c$, so that $J=(1-q)\rho_a(1-\rho_a)$ ; this is represented in orange on fig.-\ref{fig-Flow}, and requires $\rho_a<\frac{1}{2}$ and $\rho_a<1-\rho_b$ ; this is called the Low Density phase: the left boundary imposes its (low) density to the whole system, apart from an exponential boundary layer at the right boundary.

\item $\rho_b=1-\rho_c$ and $\rho_a>\rho_c$, so that $J=(1-q)\rho_b(1-\rho_b)$ ; this is represented in red on fig.-\ref{fig-Flow}, and requires $\rho_b>\frac{1}{2}$ and $\rho_a>1-\rho_b$ ; this is called the High Density phase: the right boundary imposes its (high) density to the whole system, apart from an exponential boundary layer at the left boundary ; all this can be obtained from the Low Density phase through a left$\leftrightarrow$right and particle$\leftrightarrow$hole symmetry.

\item $\rho_a=\rho_c$ and $\rho_b=1-\rho_c$, so that $J=(1-q)\rho_a(1-\rho_a)=(1-q)\rho_b(1-\rho_b)$ ; this is represented in green on fig.-\ref{fig-Flow}, and requires $\rho_a=1-\rho_b<\frac{1}{2}$ ; this is called the Shock Line, since the domain wall going from $\rho_c$ to $1-\rho_c$ is a shock, which can be positioned anywhere in the system ; the steady state is in fact a superposition of all possible shock positions, and the average density is linear from $\rho_a$ to $\rho_b$.

\item The only situation which is not accounted for by these three cases is $\rho_a>\frac{1}{2}$ and $\rho_b<\frac{1}{2}$ ; since $\rho$ cannot decrease between $\rho_c$ and $1-\rho_c$, this requires $\rho_c=1-\rho_c=\frac{1}{2}$, so that $J=\frac{1-q}{4}$ ; this being the largest current allowed, this is called the Maximal Current phase ; the density is close to $\frac{1}{2}$ in the whole system, apart from {\it algebraic} boundary layers on both sides.

\end{itemize}

\begin{figure}[ht]
\begin{center}
 \includegraphics[width=0.6\textwidth]{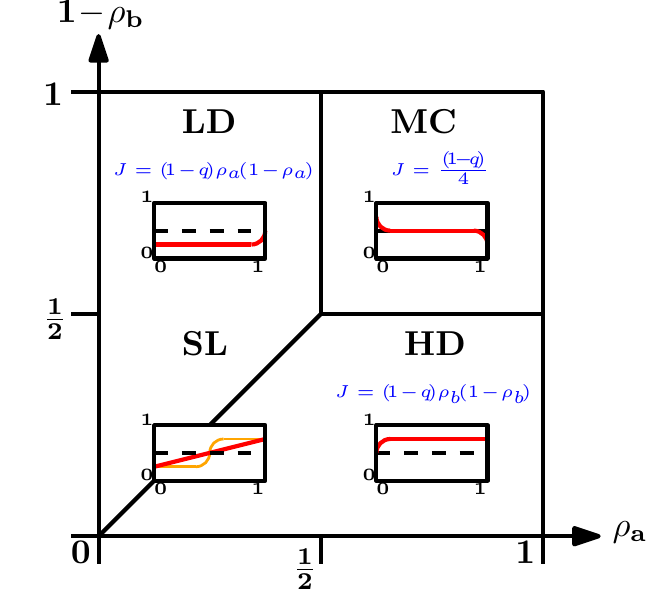}
  \caption{Phase diagram of the open ASEP. The values of the mean current in each phase are given in blue, and the mean density profiles are represented in the insets.}
\label{fig-Diag}
 \end{center}
 \end{figure}

We can summarise this by drawing the phase diagram of the system (fig.-\ref{fig-Diag}). The transitions between the MC phase and the HD and LD phases are continuous in both the current and the density profiles. The transition over the SL, however, is discontinuous in the profiles (the mean density goes from $\rho_c$ to $1-\rho_c$), but still continuous in the current.

\subsubsection{Matrix Ansatz}
\label{II-2-b}

In this section, we present the exact form of the steady state which we approximated in the previous one. It is formulated in terms of the famous matrix Ansatz, devised by Derrida, Evans, Hakim and Pasquier in \cite{derrida1993exact}, using the recursion relations found by Derrida, Domany and Mukamel in \cite{derrida1992exact} for the TASEP. Since the techniques involved here are rather specific to the ASEP, as they are related to it being integrable (although matrix Ansatze have been used for models which are not known to be integrable, such as the ABC model \cite{Evans1998}), we will be much less precise here, and will merely give the main results as well as a few indications on the methods appropriate to approach them. For more details, one may refer to \cite{derrida1993exact} or to section II.2.1 of \cite{Lazarescu2013}.

The statement of the matrix Ansatz is the following: for any configuration $\mathcal{C}=\{n_i\}$, with $n_i\in \{0,1\}$, the steady state probability $P^\star({\mathcal C})$ can be written in terms of a product of $L$ matrices which may take two values $D$ and $E$, sandwiched between two vectors $\langle\!\langle W|\!|$ and $|\!|V\rangle\!\rangle$, up to a normalisation (those vectors are written using double bras and kets to distinguish the inner space of $D$ and $E$ from the physical space on which $M$ acts). The product of matrices corresponding to $\mathcal{C}$ is obtained by multiplying matrices $D$ for each particle, and $E$ for each hole, in the same order as they appear in the configuration. In other terms:
\begin{equation}\label{II-2-Pstar}
P^\star({\mathcal C}) = \frac{1}{Z_L}\langle\!\langle W|\!|\prod_{i=1}^L\left(n_i D+(1-n_i)E\right)|\!|V\rangle\!\rangle
\end{equation}
with the normalisation factor equal to the sum of all possible products, which is to say
\begin{equation}
Z_L=\langle\!\langle W|\!|(D+E)^L|\!|V\rangle\!\rangle
\end{equation}
so that $\sum\limits_{{\cal C}}P^\star({\mathcal C})=1$. For instance, the stationary probability of configuration $101101$ for a system of size $6$ is given by $\frac{\langle\!\langle W|\!|DEDDED|\!|V\rangle\!\rangle}{\langle\!\langle W|\!|(D+E)^6|\!|V\rangle\!\rangle}$.

~

These matrices and vectors are of course not arbitrary, but must verify the following conditions:
\begin{empheq}[box=\fbox]{align}
\langle\!\langle W|\!|(\alpha E-\gamma D) &= (1-q )\langle\!\langle W|\!| \label{II-2-DEb},\\
DE-q~ED &= (1-q)\left(D+E\right) \label{II-2-DEa},\\
(\beta D-\delta E)|\!|V\rangle\!\rangle &= (1-q)|\!|V\rangle\!\rangle\label{II-2-DEc}.
\end{empheq} 
Given these, it is rather straightforward to check that $M|P^\star\rangle=0$.

We can obtain the stationary current from any of the equations (\ref{II-1-Ji1}~-~\ref{II-1-Ji3}), in combination with the corresponding equation from (\ref{II-2-DEb}~-~\ref{II-2-DEb}). For instance, equation (\ref{II-1-Ji2}) writes
\begin{equation}
J= \frac{1}{Z_L}\langle\!\langle W|\!| (D+E)^{i-2} DE (D+E)^{L-i} |\!|V\rangle\!\rangle-q\frac{1}{Z_L}\langle\!\langle W|\!| (D+E)^{i-2} ED(D+E)^{L-i}  |\!|V\rangle\!\rangle=(1-q)\frac{Z_{L-1}}{Z_L}
\end{equation}
using equation (\ref{II-2-DEa}) to simplify the expression. In all cases, we obtain
\begin{equation}\label{II-2-J}\boxed{
J=(1-q)\frac{Z_{L-1}}{Z_L}
}\end{equation}
and we are left having only to calculate $Z_L$ for any $L$.

~

 This calculation was first done in \cite{sasamoto1999one}, while the simpler equivalent for the TASEP can be found in \cite{derrida1993exact}. Since $Z_L$ is a projection of $(D+E)^L$, the natural procedure is to diagonalise $(D+E)$. Defining two new matrices $d$ and $e$ such that $D=1+d$ and $E=1+e$, the bulk algebra (\ref{II-2-DEa}) becomes 
\begin{equation}
de-q~ed=(1-q)
\end{equation}
which is that of a $q$-deformed harmonic oscillator \cite{chaichian1996introduction}, where $d$ is the annihilation operator, and $e$ the creation operator. The eigenstates of $(D+E)$ can then be found to be $q$-deformed coherent states, with a unitary complex parameter $z$. One can then write the corresponding representation of the identity into $Z_L$, and obtain, after a few lines of calculations,
\begin{equation}\label{II-2-ZLint}\boxed{
Z_L= \frac{1}{2}\oint_{S} \frac{dz}{2 i \pi z}\frac{(1+z)^L(1+z^{-1})^L(z^2,z^{-2})_{\infty}}{ (a z,a/z,\tilde{a}z,\tilde{a}/z,b z,b/z,\tilde{b}z,\tilde{b}/z)_{\infty}},
}\end{equation}
where $(\cdot)_\infty$ is the $q$-Pochhammer symbol
\begin{equation}
(x)_{\infty}=\prod_{k=0}^{\infty}(1-q^k x)
\end{equation}
with the notation convention $(x,y)_{\infty}=(x)_{\infty}(y)_{\infty}$. The two parameters $\tilde{a}$ and $\tilde{b}$ are defined similarly to $a$ and $b$, but with a {\it minus} sign in front of the square roots, and have absolutely no importance in any of the calculations that we will do.

For $a<1$ and $b<1$, the domain of integration is the unit circle, which is to say that we take the residues at all the poles of the integrand which are in $S=\{0;aq^k,\tilde{a}q^k,bq^k,\tilde{b}q^k\}_{k\in\mathbb{N}}$, and not at the other ones (which are their inverses). Since $Z_L$ is analytic in all the parameters, the poles of $F$ at which we have to take a residue are always the same, even if one of them leaves the unit circle. For that reason, the integral in (\ref{II-2-ZLint}) must be done around $S$ rather than the unit circle.

We may rapidly remark on the origin of each term in that expression: $(1+z)^L(1+z^{-1})^L$ is the eigenvalue of $(D+E)$, to the power $L$ ; $(z^2,z^{-2})_{\infty}$ comes from the normalisation of the eigenvectors of $(D+E)$ ; $(a z,a/z,\tilde{a}z,\tilde{a}/z)_{\infty}$ and $(b z,b/z,\tilde{b}z,\tilde{b}/z)_{\infty}$ come, respectively, from the scalar product between $\langle\!\langle W|\!|$ and the right eigenvector of $(D+E)$, and between $|\!|V\rangle\!\rangle$ and  the left eigenvector of $(D+E)$.

~

\begin{figure}[ht]
\begin{center}
\includegraphics[width=0.6\textwidth]{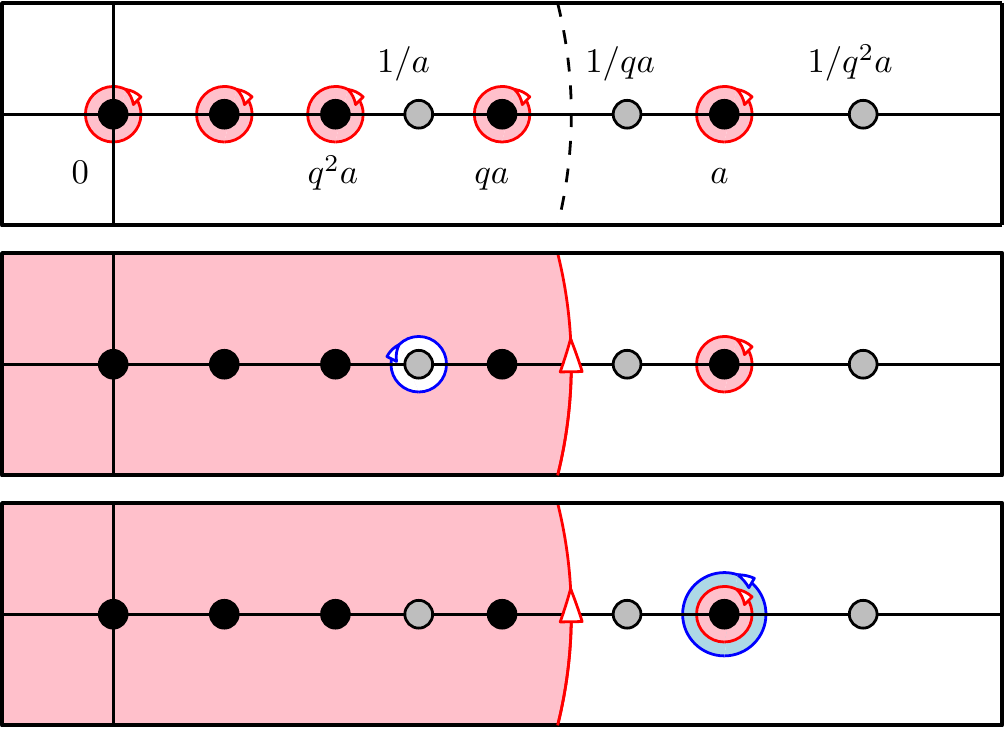}
\caption{Only the pole at $0$ and those related to $a$ are represented~; the dashed line in the first figure is part of the unit circle~; the three sets of contours are equivalent.}
\label{IV-fig-int}
\end{center}
\end{figure}

We can finally take $L$ to infinity in $Z_L$, to obtain the average current for large sizes. First of all, since the contour integral is done around an infinite number of poles, we will keep things as simple as possible: for any finite value of $a$ and $b$, $S$ contains all the poles inside the unit circle, plus a few poles outside of it (those for which $aq^k>1$ or $bq^k>1$), minus the inverse of those poles. Because of the symmetry of the integrand, the poles to not be taken in the unit circle have the opposite residue of those to be taken out of the unit circle, so that, all in all, the integral can be written around the unit circle, plus twice the residues around every pole of the form $aq^k>1$ or $bq^k>1$ outside of the circle. The poles related to $\tilde{a}$ and $\tilde{b}$ always stay inside of the unit circle, which is why they do not matter. This is summarised on fig.-\ref{IV-fig-int}.

To find the large $L$ behaviour of the integral, we need to note that, be it on the unit circle, or on the line $[1,+\infty[$ on which all the poles sit, the dominating part of the integrand is $(1+z)^L(1+z^{-1})^L$ which is largest on the real axis and increases with $z$. If any of the poles in $S$ are outside of the unit circle, the one which is the furthest from $1$ dominates, so that $Z_L\sim \max[(1+a)^L(1+a^{-1})^L,(1+b)^L(1+b^{-1})^L]$. Otherwise, the integral on the unit circle is dominated by the value at $z=1$ and $Z_L\sim 4^L$.

It is left to the reader, as an exercise, to replace these expressions in eq.(\ref{II-2-J}) and recover the phase diagram of $J$.

~

Using similar techniques, one can compute the average density, as well as correlation functions \cite{Derrida1993a,Derrida2004a,uchiyama2005correlation}, and validate the mean field calculations performed earlier. Note that we will treat contour integrals of the kind we just encountered in more detail in the next section, where we will see that all the cumulants of the current involve integrals of powers of the very same integrand.

\newpage

\section{Exact cumulants and large size limit}
 \label{III}

In the previous section, we saw what could be said of the average current in the open ASEP. In this section, we use the methods presented in \ref{I-2} to extend our analysis to the fluctuations of the current as well, through the computation of the generating function of its cumulants, which, as we saw, is the largest eigenvalue of a deformed Markov matrix. We will first construct that matrix, and remark on some of its symmetries, which will help simplify the problem. We will then see how to obtain exact expressions for its largest eigenvalue, using techniques relying on the integrability of the model: this will be done for the periodic TASEP, using the coordinate Bethe Ansatz \cite{derrida1998exact}, and for the open ASEP, using the Q-operator method \cite{Lazarescu2014}, at only a moderate level of detail, as these methods are entirely specific to the ASEP and are useful for other integrable models but not for other driven lattice gases. Once these exact expressions have been obtained, we will see how we can take the limit of large sizes and obtain the behaviour of the large deviation function of the current for small fluctuations.

\subsection{Deformed Markov matrix for the current}
\label{III-1}

The very first thing we need to define is precisely which current we intend to analyse. As we saw in section \ref{II-2}, we can define a current $j_i$ for each bond in the system. Their averages $J_i$ need to be equal in the steady state, but we can in principle choose which current to monitor, take any linear combination of them, or even keep them all as independent variables.

\begin{figure}[ht]
\begin{center}
 \includegraphics[width=0.8\textwidth]{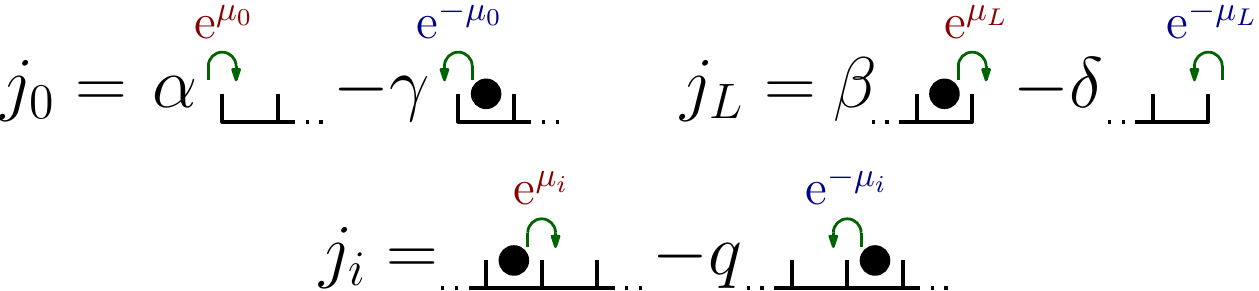}
  \caption{}
\label{III-fig-Currents}
 \end{center}
 \end{figure}

Each current defines a time-additive jump observable with $U(\mathcal{C}',\mathcal{C})=\pm 1$, in the notations of eq.(\ref{I-2-F}), if the transition increases or decreases that specific current, or $0$ if the jump is at a different location. We can then deform our Markov matrix with respect to all these currents, each with a conjugate parameter $\mu_i$ (fig.-\ref{III-fig-Currents}).

We obtain the following matrix
\begin{equation}\label{III-1-Mmu}
M_{\{\!\mu_i\!\}}=m_0(\mu_0)+\sum_{i=1}^{L-1} M_{i}(\mu_i)+m_L(\mu_l)
\end{equation}
with
\begin{equation}\label{III-1-Mmui}
m_0(\mu_0)=\begin{bmatrix} -\alpha & \gamma{\rm e}^{-\mu_0} \\ \alpha{\rm e}^{\mu_0} & -\gamma  \end{bmatrix}~,~ M_{i}(\mu_i)=\begin{bmatrix} 0 & 0 & 0 & 0 \\ 0 & -q & {\rm e}^{\mu_i} & 0 \\ 0 & q{\rm e}^{-\mu_i}& -1 & 0 \\ 0 & 0 & 0 & 0 \end{bmatrix}~,~m_L(\mu_L)=\begin{bmatrix} -\delta & \beta{\rm e}^{\mu_L} \\  \delta{\rm e}^{-\mu_L} & -\beta  \end{bmatrix}
\end{equation}
(where, as before, it is implied that $m_0$ acts as written on site $0$ in the basis $\{0,1\}$ and as the identity on the other sites, and the same goes for $m_L$ on site $L$; similarly, $M_i$ is expressed by its action on sites $i$ and $i\!+\!1$ in the basis $\{00,01,10,11\}$ and acts as the identity on the rest of the system).

The largest eigenvalue $E({\{\!\mu_i\!\}})$ of that matrix is the joint generating function of the cumulants of all the local currents, and the left and right eigenvectors carry the probabilities of configurations conditioned on the values of the integrated currents going to or coming from the steady state (as explained in section \ref{I}).

~

Before proceeding to analyse that deformed Markov matrix, we should remark on a rather useful symmetry. If one considers the diagonal matrix $R_i(\lambda)$ (with $1\leq i\leq L$) with an entry ${\rm e}^{\lambda}$ for all configurations for which site $i$ is occupied, and $1$ otherwise, one may easily check that the matrix similarity $R_i(\lambda)^{-1}M_{\{\mu_i\}}R_i(\lambda)$ simply replaces $M_{i-1}(\mu_{i-1})$ and $M_{i}(\mu_{i})$ by, respectively, $M_{i-1}(\mu_{i-1}-\lambda)$ and $M_{i}(\mu_{i}+\lambda)$, and leaves the rest of $M_{\{\mu_i\}}$ unchanged. That is to say that part of the deformation is transferred from $M_{i-1}(\mu_{i-1})$ to $M_{i}(\mu_{i})$. Using combinations of this transformations for any sites and parameters $\lambda$, we conclude that {\it all the Markov matrices deformed with respect to the currents} are similar, and therefore {\it have the same eigenvalues}, as long as {\it the sum of the deformation parameters} $\mu=\sum_{i=0}^{L}\mu_i$ {\it is fixed}. Note that the eigenvectors, however, are different, but related to each other through those simple transformations.

We will therefore write $E(\mu)$ instead of $E({\{\!\mu_i\!\}})$. Since we are mainly interested in the eigenvalues of $M_{\{\mu_i\}}$ rather than its eigenvectors, we may choose a specific combination of $\mu_i$'s to simplify our calculations. We will therefore put all the deformation on the first jump matrix $m_0(\mu)$ and leave the others un-deformed, unless specified otherwise. We will write
\begin{equation}\label{III-1-M0mu}
M_{\mu}=m_0(\mu)+\sum_{i=1}^{L-1} M_{i}+m_L.
\end{equation}

~

We may make one last remark on the deformed Markov matrix. There is a particular set of weights $\{\mu_i\}$ defined by
\begin{equation}
\{\mu_0=\nu\log{\biggl(\frac{\alpha}{\gamma}\biggr)},~~\mu_i=\nu\log{(1/q)},~~\mu_L=\nu\log{\biggl(\frac{\beta}{\delta}\biggr)} \}
\end{equation}
for which $M_{\{\mu_i\}}$ becomes:
\begin{equation}\label{MLambda}
m_0=\begin{bmatrix} -\alpha & \gamma^{1+\nu}\alpha^{-\nu} \\ \alpha^{1+\nu}\gamma^{-\nu} & -\gamma  \end{bmatrix}~,~ M_{i}=\begin{bmatrix} 0 & 0 & 0 & 0 \\ 0 & -q & q^{-\nu} & 0 \\ 0 & q^{1+\nu}& -1 & 0 \\ 0 & 0 & 0 & 0 \end{bmatrix}~,~m_L=\begin{bmatrix} -\delta & \beta^{1+\nu}\delta^{-\nu} \\  \delta^{1+\nu}\beta^{-\nu} & -\beta  \end{bmatrix}
\end{equation}
which is the deformed Markov matrix measuring the entropy production. We see immediately, as before, that
\begin{equation}
M_{-1-\nu}= ~^t\!M_\nu
\end{equation}
which implies the Gallavotti-Cohen symmetry for the eigenvalues and between the left and right eigenvectors of $M_\nu$ with respect to the transformation $\nu\leftrightarrow(-1\!-\!\nu)$.

Considering that $\mu=\nu \log{\Bigl(\frac{\alpha\beta}{\gamma \delta q^{L-1}}\Bigr)}$, we also obtain the Gallavotti-Cohen symmetry related to the current, namely
\begin{equation}\boxed{
E(\mu)=E\biggl(-\log{\Bigl(\frac{\alpha\beta}{\gamma \delta q^{L-1}}\Bigr)}-\mu\biggr)
}
\end{equation}
which is also valid for the other eigenvalues of $M_\mu$, and the corresponding relations between the right and left eigenvectors, as well as a simple relation between the microscopic entropy production $s$, conjugate to $\nu$, and the macroscopic current $j$, conjugate to $\mu$:
\begin{equation}\label{sj}
s= j\log{\Bigl(\frac{\alpha\beta}{\gamma \delta q^{L-1}}\Bigr)}.
\end{equation}

There are several points to be noted here. First of all, those weights are ill-defined for the TASEP: micro-reversibility (i.e. the fact that for any allowed transition, the reverse transition is also allowed) is essential to have a fluctuation theorem. Moreover, if we take either the $q\rightarrow 0$ or the $L\rightarrow\infty$ limit, the centre of the Gallavotti-Cohen symmetry $\mu=-\frac{1}{2}\log{\Bigl(\frac{\alpha\beta}{\gamma \delta q^{L-1}}\Bigr)}$ is rejected to $-\infty$, so that the `negative current' part of the fluctuations is lost.

Finally, we may consider the detailed balance case, where $\frac{\alpha\beta}{\gamma \delta q^{L-1}}=1$. In that case, we saw in section \ref{I-2-4} that there is no entropy production whatsoever, i.e. that $E(\nu)=0$ (where we use the letter $E$ abusively, since it is not the same function as $E(\mu)$). We see, indeed, from eq.(\ref{sj}), that $s=0$. This does not mean, however, that $j=0$: the deformations through $\mu$ and $\nu$ are in that case not equivalent, and $E(\mu)\neq 0$. The only implication this has on $E(\mu)$ is that it is an even function: $E(\mu)=E(-\mu)$, all the odd cumulants are zero, and positive and negative currents of the same amplitude are equiprobable.

\subsection{Exact cumulants of the current from integrability}
\label{III-2}

As we have mentioned earlier, the deformed Markov matrix of which we seek to extract the largest eigenvalue is integrable: its algebraic properties make it tractable in a systematic way, at least in principle, much like a quadratic one-dimensional spin chain Hamiltonian can always be diagonalised through free fermion techniques, although diagonalising it in practice is still a difficult problem. A wide variety of methods have been developed to tackle integrable models, such as the Bethe Ansatz and its variants (coordinate \cite{Gaudin1983}, algebraic \cite{Faddeev1996}, functional \cite{prolhac2010tree}, off-diagonal \cite{Cao2013a,Wen2015}, modified \cite{SamuelBelliard2014}), the Q-operator approach \cite{Baxter1982}, the q-Onsager algebra approach \cite{Baseilhac2006,Baseilhac2012}, the separation of variables approach \cite{Faldella2014}, and more \cite{Crampe2014}.

In this section, we will focus on two of these approaches, which are well adapted to our endeavour. We will first examine the coordinate Bethe Ansatz for the periodic TASEP \cite{derrida1998exact}, which is one of the simplest methods among the ones we mentioned, but requires the number of particles in the system to be fixed. We will then look at the Q-operator approach for the open ASEP \cite{Lazarescu2014}, which allows to calculate the generating function of the cumulants of the current exactly, although the analytical form of the solution has a few drawbacks, as we shall see.

Considering that these methods and calculations are rather specific to the ASEP and its being integrable, and are of little use for other bulk-driven particle gases, we will only give the minimum level of detail necessary to understand how the results are attained and why they have their peculiar structure. References will be given along the way for the readers in need of more detail.

\subsubsection{Periodic case: coordinate Bethe Ansatz}
\label{III-2-a}

The coordinate Bethe Ansatz is perhaps the simplest way to approach the ASEP \cite{Bethe1931,Gaudin1983}, but requires the number of particles to be fixed. It relies on the fact that integrable systems can be understood as a generalisation of free fermions, where the anti-commutation rule depends on the particles being exchanged. As a result, the eigenstates of the system can be written as generalised determinants of one-particle eigenstates, where the coefficient of each term is a function of the permutation instead of its sign.

We will see here how this can be used for the totally asymmetric case, as was first done in \cite{derrida1998exact}, and which, while being much simpler than the general case, has essentially the same behaviour. These results were later extended to the general periodic case in \cite{prolhac2010tree}. 

~

To make our calculations simpler, we will count the current on every bond in the system with equal weight, so that the system retains its translational invariance. The deformed Markov matrix we are considering is therefore
\begin{equation}\label{III-1-MperMu}
M_{\mu}=\sum_{i=1}^{L} M_{i}(\mu/L)
\end{equation}
with
\begin{equation}\label{III-1-Mmui}
 M_{i}(\mu/L)=\begin{bmatrix} 0 & 0 & 0 & 0 \\ 0 &0 & {\rm e}^{\mu/L} & 0 \\ 0 & 0& -1 & 0 \\ 0 & 0 & 0 & 0 \end{bmatrix},
\end{equation}
acting on sites $i$ and $i+1$, with $L+1\equiv 1$. Note that the number of particles, $N$, is conserved.

~

Let us first take $N=1$. The system is then simply a totally biased random walk on a circle. Its eigenvectors are of the form
\begin{equation}
|\psi^{(1)}(z)\rangle=\sum\limits_{x=1}^{L}z^x|x\rangle,
\end{equation}
where $|x\rangle$ is the configuration with the particle at site $x$. The periodicity condition imposes that $z^L=1$, which has $L$ solutions. The eigenvalue associated with each of these solutions is $\Lambda(z)=\bigl(\frac{{\rm e}^{\mu/L}}{z}-1\bigr)$.

~

We now consider $N=2$. Around configurations where the two particles are far from each other, at positions $x_1<x_2$ (with an arbitrary site being labelled as $1$), the system looks like two independent random walks, so that states such as
\begin{equation}
\sum\limits_{x_1<x_2}z_1^{x_1}z_2^{x_2}|x_1,x_2\rangle
\end{equation}
are locally stable, with an eigenvalue $\Lambda(z_1,z_2)=\bigl(\frac{{\rm e}^{\mu/L}}{z_1}+\frac{{\rm e}^{\mu/L}}{z_2}-2\bigr)$. This cannot be an eigenstate, since the two particles are in fact not independent, so that configurations where the two particles are next to each other would not be multiplied by $\Lambda(z_1,z_2)$: the contributions that would have come from the configuration where both particles are at the same site are missing. However, we see that the eigenvalue is invariant under exchange of $z_1$ and $z_2$, so that there are (at least) two locally stable states with that same eigenvalue. Taking a suitable combination of the two allows to make those missing terms cancel, and we obtain a true eigenstate. This is somewhat similar to the `method of images' which is generally used for random walks with walls. Let us therefore consider
\begin{equation}
|\psi^{(2)}(z_1,z_2)\rangle=\sum\limits_{x_1<x_2}a~z_1^{x_1}z_2^{x_2}+b~z_2^{x_1}z_1^{x_2}|x_1,x_2\rangle.
\end{equation}
Writing the eigenvalue equation for a configuration $|x_1,x_1+1\rangle$, the coefficients $a$ and $b$ need to be such that
\begin{equation}
\Lambda(z_1,z_2)(a~z_1^{x_1}z_2^{x_1+1}+b~z_2^{x_1}z_1^{x_1+1})={\rm e}^{\mu/L}(a~z_1^{x_1-1}z_2^{x_1+1}+b~z_2^{x_1-1}z_1^{x_1+1})-(a~z_1^{x_1}z_2^{x_1+1}+b~z_2^{x_1}z_1^{x_1+1})
\end{equation}
which simplifies to
\begin{equation}\label{Betheab}
a({\rm e}^{\mu/L}-z_2)+b({\rm e}^{\mu/L}-z_1)=0.
\end{equation}

Moreover, due to the periodicity of the system, and because the eigenstates of $M_\mu$ cannot depend on the arbitrary labelling of the sites which we chose, we need to have
\begin{equation}
a~z_1^{x_1}z_2^{x_2}+b~z_2^{x_1}z_1^{x_2}=a~z_1^{x_2}z_2^{x_1+L}+b~z_2^{x_2}z_1^{x_1+L}
\end{equation}
for any $x_1$ and $x_2$, which is to say
\begin{equation}\label{abz1z2}
a=b~z_1^L~~~~{\rm and}~~~~b=a~z_2^L.
\end{equation}
Since we are only interested in the eigenvalues of $M_\mu$, we do not need to worry about $a$ and $b$. Eliminating them from eq.(\ref{abz1z2}), we get the `Bethe equations'
\begin{equation}
z_i^L=\frac{(z_i-{\rm e}^{\mu/L})}{({\rm e}^{\mu/L}-z_j)}~~~~{\rm for}~~~~i=1,2~;~j\neq i
\end{equation}
of which the solution can then be used to obtain $\Lambda(z_1,z_2)$.

~

This simple case can then be extended to more particles without much effort: the integrability of the model ensures that triple collisions (when three particles are on adjacent sites) are merely combinations of double collisions, and do not add extra constraints to the $z_i$'s. It is then straightforward to generalise eq.(\ref{Betheab}) to a relation between coefficients of terms with only two adjacent $z_i$'s being exchanged. Eliminating them from the relations imposed by periodicity, we obtain the Bethe equations for $N$ particles (and we recall the expression of the eigenvalues):
\begin{equation}\label{BethePer}\boxed{
\Lambda(\{z_i\})=\sum\limits_{i=1}^N\biggl(\frac{{\rm e}^{\mu/L}}{z_i}-1\biggr)~~~~{\rm with}~~~~z_i^L=\frac{(z_i-{\rm e}^{\mu/L})^{N-1}}{\prod\limits_{j\neq i}({\rm e}^{\mu/L}-z_j)}~~~~{\rm for}~~~~i=1..N.
}\end{equation}
For more details on how to obtain these equations, one may refer to chapter 8 of \cite{Baxter1982} or to \cite{Lazarescu2013}.

~

The Bethe equations are a system of $N$ coupled non-linear equations, which cannot be systematically solved in principle. Moreover, since the number of solutions to these equations is unknown {\it a priori}, it is not obvious that every eigenstate of the model corresponds to one of these solutions.

That being said, the fact that we are looking for the largest eigenvalue of $M_\mu$, which has non-negative entries save for its diagonal, ensures that we can attain our goal: the Perron-Frobenius theorem \cite{Gantmacher2000} tells us that the largest eigenvalue $E(\mu)$ of $M_\mu$ is always non-degenerate, which means that we can follow it, and the corresponding eigenvectors, continuously by varying $\mu$. Moreover, we know that $E(0)=0$, and it is easy to check that the right eigenvector for that eigenvalue at $\mu=0$ is uniform in the chosen occupancy sector (i.e. all the weights of configurations with $N$ particles are equal, and all the others are $0$) ; we will write it as $|1_N\rangle$. It corresponds to a Bethe state with all $z_i$'s equal to $1$. This information, combined with the Bethe equations, is enough to obtain an expression for $E(\mu)$.

To that purpose, we first need to change variables, for a simple matter of convenience: we will write $z_i={\rm e}^{\mu/L}(1+y_i)$, so that the eigenvalues and the Bethe equations become, after a few simple manipulations,
\begin{equation}\label{BethePerY}
\Lambda(\{z_i\})=\sum\limits_{i=1}^N\biggl(\frac{1}{1+y_i}-1\biggr)~~~~{\rm with}~~~~{\rm e}^{\mu}\frac{(1+y_i)^L}{y_i^N}=(-1)^{N-1}\biggl(\prod\limits_{j=1}^N y_j\biggr)^{-1}~~~~{\rm for}~~~~i=1..N.
\end{equation}
The state we are interested in then corresponds to the solution with $y_i\rightarrow 0$ when $\mu\rightarrow 0$. Notice that the right-hand side of the Bethe equations is the same for every $i$. Let us define $B=(-1)^{N-1}{\rm e}^{\mu}\prod\limits_{j=1}^N y_j$, and
\begin{equation}\label{hBethe}
h(y)=\frac{(1+y)^L}{y^N}.
\end{equation}
The Bethe equations then tell us that all the $y_i$'s are roots of $1-Bh(y)$, with the self-consistency condition given by the definition of $B$. Moreover, it is easy to check that, for $\mu$ small enough, and $B$ smaller than $1$, $1-Bh(y)$ has exactly $N$ roots inside of the unit circle among its $L$ roots. Since we know that the $y_i$'s should be small, these are the roots we are looking for.

Summarising all this, we have three relations which we need to combine in order to obtain an expression of $E(\mu)$:
\begin{itemize}
\item the $y_i$'s are the roots of $1-Bh(y)$ which are inside of the unit circle ;
\item $E(\mu)=\sum\limits_{i=1}^N\Bigl(\frac{1}{1+y_i}-1\Bigr)$ ;
\item $\mu=\sum\limits_{i=1}^N\log\bigl(-B^{\frac{1}{N}}/y_i\bigr)$.
\end{itemize}
Since both $E$ and $\mu$ are of the form $\sum\limits_{i=1}^N f(y_i)$, we can write them as contour integrals around the unit circle $c_1$, using the first relation:
\begin{equation}
\sum\limits_{i=1}^N f(y_i)=\oint_{c_1}\frac{dz}{\imath2\pi }f(z)\frac{(-Bh'(z))}{1-Bh(z)}
\end{equation}
which, after an integration by parts, becomes
\begin{equation}
\oint_{c_1}\frac{dz}{\imath2\pi }f'(z)\log\bigl(1-Bh(z)\bigr)
\end{equation}
with $f'(z)=-(1+z)^{-2}$ for $E$ and $f'(z)=-z^{-1}$ for $\mu$. We can also check that there is no extra constant part coming from the integration by parts of a logarithm: for $\mu\rightarrow0$, both $E$ and $\mu$ have to vanish, which is the case here through $B\rightarrow0$.

We therefore have an implicit expression for $E(\mu)$ in terms of $B$, given by
\begin{equation}\boxed{
\mu=\oint_{c_1}\frac{dz}{\imath2\pi z}\log\bigl(1-Bh(z)\bigr)
}\end{equation}
and
\begin{equation}\boxed{
E(\mu)=\oint_{c_1}\frac{dz}{\imath2\pi(1+z)^2}\log\bigl(1-Bh(z)\bigr),
}\end{equation}
which we may expand as series in $B$, and calculate the coefficients (which are all binomial numbers ; it is left as an exercise to the reader to check it):
\begin{equation}
\mu=-\sum_{k=1}^{\infty}C_{k}\frac{B^k}{k}~~~~{\rm with}~~~~C_k=\binom{kL}{kN}
\end{equation}
and
\begin{equation}
E(\mu)=-\sum_{k=1}^{\infty}D_{k}\frac{B^k}{k}~~~~{\rm with}~~~~D_k=\binom{kL-2}{kN-1}.
\end{equation}

The periodic ASEP, with $q\neq 0$, yields results with a similar structure, as was found in \cite{prolhac2010tree}.

~

From these formulae, one can, in principle, calculate any cumulant in a finite number of steps, by inverting $\mu(B)$ order by order and injecting it into $E(B)$. Since we want to take a Legendre transform of $E(\mu)$ to obtain the large deviation function of the current, we are rather interested in an closed expression of $E(\mu)$, at least in some limit. We will see how to obtain that shortly, but we will first have a look at the open ASEP, where the coordinate Bethe Ansatz only applies in special cases \cite{Crampe2010,crampe2011matrix,simon2009construction}.

\subsubsection{Open case: Q-operator method}
\label{III-2-b}

We now go back to the open ASEP, with a deformation on $m_0$ as in eq.(\ref{III-1-M0mu}). Because the number of particles is not fixed, the coordinate Bethe Ansatz cannot be applied for generic boundary parameters, and we will have to use a different method: that of the so-called `Q-operator', sometimes called `auxiliary operator' \cite{Korff2005}, which gives a relatively simple way to obtain the expressions we seek. The main difference between this method and that of the Bethe Ansatz is that it does not require an Ansatz for the eigenvectors to access the eigenvalues. Moreover, it is entirely algebraic: it yields relations verified by the matrix $M_\mu$, rather than by parameters entering an Ansatz for its eigenvectors (which is the case for the Bethe equations), so that the completeness of the solution does not need to be proven {\it a posteriori}.

The calculations and proofs involved in applying the Q-operator method are rather lengthy and technical, and we will not dwell on them here. All that we will see here is based on \cite{Lazarescu2013,Lazarescu2014}, to which the reader may refer for more details.

~

At the core of this method is a generalisation of the $d$ and $e$ matrices which we used in section \ref{II-2-b} to build the Matrix Ansatz. Let us consider similar matrices, with a slightly different q-commutation rule, as well as a third matrix $A$, defined as
\begin{equation}\label{dexy}
de-q~ed=(1-q)(1-x^2 A^2)~~~~,~~~~dA-q~Ad=0~~~~,~~~~Ae-q~eA=0,
\end{equation}
where $x$ is a free parameter.

As before, one can think of $d$ and $e$ as the annihilation and creation operators of a q-deformed harmonic oscillator with an extra parameter, and $A$ as the q-counting matrix $q^n$ (where $n$ is the number of excitations). In this usual representation, they can be written down as
\begin{equation}
d=\sum\limits_{n=1}^{\infty}(1-q^n)|\!|n-1\rangle\!\rangle\langle\!\langle n|\!|~~,~~e=\sum\limits_{n=0}^{\infty}(1-x^2 q^n)|\!|n+1\rangle\!\rangle\langle\!\langle n|\!|~~,~~A=\sum\limits_{n=0}^{\infty} q^n|\!|n\rangle\!\rangle\langle\!\langle n|\!|.
\end{equation}
In section \ref{II-2-b}, we used these objects to build a vector, with $1+d$ for a particle, and $1+e$ for a hole. Here, we need them to construct a transfer matrix, in a similar way. Let us define a $2\times 2$ matrix $X(x)$ with those objects as entries:
\begin{equation}
X(x)=\begin{bmatrix} 1+x A & e\\ d &1+x A\end{bmatrix},
\end{equation}
expressed in basis $\{0,1\}$. We also need to define a matrix $A_\mu$ such that
\begin{equation}
dA_\mu-{\rm e}^{-\mu}~A_\mu d=0~~~~,~~~~A_\mu e-{\rm e}^{-\mu}~eA_\mu=0,
\end{equation}
which is to say $A_\mu={\rm e}^{-n\mu}$, with the same structure as $A$ but with $q$ replaced by ${\rm e}^{-\mu}$. Moreover, just as for the matrix Ansatz in section \ref{II-2-b}, we need to define special vectors to be placed at the boundaries, containing the parameters from $m_0$ and $m_L$. In the present case, we need four vectors $|\!| V \rangle\!\rangle$, $\langle\!\langle W|\!|$, $|\!|\tilde{V} \rangle\!\rangle$ and $\langle\!\langle  \tilde{W}|\!|$, such that
\begin{align}
[\beta (d+1+x A) - \delta (e+1+x A) -(1-q)] ~|\!| V \rangle\!\rangle &= 0 \label{V-2-V},\\
\langle\!\langle W|\!|~ [\alpha(e+1+x A)- \gamma (d+1+x A)-(1-q)] &= 0\label{V-2-W},\\
[\beta (d-1-x  A)-\delta (e-1-x A) +(1-q)x A]~|\!|  \tilde{V} \rangle\!\rangle  &=0\label{V-2-Vt}, \\
\langle\!\langle \tilde{W}|\!|~ [\alpha(e-1-x A) - \gamma (d-1-x A)+(1-q)x A] &=0\label{V-2-Wt}.
\end{align}
As one can easily check, the first two are generalisations of the ones which we used for the Matrix Ansatz (which we recover by taking $x=0$).

Using all those elements, we build two transfer matrices $U_\mu(x)$ and $T_\mu(x)$ (each corresponding to a pair of boundary vectors), with a structure very similar to that of the matrix Ansatz: instead of a product of matrices $D$ and $E$ for particles or holes, we use the elements of $X(x)$ for transitions between configurations (i.e. $X_{i,j}$ at site $k$ if the initial configuration has occupancy $j$ at site $k$, and the final one has occupancy $j$ at site $k$). We also need to add a matrix $A_\mu$ at the place where we are counting the current, which is between the left border and the first site in this case (if we had more general $\mu_i$'s, we would need a matrix $A_{\mu_i}$ for each of those). For instance, the entry of $U_\mu(x)$ from configurations $001010$ to the right to $010111$ to the left is given by $\langle\!\langle W|\!| A_\mu (1+xA)~d~e~d~(1+xA)~d |\!| V \rangle\!\rangle$.

We write these matrices in the following way
\begin{align}
U_\mu(x)&=\langle\!\langle W|\!| A_\mu \prod_{i=1}^{L}X^{(i)}(x) |\!| V \rangle\!\rangle,\\
T_\mu(x)&=\langle\!\langle  \tilde{W}|\!|  A_\mu \prod_{i=1}^{L}X^{(i)}(x)|\!|  \tilde{V} \rangle\!\rangle,
\end{align}
with an exponent $(i)$ on each $X$ which serves only to signify that it acts on site $i$. Note that those transfer matrices are in principle defined up to a non-trivial normalisation, which can be a function of $x$, and is implicitly included in the boundary vectors. That normalisation plays a part in particular in equation (\ref{V-2-PQs}), which is not homogeneous in $U_\mu$ and $T_\mu$, and we choose it so that eq.(\ref{t1F}) holds.

~

The main property of these matrices is that any product $U_\mu(x)T_\mu(y)$ commutes with any other product $U_\mu(x')T_\mu(y')$, and with $M_\mu$:
\begin{equation}\label{V-2-MUTs}
[U_\mu(x)T_\mu(y), U_\mu(x')T_\mu(y')]=0~~~~,~~~~[U_\mu(x)T_\mu(y),M_\mu]=0,
\end{equation}
which can be shown using so-called `R-matrices', as is customary for transfer matrix methods \cite{Faddeev1996}. Note that the dependence in $x$ of each of the matrices is not only in the diagonal entries of $X$ and in the relations defining the boundary vectors, but also implicitly in $d$ and $e$ through eq.(\ref{dexy}).

Those two transfer matrices do not commute with one-another, but we may use them to define two that do: $P(x)$ and $Q(x)$ defined by
\begin{equation}
P(x)=U_\mu(x)\Bigl[U_\mu(0)\Bigr]^{-1}~~~~,~~~~Q(x)=(1-{\rm e}^{-\mu})^{-1}U_\mu(0)T_\mu(x)
\end{equation}
such that:
\begin{equation}
U_\mu(x)T_\mu(y)=(1-{\rm e}^{-\mu})P(x)Q(y).
\end{equation}
The downside of these is that, unlike $U_\mu$ and $T_\mu$, $P$ is not defined constructively because it involves $[U_\mu(0)]^{-1}$ (which exists for a generic $\mu$), and does not have the same product structure.

~

Apart from their commuting with $M_\mu$ (which means they have the same eigenvectors), the operators $P$ and $Q$ have another useful property: given certain constraints on their variables, their product decouples into a transfer matrix with a finite auxiliary space and a shifted version of the product. More precisely, for every positive integer $k$,
\begin{equation}\label{V-2-PQs}
P(x)Q(1/q^{k-1}x)=t^{(k)}(x)+{\rm e}^{-2k\mu}P(q^k x)Q(q/x).
\end{equation}
The matrix $t^{(k)}(x)$ is the transfer matrix of an integrable vertex model with a $k$-dimensional auxiliary space. For instance, $t^{(2)}(x)$ is the transfer matrix of the six-vertex model \cite{Baxter1982}. Moreover, $t^{(1)}(x)$ is a scalar matrix, which we can calculate, choosing the appropriate normalisation for $U_\mu$ and $T_\mu$:
\begin{equation}\label{t1F}
t^{(1)}(x)\equiv F(x)=\frac{(1+x)^L(1+x^{-1})^L(x^2,x^{-2})_{\infty}}{ (a x,a/x,\tilde{a}x,\tilde{a}/x,b x,b/x,\tilde{b}x,\tilde{b}/x)_{\infty}},
\end{equation}
where the identity operator on configuration space is implicit. We already encountered that function inside of a contour integral in eq.(\ref{II-2-ZLint}). $t^{(2)}(x)$ and $F(x)$ are strongly connected to $M_\mu$, through the usual relation between the six-vertex transfer matrix and the Hamiltonian of the spin-$\frac{1}{2}$ XXZ chain (which is equivalent to the ASEP):
\begin{equation}
M_\mu=\frac{1}{2}(1-q)\frac{d}{d x} \log\biggl(\frac{t^{(2)}(x)}{F(qx)}\biggr)\bigg|_{x=-1}.
\end{equation}
Combining eq.(\ref{V-2-PQs}) for $k=1$ and $k=2$, we can eliminate $P$ and obtain the `T-Q equation':
\begin{equation}\label{V-2-TQs}
t^{(2)}(x)Q(1/x)=F(x)Q(1/qx)+{\rm e}^{-2\mu}F(qx)Q(q/x)
\end{equation}
which allows us to express $M_\mu$ in terms of $Q$ alone:
\begin{equation}\label{V-2-Mmut2}\boxed{
M_\mu=\frac{1}{2}(1-q)\frac{d}{d x} \log\biggl(\frac{Q(q/x)}{Q(1/x)}\biggr)\bigg|_{x=-1}.
}\end{equation}
This relation between $M_\mu$ and $Q$ (which applies to the whole spectrum of $M_\mu$), along with eq.(\ref{V-2-PQs}) for $k=1$:
\begin{equation}\label{PQF}\boxed{
P(x)Q(1/x)=F(x)+{\rm e}^{-2\mu}P(q x)Q(q/x)
}\end{equation}
is enough to obtain $E(\mu)$: they are a more complex version of the relations between $E$, $\mu$ and the Bethe roots we used in the periodic case (in fact, applying the Q-operator method to the periodic case, we find that the Bethe roots are the inverses of the roots of the eigenvalues of $Q$).

~

From this point on, we need to use the `functional Bethe Ansatz', as was done in \cite{prolhac2010tree} for the periodic ASEP, in order to unravel these relations and obtain an expression of $E(\mu)$. We first need to define a matrix $B$ which plays the same role as the intermediate variable we used in the previous section:
\begin{equation}
B=-{\rm e}^{2\mu}\bigl(Q(0)\bigr)^{-1}=-{\rm e}^{2\mu}(1-{\rm e}^{-\mu})\bigl(U_\mu(0)T_\mu(0)\bigr)^{-1}.
\end{equation}

Moreover, just as in the previous section, we need to know the behaviour of $B$ and $Q$ in the $\mu\rightarrow 0$ limit. We find, from analytical calculations which we will not go into here but can be found in \cite{Lazarescu2013}, that the first eigenvalue of $B$ goes to $0$, while the others remain finite, and that the roots of the first eigenvalue of $Q(1/x)$ are inside of the unit circle $c_1$ if $a<1$ and $b<1$ (as was needed in section \ref{II-2-b}), while those of $P$ are outside of it. Restricting ourselves to that first eigenspace from now on, this allows us to separate $P$ and $Q$ in eq.(\ref{PQF}) using a contour integral, and express $M_\mu$ in eq.(\ref{V-2-Mmut2}) only in terms of $F(x)$.

~

In the following, the notations $F$, $P$, $Q$, and $B$ will refer to the eigenvalues of the operators in the dominant eigenspace, and not to the operators themselves.

Let us define a function $W(x)$ as:
\begin{equation}\label{IV-2-W}
W(x)=-\frac{1}{2}\log\biggl(\frac{P(x)Q(1/x)}{{\rm e}^{-2\mu}P(q x)Q(q/x)}\biggr),
\end{equation}
and a convolution kernel $K$, as:
\begin{equation}\label{IV-2-K}
K(z,\tilde{z})=2\sum_{k=1}^{\infty}\frac{q^k}{1-q^k}\Bigl((z/\tilde{z})^k+(z/\tilde{z})^{-k}\Bigr)
\end{equation}
along with the associated convolution operator $X$:
\begin{equation}\label{IV-2-X}
X[f](z)=\oint_{c_1}\frac{d\tilde{z}}{\imath2\pi\tilde{z}}f(\tilde{z})K(z,\tilde{z}).
\end{equation}

Expanding $\log(P(x))$ and $\log(Q(1/x))$ in powers of $x$ and $1/x$ respectively, we can easily check that 
\begin{equation}
-\log\bigl(P(q x)Q(q/x)/Q(0)\bigr)=X[W](x),
\end{equation}
which allows us to rewrite eq.(\ref{PQF}) as a functional equation of only one unknown function $W$:
\begin{equation}\label{IV-2-WW}\boxed{
W(x)=-\frac{1}{2}\ln\Bigl(1-B F(x) e^{X[W](x)}\Bigr).
}\end{equation}

The last step is to take eq.(\ref{V-2-Mmut2}) in the first eigenspace of $M_\mu$, and eq.(\ref{IV-2-W}) at $x=0$, to find:
\begin{equation}
E(\mu)=\frac{1}{2}(1-q)\frac{d}{dx}\log\biggl(\frac{Q(q/x)}{Q(1/x)}\biggr)\biggl|_{x=-1}=\frac{1}{2}(1-q)\oint_{c_1}\frac{dz}{\imath2\pi(1+z)^2}\log\biggl(\frac{Q(q/z)}{Q(1/z)}\biggr)~~~~,~~~~\mu=-W(0)
\end{equation}
where the contour integral form can be checked formally by expanding $\log(Q(1/x))$ in powers of $1/x$.

Considering that, for $\mu$ small enough, $P$ is holomorphic inside of the unit circle, we can replace $\frac{1}{2}\log\Bigl(\frac{Q(q/z)}{Q(1/z)}\Bigr)$ by $-W(z)$ when expressing $E(\mu)$ as a contour integral (since $P$ will not contribute), and obtain:
\begin{equation}\label{IV-2-muB}\boxed{
\mu=-\oint_{c_1}\frac{dz}{\imath2\pi z}W(z)
}\end{equation}
and
\begin{equation}\label{IV-2-EB}\boxed{
E(\mu)=-(1-q)\oint_{c_1}\frac{dz}{\imath2\pi(1+z)^2}W(z).
}\end{equation}
All this was done for $a<1$ and $b<1$, but can then be generalised to any $a$ and $b$ through the same reasoning as in section \ref{II-2-b} for the mean current, replacing the unit circle $c_1$ by small contours around $S=\{0,q^k a,q^k \tilde{a},q^k b,q^k \tilde{b}\}$.

~

As we see, the form of $E(\mu)$ is similar to what we found for the periodic TASEP. It is even more similar to the periodic ASEP case \cite{prolhac2010tree}: the expressions only differ by the factors $2$ in $K$ and $\frac{1}{2}$ in $W(x)$, and by the function $F(x)$ which is exchanged with $h(x)$ as defined in eq.(\ref{hBethe}).

Moreover, if we choose $a=1$, $\tilde{a}=-q$, $b=\sqrt{q}$ and $\tilde{b}=-\sqrt{q}$, which is to say $\alpha=\frac{1}{2}$, $\gamma=\frac{q}{2}$, $\beta=1$ and $\delta=q$, special cancellations occur in $F(x)$ which reduces to $(1+x)^{L+1}(1+x^{-1})^{L+1}$. This is the same as the function $h$ for the periodic ASEP with $2L+2$ sites and $L+1$ particles. Because of the extra factors $2$ and $\frac{1}{2}$, the generating function of the cumulants of the current is half that which we found in the periodic case, taken at $2\mu$. This also works if we exchange $a$ with $b$ and $\tilde{a}$ with $\tilde{b}$. Those two special points correspond to $\rho_a=\frac{1}{2}$ and $1-\rho_b=\frac{1}{1+q}$, or the opposite, and are on the transition lines between the MC phase and the LD or HD phase. We will come back to this remark later.

~

The expressions we just obtained are exact, and valid for any values of the parameters of the system, including its size. They are, however, somewhat unwieldy, especially if we want to perform a Legendre transform in order to obtain the large deviation function of the current, which is our goal. In the next section, we take the limit of large sizes and see how we can obtain closed expressions for $E(\mu)$ and $g(j)$ for small fluctuations of the current.

\subsection{Large size limit}
\label{III-3}

In this section, we will need to take the result we just obtained for the generating function of the cumulants of the current in the open ASEP, and extract its large size behaviour, through various approximations. We will not go into every fine detail of the necessary calculations, but we will endeavour to make them as easy to follow as possible. That being said, the calculations themselves might not be of interest to every reader, and we will make the results clearly visible at the end of each sub-section (i.e. they will be boxed).

~

Before taking any limit, we will need to examine eqs.(\ref{IV-2-muB}) and (\ref{IV-2-EB}) a little closer. The function $W(z)$ that they contain is defined, in eq.(\ref{IV-2-WW}), through a self-consistency equation. Expanding the logarithm in powers of $B$, we can express every coefficient by calculating $W(x)$ perturbatively in $B$. The coefficient of $B^k$ will be a combination of $k$ functions $F$, either at the same point or convolved together through $K$. For instance, the coefficient of $B^2$ in $\mu$ is
\begin{equation}\label{III-3-C2}
-\frac{1}{4}\oint_S\frac{dz}{\imath2\pi z}F(z)^2~-~\frac{1}{4}\oint_S\frac{dz_{1}}{\imath2\pi z_{1}}\oint_S\frac{dz_{2}}{\imath2\pi z_{2}}F(z_{1})F(z_{2})K(z_{1},z_{2})
\end{equation}

As was painstakingly verified in \cite{Lazarescu2013}, it turns out that, in the large size limit, the term containing no convolutions is dominant in every case, and that, when all is said and done, the behaviour of $E(\mu)$ differs from the TASEP case (where $q=0$ so that $K=0$) only by a global factor $(1-q)$. We will therefore save ourselves some trouble here and only consider that simpler case: $q=\gamma=\delta=0$.

~

The simplified expressions we will have to examine are the following:
\begin{equation}\label{muTASEP}
\mu=-\sum_{k=1}^{\infty}C_{k}\frac{B^k}{k}~~~~{\rm with}~~~~C_k=\frac{1}{2}\oint_{\{0,a,b\}}\frac{dz}{\imath2\pi z}F^k(z),
\end{equation}
\begin{equation}\label{ETASEP}
E(\mu)=-\sum_{k=1}^{\infty}D_{k}\frac{B^k}{k}~~~~{\rm with}~~~~D_k=\frac{1}{2}\oint_{\{0,a,b\}}\frac{dz}{\imath2\pi (1+z)^2}F^k(z),
\end{equation}
where
\begin{equation}\label{FTASEP}
F(x)=\frac{(1+x)^L(1+x^{-1})^L(1-x^2)(1-x^{-2})}{ (1-a x)(1-a/x)(1-b x)(1-b/x)}.
\end{equation}

\begin{figure}[ht]
\begin{center}
\includegraphics[width=0.5\textwidth]{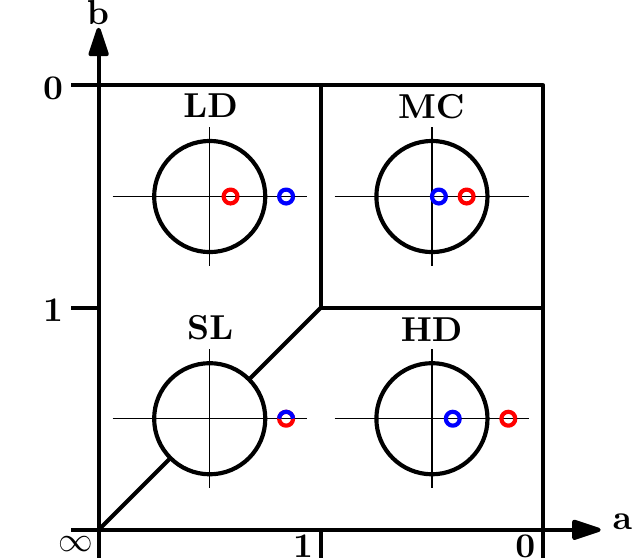}
\caption{Positions of $a$ (blue circles) and $b$ (red circles) with respect to the unit circle (black circles) in each phase of the open ASEP.}
\label{IV-fig-DiagCont}
\end{center}
\end{figure}

As for the calculation of the mean current that we saw in section \ref{II-2-b}, the asymptotic form of these contour integrals, for $L$ large, will depend on the position of $a$ and $b$ with respect to the unit circle (see fig.\ref{IV-fig-DiagCont}).

Moreover, the correct way to handle that large size limit is not entirely straightforward, due to $E(\mu)$ being expressed implicitly: each cumulant is a combination of $C_k$'s and $D_k$'s of same or lower order, such as
\begin{equation}
E_{3}=\frac{3D_{1}C_{2}^2-2D_{1}C_{1}C_{3}-3D_{2}C_{1}C_{2}+2D_{3}C_{1}^2}{C_{1}^5}
\end{equation}
but the leading order of $E_k$ in $L$ might not be obtained by taking the leading orders of each term, as this may simply give $0$ due to certain cancellations. As it turns out, this is in fact always the case: at leading order in $L$, in every phase, one finds $D_k\sim JC_k$, where $J$ is the average current, so that $E(\mu)\sim J\mu$. This cancels out in every $E_k$ other than $E_1=J$. For that reason, we will sometimes be calculating equivalents of 
\begin{equation}
E(\mu)-J\mu=-\sum_{k=1}^{\infty}\tilde D_{k}\frac{B^k}{k}~~~~{\rm with}~~~~\tilde D_k=D_k-JC_k
\end{equation}
instead, which we will find to be sufficient in every case we will examine. We will only ignore one point of the phase diagram: $\alpha=\beta=\frac{1}{2}$, which is to say $a=b=1$, at which the $k$ first orders of $E_k$ vanish.

Finally, one should note that, because we will make approximations on the coefficients of $E(\mu)$ expanded (implicitly) around $\mu=0$, the expressions we will obtain for $E(\mu)$ or $g(j)$ will in principle be valid only for a small $\mu$ and a small fluctuation $(j-J)$.

\subsubsection{Low/high density phases}
\label{III-3-a}

We start with the low density phase: $a>1$ and $a>b$. The corresponding results for the high density phase are obtained by exchanging $a$ and $b$.

As we recall from section \ref{II-2-b}, we can replace the integration set $\{0,a,b\}$ by the unit circle plus twice the poles which are outside of it. The dominant part of each contour integral is then the residue around $a$, which is the one where $F(x)$ is maximal. That residue is of order $k$ in both $C_k$ and $D_k$. We can therefore write
\begin{align}\label{IV-1-muELD}
\mu&=-\sum_{k=1}^\infty \frac{ B^k}{k!} \frac{d^{k-1}}{dz^{k-1}} \Big \{ \frac{\phi^k(z)}{z} \Big\}  \Big|_{z =a} ,\\
E(\mu)&=-\sum_{k=1}^\infty \frac{ B^k}{k!} \frac{d^{k-1}}{dz^{k-1}} \Big \{ \frac{\phi^k(z)}{(1+z)^2} \Big\}  \Big|_{z =a},
\end{align}
with
\begin{equation}\label{IV-1-FphiLD}
\phi(z)=(z-a)F(z)=z\frac{(1+z)^{L}(1+z^{-1})^{L}(1-z^2)(1-z^{-2})}{(1-a z)(1-b z)(1-b /z)}.
\end{equation}

One may recognise there a structure related to the Lagrange inversion formula \cite{Sedgewick2009}. Considering two variables $w$ and $z$ such that
\begin{equation}\label{IV-1-wzB}
 w = z + B \phi(w)
\end{equation}
we can express a function $f$ taken at $w$ by expanding it around $z$, as:
 \begin{equation}\label{IV-1-lagrange}
 f(w) -f(z)=\sum\limits_{k=1}^{\infty}\frac{B^k}{k!} \frac{d^{k-1}}{dz^{k-1}}\Bigl(\phi^k(z) f'(z)\Bigr).
 \end{equation}
Choosing $f(z)=-\log(z)$ for $\mu$ and $1/(1+z)$ for $E(\mu)$, we can then write:
\begin{align}
 \mu ~~~&=  -\log(w) +  \log(a),\\
 E(\mu) &= \frac{1}{w+1} -  \frac{1}{a+1},
 \end{align}
where $w$ is as defined in (\ref{IV-1-wzB}). Combining those two equations, we get a closed expression for $E(\mu)$:
 \begin{equation}\label{IV-1-EmuTASEP}\boxed{
 E(\mu) = \frac{a}{a+1} \frac{ {\rm e}^\mu -1 }{{\rm e}^\mu + a}.
 }\end{equation}

It only remains for us to perform a Legendre transform to obtain the large deviation function of the current:
\begin{equation}\label{IV-2-gjASEP}\boxed{\boxed{
g(j)=(1-q)\Bigl[\rho_a-r+r(1-r)\log\Bigl(\frac{1-\rho_a}{\rho_a}\frac{r}{1-r}\Bigr)\Bigr].
}}\end{equation}
where $r$ is such that $ j=(1-q)r(1-r)$ and we recall that $\rho_a=\frac{1}{(1+a)}$. 

This expression was first obtained in \cite{Bodineau2006} using a method which we will discuss in section \ref{V}. The formula (\ref{IV-1-EmuTASEP}), with an extra factor $(1-q)$ for the general ASEP, was obtained in \cite{PhysRevLett.107.010602} through a numerical approach to the Bethe equations. Contrary to what we noted earlier, these expressions are in fact valid in their whole phase, and not only for small $\mu$, as we will see in section \ref{V}.

\subsubsection{Shock line}
\label{III-3-b}

We now look at the shock line, where $a=b>1$. Unfortunately, the method we used in the previous section does not work here: the residue at $a$ in $C_k$ and $D_k$ is of order $2k$, so that we are missing all the derivatives of even order in $\mu$ and $E$. Instead, we simply calculate the leading order of each coefficient. After simplifying every term as much as possible (see chapter IV of \cite{Lazarescu2013} for a detailed calculation), we obtain the large-size equivalent of our two series:
\begin{align}
\mu=-\frac{2}{L}\frac{a+1}{a-1}&\sum_{k=1}^{\infty}\frac{k^{2k-1}}{(2k)!} B^k\label{IV-1-muHL},\\
E(\mu)-(1-q)\frac{a}{(1+a)^2}\mu=-(1-q)\frac{2}{L^2}\frac{a}{a^2-1}&\sum_{k=1}^{\infty}\frac{k^{2k-2}}{(2k)!} B^k\label{IV-1-EHL}
\end{align}
and we find that the cumulants of the current behave as
\begin{equation}
E_k\sim(1-q)\frac{a}{a^2-1}\bigl(\frac{a-1}{a+1}\bigr)^kL^{k-2}
\end{equation}
for $k\geq 2$.

These two series can be expressed in terms of the Lambert ${\cal W}$ function \cite{Corless1996}, defined as the solution to $x={\cal W}_{\cal L}{\rm e}^{{\cal W}_{\cal L}}$. The series expansion of ${\cal W}_{\cal L}(x)$ around $0$ is:
\begin{equation}\label{IV-1-LambertW}
{\cal W}_{\cal L}(x)=-\sum\limits_{n=1}^{\infty}\frac{n^{n-1}}{n!}(-x)^n
\end{equation}
so that $\mu$ and $E$ become:
\begin{align}
\mu=\frac{2}{L}\frac{a+1}{a-1}&\Bigl[{\cal W}_{\cal L}(\sqrt{B}/2)+{\cal W}_{\cal L}(-\sqrt{B}/2)\Bigr]\label{IV-2-muHL}\\
E(\mu)-(1-q)\frac{a}{(1+a)^2}\mu=(1-q)\frac{2}{L^2}\frac{a}{a^2-1}&\Bigl[2\Bigl({\cal W}_{\cal L}(\sqrt{B}/2)+{\cal W}_{\cal L}(-\sqrt{B}/2)\Bigr)\nonumber\\
+&{\cal W}_{\cal L}(\sqrt{B}/2)^2+{\cal W}_{\cal L}(-\sqrt{B}/2)^2\Bigr].\label{IV-2-EHL}
\end{align}

~~

\begin{figure}[ht]
\begin{center}
\includegraphics[width=0.6\textwidth]{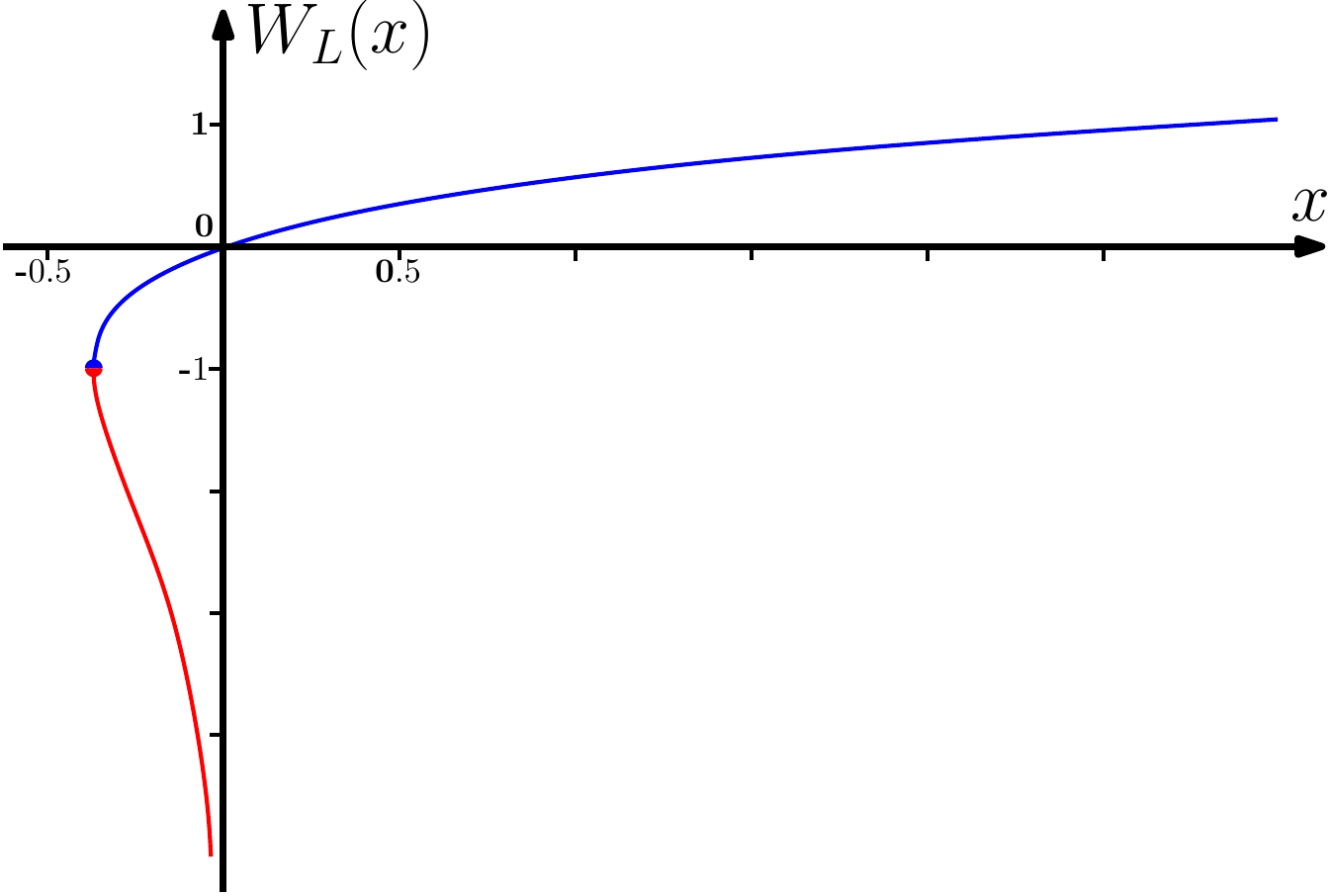}
\caption{Plot of the Lambert ${\cal W}$ function. The principal branch (blue) behaves as $\log(x)$ for $x\rightarrow\infty$. The second branch (red) behaves as $\log(-x)$ for $x\rightarrow 0^-$.}
\label{IV-fig-LambertW}
\end{center}
\end{figure}

Many things are known about ${\cal W}_{\cal L}$, including its asymptotic behaviour, which is what we need. The main branch of the function, called ${\cal W}_0$, is defined on the whole complex plane except for $]\!-\!\infty,\!-1/e]$, and behaves as $\log(x)$ (even for $x$ complex, in which case the angular part of $x$ can be neglected and we get $\log(|x|)$). This will be appropriate for $B<0$. For $B>0$, however, the functions ${\cal W}_{\cal L}(-\sqrt{B}/2)$ in our expressions have to be continued analytically to the second branch ${\cal W}_{-1}$, on which $x$ goes back from $-1/e$ to $0$ and behaves as $\log(-x)$ (fig.-\ref{IV-fig-LambertW}).

For $B\rightarrow -\infty$, we therefore have ${\cal W}_{\cal L}(\pm\sqrt{B}/2)\sim \log(|B|)/2$, so that:
\begin{align}
\mu\sim\frac{2}{L}\frac{a+1}{a-1}&\log(|B|)\label{IV-2-muHL-}\\
E(\mu)-(1-q)\frac{a}{(1+a)^2}\mu\sim(1-q)\frac{2}{L^2}\frac{a}{a^2-1}&\log(|B|)^2/2\label{IV-2-EHL-}
\end{align}
meaning that, for $\mu$ positive and small (since we are working in the $L\rightarrow\infty$ limit ; see comment below), $E$ behaves as
\begin{equation}\label{IV-2-ESL+}\boxed{
E_+(\mu)-(1-q)\frac{a}{(1+a)^2}\mu\sim(1-q)\frac{a(a-1)}{4(a+1)^3}\mu^2.
}\end{equation}

We take the Legendre transform of this and get, for $j>J=(1-q)\rho_a(1-\rho_a)$:
\begin{equation}\label{IV-2-g+}\boxed{\boxed{
g_+(j)\sim \frac{(j-J)^2}{J(1-2\rho_a)}
}}\end{equation}
(where $J=(1-q)\rho_a(1-\rho_a)$ is the average current).

The {\it a priori} domains of validity of equations (\ref{IV-2-ESL+}) and (\ref{IV-2-g+}) are not entirely obvious, and require a careful analysis. Starting from equations (\ref{muTASEP}) and (\ref{ETASEP}), which are exact for any $L$ and $B$, we first took a large $L$ limit and kept only the leading contributions, yielding equations (\ref{IV-1-muHL}) and (\ref{IV-1-EHL}), still valid for any $B$. We then took a large (negative) $B$, and obtained equations (\ref{IV-2-muHL-}) and (\ref{IV-2-EHL-}). Because of the order in which we took the limits, $L$ is taken to be large enough so that the first correction to (\ref{IV-1-muHL}) be negligible for any $B$ we consider, which is to say that $L$ goes to $\infty$ faster than $\log(|B|)$. For that reason, the equations we obtain here, and similarly in all the following calculations, are in principle valid only for $\mu$ small, which is to say $j$ close to its average. We will indeed see in section \ref{V} that these expressions correspond only to the leading order of those we find from the macroscopic fluctuation theory.

~~

To do the same for $\mu<0$, we need to be on the second branch of ${\cal W}_{\cal L}$, for which we have $B\rightarrow 0^+$. In that case, we have ${\cal W}_{\cal L}(-\sqrt{B}/2)\sim \log(B)/2$, but ${\cal W}_{\cal L}(\sqrt{B}/2)\sim 0$ (that part being still on the main branch). This gives us:
\begin{align}
\mu\sim\frac{2}{L}\frac{a+1}{a-1}&\log(|B|)/2\label{IV-2-muHL+}\\
E(\mu)-(1-q)\frac{a}{(1+a)^2}\mu\sim(1-q)\frac{2}{L^2}\frac{a}{a^2-1}&\log(|B|)^2/4\label{IV-2-EHL+}
\end{align}
so that, for $\mu$ negative and small, we have
\begin{equation}\label{IV-2-ESL-}\boxed{
E_-(\mu)-(1-q)\frac{a}{(1+a)^2}\mu\sim(1-q)\frac{a(a-1)}{2(a+1)^3}\mu^2.
}\end{equation}

We take the Legendre transform of this and get, for $j<J$:
\begin{equation}\boxed{\boxed{
g_-(j)\sim \frac{(j-J)^2}{2J(1-2\rho_a)}.
}}\end{equation}

Notice that the dependence in $L$ has vanished from both cases, so that even though all the cumulants of the current depend on $L$ at $\mu=0$, none of them do for a finite $\mu$. Notice also that $g_-$ differs from $g_+$ by a factor $\frac{1}{2}$, which comes from the fact that for $\mu>0$, both functions ${\cal W}_{\cal L}$ in $\mu$ and ${\cal W}_{\cal L}^2$ in $E(\mu)$ contribute to the limit, whereas for $\mu<0$, only one of each does. This results in $g(j)$ not being analytic at $j=J$, which is the signature of a non-equilibrium phase transition (see section \ref{V} for a description of the phase on each side of the transition).

\subsubsection{LD/MC transition line}
\label{III-3-c}

We now consider the transition line between the low density and the maximal current phase, for which $a=1$ and $b<1$. As we noted earlier, the cumulants of the current on this line are related to those for a half-filled periodic TASEP of size $2L+2$. The calculations we will need to perform here are therefore the same as can be found in \cite{derrida1999universal}, which we will reproduce, and use as a reference for the maximal current phase in the next section (which is similar but not identical).

Taking the limit of large $L$ in $\mu$ and $E$ (see chapter IV of \cite{Lazarescu2013} for a detailed calculation), we obtain
\begin{align}
\mu=-\frac{L^{-1/2}}{2\sqrt{\pi}}&\sum_{k=1}^{\infty}\frac{B^k}{k^{3/2}} \label{IV-1-muHDMC2},\\
E(\mu)-(1-q)\frac{1}{4}\mu=-(1-q)\frac{L^{-3/2}}{16\sqrt{\pi}}&\sum_{k=1}^{\infty}\frac{B^k}{k^{5/2}}\label{IV-1-EHDMC2},
\end{align}
so that the cumulants behave as
\begin{equation}
E_k\sim\pi(\pi L)^{(k-3)/2}
\end{equation}
for $k\geq 2$. Notice that the pre-factor $L^{-3/2}$ in (\ref{IV-1-EHDMC2}) is a sign of the dynamical scaling $z=\frac{3}{2}$ of the KPZ universality class.

~

These series can be written in terms of the polylogarithm ${\rm Li}_{5/2}(B)$, defined as
\begin{equation}\label{IV-2-HMC1}
H(B)=-{\rm Li}_{5/2}(B)=-\sum_{k=1}^{\infty}\frac{B^k}{k^{5/2}}=\frac{2}{\sqrt{\pi}}\int_{-\infty}^{+\infty} d\theta~\theta^2 \log\bigl[1-B{\rm e}^{-\theta^2} \bigl],
\end{equation}
so that
\begin{align}\label{IV-2-gMC}
\mu&=\frac{L^{-1/2}}{2\sqrt{\pi}}B ~H'(B)\\
E(\mu)-\frac{1-q}{4}\mu&=\frac{L^{-3/2}}{16\sqrt{\pi}}~H(B)
 \end{align}
As in the previous section, the cases $\mu<0$ and $\mu>0$ require different approaches.

~~

For $\mu>0$, we need to take $B\rightarrow-\infty$. In this case, the integrand in $H(B)$ can be approximated by:
\begin{equation}
\log\bigl[1-B{\rm e}^{-\theta^2} \bigl]\sim \log\bigl[|B|~{\rm e}^{-\theta^2} \bigl]~ \mathbb{I}\bigl[\theta^2<\log(|B|)\bigr]
\end{equation}
where $\mathbb{I}[X]$ is the indicator of $X$, equal to $1$ if $X$ is true, and $0$ if $X$ is false.

We can then estimate:
\begin{equation}\label{IV-2-HMC2}
H(B)\sim \frac{4}{\sqrt{\pi}}\int_{0}^{\log(|B|)^{1/2}} d\theta~\theta^2 \bigl(\log(|B|)-\theta^2\bigr)=\frac{8}{15\sqrt{\pi}}\log(|B|)^{5/2}
\end{equation}
and
\begin{equation}\label{IV-2-HMC3}
B~H'(B)\sim \frac{4}{3\sqrt{\pi}}\log(|B|)^{3/2}
\end{equation}
so that
\begin{equation}\label{IV-2-gMC2}\boxed{
E_+(\mu)-\frac{1-q}{4}\mu\sim(1-q)\frac{1}{20}\Bigl(\frac{3}{2}\Bigr)^{2/3} L^{-2/3}\pi^{2/3}\mu^{5/3}.
}\end{equation}

We can then take the Legendre transform of this result. We find, for $j>J=\frac{1-q}{4}$:
\begin{equation}\label{IV-2-gMC4}\boxed{\boxed{
g_+(j)\sim(j-J)^{5/2}\frac{32\sqrt{3}L}{5\pi(1-q)^{3/2}}
}}\end{equation}
and notice that, for once, it depends on $L$.

~~

\begin{figure}[ht]
\begin{center}
\includegraphics[width=0.9\textwidth]{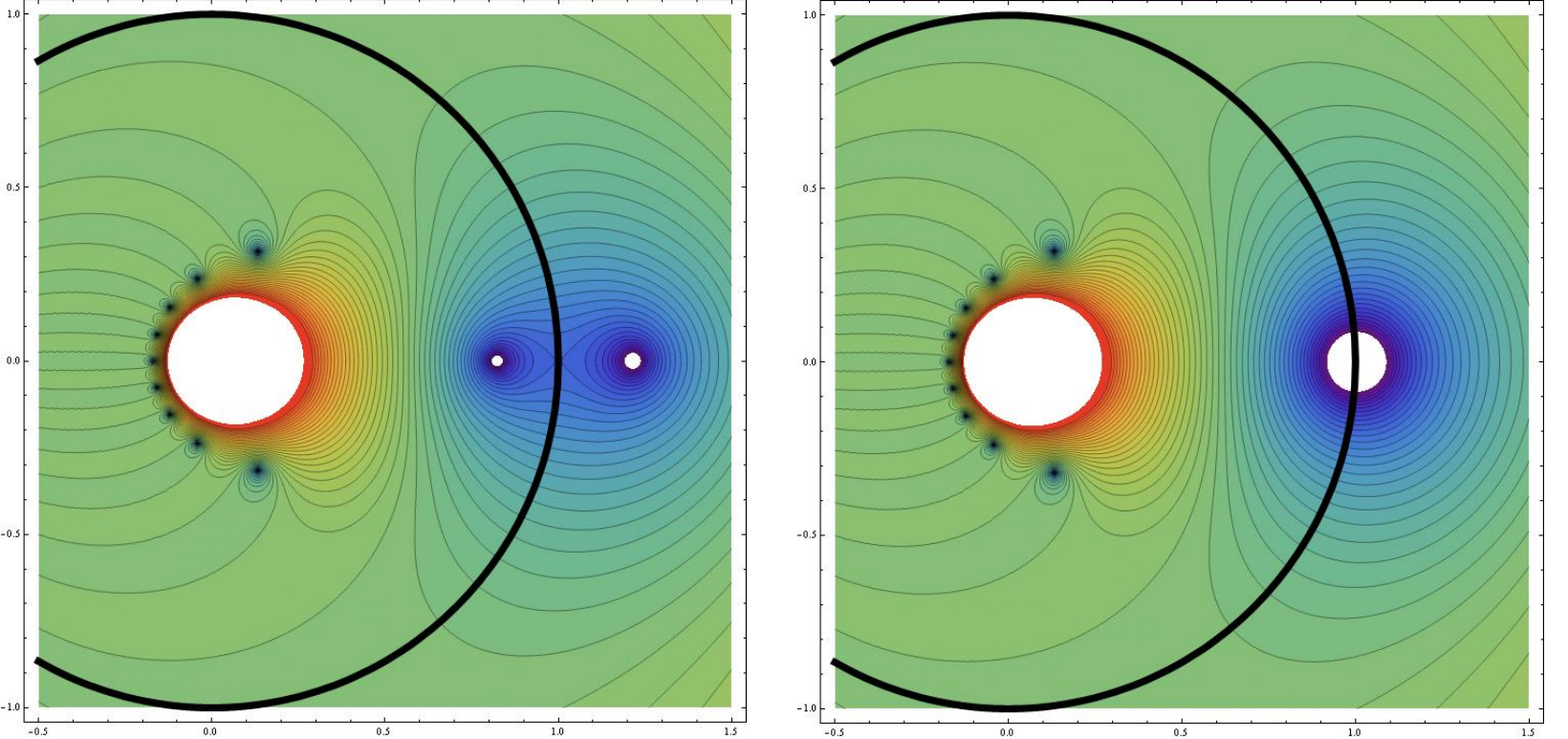}
\caption{Bethe roots for a periodic system with $20$ sites and $10$ particles. The roots are at the centres of the white discs. The unit circle is represented in black. On the left, where $B<1$, a pair of roots can be seen to approach the unit circle on the real axis. On the right, where $B=1$, those roots have merged.}
\label{IV-fig-RootsPer}
\end{center}
\end{figure}

For $\mu<0$, we have to take into account the fact that the polylogarithm $Li_{5/2}(x)$ has a branch cut at $x=1$ and is not defined for $x>1$. To remedy this, we have to go back to the expressions of $\mu$ and $E$ in terms of the roots of $(1-Bh(x))$ we saw in section \ref{III-2-a}. These expressions apply in principle to all eigenvalues of $M_\mu$, and depend on which roots are included in the integral. In the case of the steady state, for $B$ small enough, all the roots we consider are inside of the unit circle. However, it can be shown that, as $B$ gets closer to $1$, one of the roots $z_0$ goes to $1$, and merges with its counterpart $z_0^{-1}$ from outside of the unit circle (fig.-\ref{IV-fig-RootsPer}). Since we know, from the Perron-Frobenius theorem, that $E(\mu)$ never crosses any other eigenvalue of $M_\mu$, the choice of roots that consists in taking $z_0^{-1}$ instead of $z_0$ must correspond to $E(\mu)$ as well (because they coincide for $B=1$). We can therefore find the correct analytic continuations for $\mu$ and $E(\mu)$ in terms of $B$ by finding $z_0$, replacing its contribution in those series by that of $z_0^{-1}$, and taking $B$ back from $1$ to $0$. This procedure is explained in more detail in \cite{Derrida1996}.

We can find those two roots using equation (\ref{IV-2-HMC1}). An integration by parts turns it into:
\begin{equation}\label{IV-2-HMC1p}
H(B)\sim\frac{2}{3\sqrt{\pi}}\int_{-\infty}^{+\infty} d\theta~\theta^3 \frac{2\theta B {\rm e}^{-u}}{B {\rm e}^{-u}-1}.
\end{equation}
For $0<B<1$, the poles in this expression are at $\theta_{\pm}=\pm i \sqrt{-\log(B)}$. The corresponding residues are $i\frac{4\sqrt{\pi}}{3}\theta_{\pm}^3$ (as explained in \cite{Derrida1996}), and we must subtract the one corresponding to $\theta_-$ and add the other one to $H$. We get, for $B\rightarrow 0$:
\begin{align}
H(B)&=\frac{8}{3}\sqrt{\pi}\bigl[-\log(B)\bigl]^{3/2}-\sum_{k=1}^{\infty}\frac{B^k}{k^{5/2}}\sim\frac{8}{3}\sqrt{\pi}\bigl[-\log(B)\bigl]^{3/2}\\
B H'(B)&=-4\sqrt{\pi}\bigl[-\log(B)\bigl]^{1/2}-\sum_{k=1}^{\infty}\frac{B^k}{k^{3/2}}\sim-4\sqrt{\pi}\bigl[-\log(B)\bigl]^{1/2}.
\end{align}
Putting those together, we find, for $\mu<0$:
\begin{equation}\label{IV-2-gMC5}\boxed{
E_-(\mu)-\frac{1-q}{4}\mu\sim-(1-q)\frac{1}{48}\mu^{3}
}\end{equation}
and, for $j<J=\frac{1-q}{4}$,
\begin{equation}\label{IV-2-gMC7}\boxed{\boxed{
g_-(j)\sim(J-j)^{3/2}\frac{8}{3(1-q)^{1/2}}.
}}\end{equation}

Once more, we find a result which is independent of $L$. Moreover, we see that the phase transition that takes place here at $\mu=0$ is of a different nature than the one on the shock line: the behaviour of $g(j)$ with respect to $L$ changes from one side to the other.

\subsubsection{Maximal current phase}
\label{III-3-d}

Finally, we inspect the maximal current phase, for which $a<1$ and $b<1$. Once more, we start by calculating the large $L$ behaviour of every $C_k$ and $\tilde D_k$:
\begin{align}
\mu=-\frac{L^{-1/2}}{2\sqrt{\pi}}&\sum_{k=1}^{\infty}\frac{(2k)!}{k!k^{(k+3/2)}} B^k\label{IV-1-muMC2}\\
E(\mu)-(1-q)\frac{1}{4}\mu=-(1-q)\frac{L^{-3/2}}{16\sqrt{\pi}}&\sum_{k=1}^{\infty}\frac{(2k)!}{k!k^{(k+5/2)}} B^k\label{IV-1-EMC2},
\end{align}
so that, as in the previous section,
\begin{equation}
E_k\sim(1-q)\pi(\pi L)^{(k-3)/2}
\end{equation}
for $k\geq 2$.

We will use the same method as for the LD/MC line. We may still write
\begin{align}
\mu&=\frac{L^{-1/2}}{2\sqrt{\pi}}B ~H'(B)\\
E(\mu)-\frac{1-q}{4}\mu&=\frac{L^{-3/2}}{16\sqrt{\pi}}~H(B)
 \end{align}
but this time, $H(B)$ is defined as
\begin{equation}\label{IV-2-HLMC1}
H(B)=-\sum_{k=1}^{\infty}\frac{(2k)!}{k!k^{(k+5/2)}}\Bigl(\frac{B}{4}\Bigr)^k=\frac{2}{\sqrt{\pi}}\int_{-\infty}^{+\infty}d\theta~(\theta^2-1) \log\bigl[1-B\theta^2{\rm e}^{-\theta^2} \bigl].
\end{equation}

~~

For $\mu>0$, i.e. $B\rightarrow-\infty$, we have:
\begin{equation}
\log\bigl[1-B\theta^2{\rm e}^{-\theta^2} \bigl]\sim \log\bigl[|B|\theta^2{\rm e}^{-\theta^2} \bigl]~ \mathbb{I}\bigl[|B|\theta^2{\rm e}^{-\theta^2}>1\bigr].
\end{equation}

The upper bound of the integral can therefore be set at $\theta_B$ such that $|B|\theta_B^2{\rm e}^{-\theta_B^2}=1$, in which we recognise the square root of the Lambert $W$ function: $\theta_B=\sqrt{-W_{-1}(-1/B)}$. For large $B$, it behaves as $\log(|B|)^{1/2}$, as we saw in section \ref{III-3-b}. After estimating $H(B)$ in the same way as in the previous section we find the exact same expressions for $\mu$ and $E$, leading to
\begin{equation}\label{IV-2-gLMC2}\boxed{
E_+(\mu)-\frac{1-q}{4}\mu\sim(1-q)\frac{1}{20}\Bigl(\frac{3}{2}\Bigr)^{2/3} L^{-2/3}\pi^{2/3}\mu^{5/3}
}\end{equation}
and, for $j>J=\frac{1-q}{4}$,
\begin{equation}\label{IV-2-gLMC4}\boxed{\boxed{
g_+(j)\sim(j-J)^{5/2}\frac{32\sqrt{3}L}{5\pi(1-q)^{3/2}}.
}}\end{equation}
This tells us that the phase which is reached by selecting positive fluctuations of the current is the same whether one starts from the inside of the MC phase or from its boundary.

~~

\begin{figure}[ht]
\begin{center}
\includegraphics[width=0.9\textwidth]{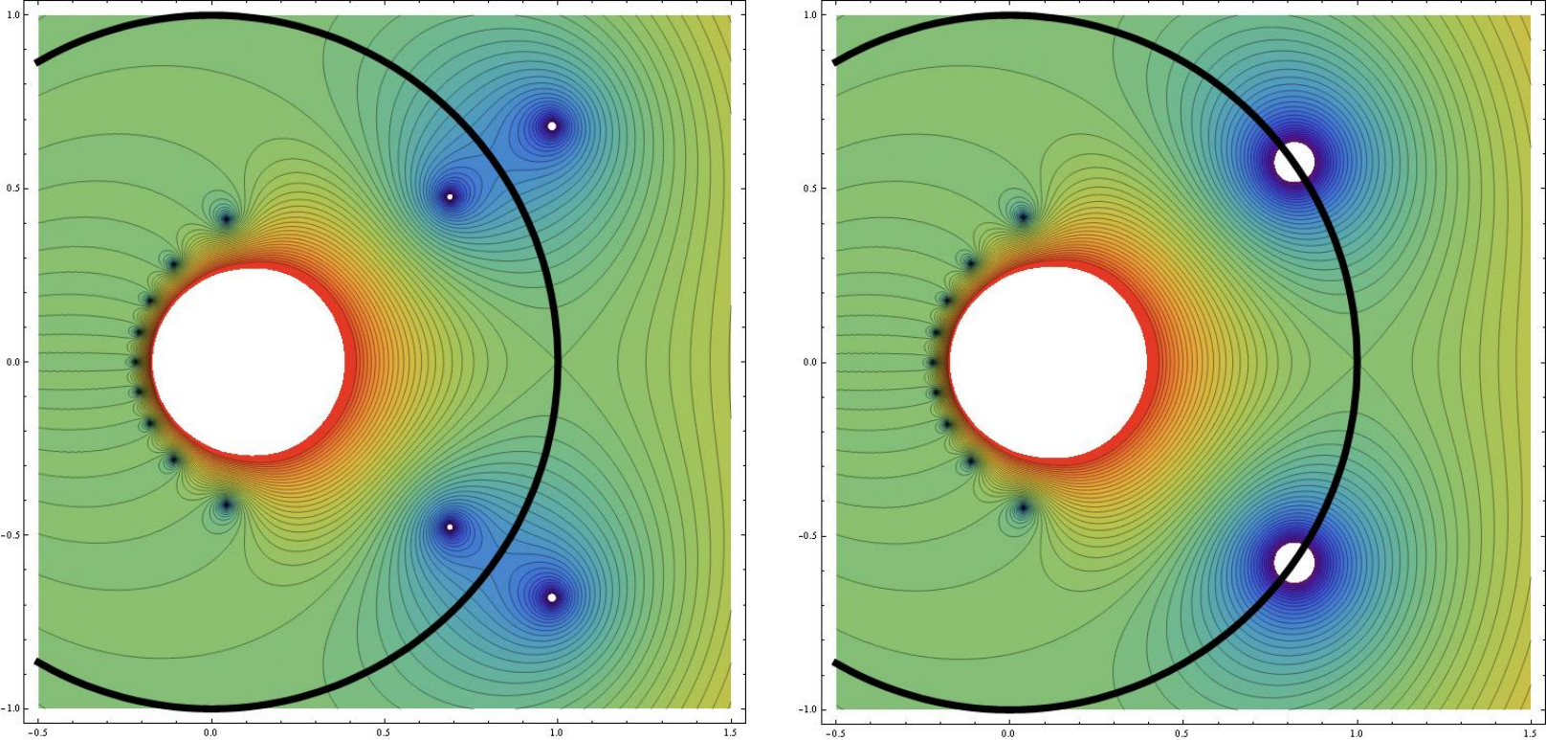}
\caption{Bethe roots for an open system with $9$ sites. The unit circle is represented in black. This time, there are two pairs of roots that merge for a critical value of $B$. Those roots get closer to the real axis as $L$ becomes large.}
\label{IV-fig-RootsOpen}
\end{center}
\end{figure}

For $\mu<0$, the situation is slightly different from that of the previous section. The roots of $(1-BF(x))$ behave similarly, but this time, there are two pairs of roots crossing the unit circle instead of one (fig.-\ref{IV-fig-RootsOpen}), both close to the real axis. Using the same procedure as before, we find them to have exactly the same behaviour with respect to $B$, the only difference being that we have twice as many residues as we had then.

This gives us:
\begin{equation}\label{IV-2-HLMC12}
H(B)\sim\frac{16}{3}\sqrt{\pi}\bigl[-\log(B)\bigl]^{3/2}
\end{equation}
and
\begin{equation}\label{IV-2-HLMC32}
B H'(B)\sim-8\sqrt{\pi}\bigl[-\log(B)\bigl]^{1/2}
\end{equation}
so that, for $\mu<0$,
\begin{equation}\label{IV-2-gLMC5}\boxed{
E_-(\mu)-\frac{1-q}{4}\mu\sim-(1-q)\frac{1}{192}\mu^{3}
}\end{equation}
and, for $j<J=\frac{1-q}{4}$,
\begin{equation}\label{IV-2-gLMC7}\boxed{\boxed{
g_-(j)\sim(J-j)^{3/2}\frac{16}{3(1-q)^{1/2}}.
}}\end{equation}

Note that we do not find the same behaviour for negative fluctuations of the current inside of the MC phase and on its boundaries, which indicates a phase transition between those two regions for $\mu<0$. We will explore that transition, and all those we mentioned in the previous sections, in more detail in section \ref{V}. But before that, we will see what may be said of $E(\mu)$ and $g(j)$ in the limit of large fluctuations.

\newpage

\section{Asymptotic limits for extreme currents}
\label{IV}

In the previous section, we obtained the behaviour of the generating function of the cumulants and of the large deviation function of the current for small fluctuations. In particular, we found evidence for a number of dynamical phase transitions, in the different forms these functions take for positive or negative fluctuations. In this section, we examine the opposite limit: that of extreme fluctuations. We will do this by taking $\mu$ to $\pm \infty$ in the deformed Markov matrix $M_\mu$, and diagonalising the limiting matrix directly, without use of integrability-related methods.

~

There are a few {\it caveats} which need to be pointed out before we start calculating. First of all, as we saw in section \ref{III-1}, the current in the open ASEP is proportional to the entropy production, and satisfies the Gallavotti-Cohen symmetry, which means that the $\mu\rightarrow-\infty$ limit, corresponding to a large negative current, can be deduces from the $\mu\rightarrow+\infty$ limit, corresponding to a large positive current. However, if we consider the totally asymmetric case instead, the Gallavotti-Cohen symmetry is destroyed, and the $\mu\rightarrow-\infty$ limit becomes that of a null current (simply because negative currents are strictly impossible). Whether the behaviour of the TASEP for $\mu\rightarrow-\infty$ and that of the ASEP for $j\rightarrow0^+$ (which corresponds to $\mu\rightarrow-\frac{1}{2}\log(\alpha\beta/\gamma \delta q^{L-1})^+$)are the same is not entirely obvious, but we will see evidence for it in section \ref{V}. In the meantime, we will focus on the TASEP when taking the $\mu\rightarrow-\infty$ limit. As for the $\mu\rightarrow+\infty$ limit, we will immediately see that considering the ASEP or the TASEP makes no difference whatsoever, and we will focus on the latter simply for the sake of consistency.

As for dealing with these limits in a proper way, one needs to be careful. Nothing guarantees that the equivalent of $M_\mu$ will be diagonalisable, and it will in fact not be in general. Consider for instance the case where we monitor the current on the first bond, and take $\mu$ to infinity. We are left with the matrix
\begin{equation}
M_\mu\sim m_0(\mu)\sim\begin{bmatrix}0 &0 \\ \alpha{\rm e}^{\mu} & 0  \end{bmatrix}
\end{equation}
acting only on site $1$, which is not diagonalisable. This comes from the fact that the eigenvectors do not have good limits in that basis: some entries, to which vanishing elements of $M_\mu$ apply, will diverge, which makes it inappropriate to neglect those elements. However, if we find a basis in which the equivalent of $M_\mu$ is diagonalisable, then the eigenelements of the limit are the limits of the eigenelements. Luckily, we already have a set of basis transformations at our disposal, which consist in changing the way in which we measure the current (see section \ref{III-1}). It turns out that putting a weight $\mu_i=\frac{\mu}{L+1}$ on each bond is the simplest distribution which will yield diagonalisable limits.

In this section, we will therefore consider the deformed Markov matrix defined as
\begin{equation}\label{IV-3-Mmu}
M_\mu=m_0(\mu_0)+\sum_{i=1}^{L-1} M_{i}(\mu_i)+m_L(\mu_l),
\end{equation}
where
\begin{equation}\label{IV-3-Mmui}
m_0(\mu_0)=\begin{bmatrix} -\alpha &0 \\ \alpha{\rm e}^{\mu_0} & 0  \end{bmatrix}~,~ M_{i}(\mu_i)=\begin{bmatrix} 0 & 0 & 0 & 0 \\ 0 & 0& {\rm e}^{\mu_i} & 0 \\ 0 &0& -1 & 0 \\ 0 & 0 & 0 & 0 \end{bmatrix}~,~m_L(\mu_l)=\begin{bmatrix} 0 & \beta{\rm e}^{\mu_L} \\  0 & -\beta  \end{bmatrix},
\end{equation}
with $\mu_i=\frac{\mu}{L+1}$ for all $i$'s, as our starting point.

\subsection{Low current limit}
\label{IV-1}

We first examine the $\mu\!\rightarrow\!-\infty$ limit. Noting $\varepsilon\!=\!{\rm e}^{\mu/(L+1)}\rightarrow 0$, we can decompose $M_\mu$ into its diagonal elements, which are finite, and its non-diagonal elements, which vanish:
\begin{equation}\label{IV-3-Mlc}
M_\mu=M_d+\varepsilon M_j
\end{equation}
where $M_d$ is the matrix containing the diagonal (escape) rates, and $M_j$ is the matrix containing the non-diagonal (jumping) rates. The entries of $M_d$ are given by:
\begin{equation}\label{IV-3-Md}
M_d({\cal C},{\cal C})=-(1-n_1)\alpha-\sum\limits_{i=1}^{L-1} n_i(1-n_{i+1})-n_L \beta
\end{equation}
where $n_i$ is the occupancy of site $i$ in ${\cal C}$. At lowest order in $\varepsilon$, those are the eigenvalues of $M_\mu$, since the non-diagonal part vanishes.

~

\begin{figure}[ht]
\begin{center}
 \includegraphics[width=0.6\textwidth]{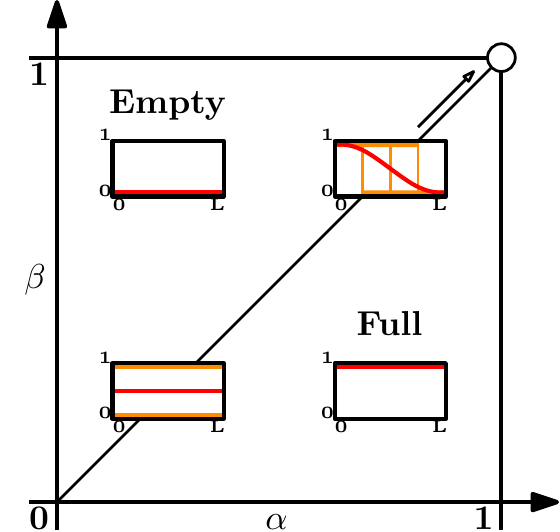}
  \caption{Phase diagram of the open ASEP for very low current. The mean density profiles are represented in red the insets. The profiles in orange are the individual configurations which compose the steady state.}
\label{fig-DiagLC}
 \end{center}
 \end{figure}

Since we are looking for the highest eigenvalue of $M_\mu$, we see that there are four possible situations (assuming that $\alpha$ and $\beta$ are limited to $[0,1]$):
\begin{itemize}
\item if $\alpha<\beta$, then the best configuration is empty ($n_i=0$ for all $i$'s), with an eigenvalue of $E=-\alpha$. If $\beta<\alpha$, we have the same in reverse: the best configuration is full ($n_i=1$ for all $i$'s) and $E=-\beta$ (those two first cases are symmetric to one another, and we will only be considering the first one) ;
\item If $\alpha=\beta<1$, then $E=-\alpha$, and we have two competing configurations: empty or full ;
\item if $\alpha=\beta=1$, then any configuration with a block of $1$'s followed by a block of $0$'s has an eigenvalue of $E=-1$, which is the highest, and there are $L+1$ of those.
\end{itemize}
The phase diagram of the model in that limit thus consists of two phases, one transition line, and one special point (fig.-\ref{fig-DiagLC}).

However, we will be looking to perform a Legendre transform on $E(\mu)$, so that this first order of the largest eigenvalue, which is independent of $\mu$, will not be enough: we will have to calculate the following order as well. We will therefore treat the non-diagonal part of $M_\mu$ perturbatively to extract its first non-trivial contribution to the largest eigenvalue.

\subsubsection{Empty/full phases}
\label{IV-1-a}

We first consider the case where $\alpha <\beta$, where the dominant eigenvalue of $M_\mu$ is equal to $-\alpha$ at leading order in $\varepsilon$. We expand that eigenvalue and its corresponding eigenvector as a series in $\varepsilon$:
\begin{equation}\label{IV-3-EMP}
E(\mu)=E_0+\sum E_k \varepsilon^k~~,~~|P_\mu\rangle=|P_0\rangle+\sum \varepsilon^k |P_k\rangle
\end{equation}
where
\begin{equation}\label{IV-3-EM}
E(\mu)|P_\mu\rangle=M_\mu|P_\mu\rangle.
\end{equation}
We already know that
\begin{equation}\label{IV-3-P0}
|P_0\rangle=|00\dots00\rangle~~,~~E_0=-\alpha.
\end{equation}
The first order in $\varepsilon$ in (\ref{IV-3-EM}) gives:
\begin{equation}\label{IV-3-P1}
E_1|P_0\rangle+E_0|P_1\rangle=M_j|P_0\rangle+M_d|P_1\rangle.
\end{equation}

Since the only state that can be reached from $|P_0\rangle$ through $M_j$ is $|10\dots00\rangle$, which has no overlap with $|P_0\rangle$, we get:
\begin{equation}\label{IV-3-E1}
|P_1\rangle=\frac{1}{E_0-M_d}M_j|P_0\rangle\sim|10\dots00\rangle ~~,~~E_1=0
\end{equation}
so that the first correction to $E(\mu)$ is $0$. The second order in $\varepsilon$ in (\ref{IV-3-EM}) gives:
\begin{equation}\label{IV-3-P2}
E_2|P_0\rangle+E_0|P_2\rangle=M_j|P_1\rangle+M_d|P_2\rangle.
\end{equation}

Again, the only state that can be reached from $|P_1\rangle$ through $M_j$ is $|010\dots00\rangle$, which has no overlap with $|P_0\rangle$:
\begin{equation}\label{IV-3-E2}
|P_2\rangle=\Bigl(\frac{1}{E_0-M_d}M_j\Bigr)^2|P_0\rangle\sim|010\dots00\rangle ~~,~~E_2=0
\end{equation}
and once more, the correction to $E(\mu)$ is $0$.

The first possible non-zero correction to $E(\mu)$ that we might get is when we find a $|P_k\rangle$ which has an overlap with $|P_0\rangle$. The shortest way to go back to $|P_0\rangle$ through jumps is to have one particle enter the system (the first step being $|P_1\rangle$), and then travel all the way to the other end, and exit the system. This can be done in $L+1$ steps, so that $E_k=0$ for $k:1..L$, and
\begin{equation}\label{IV-3-P22}
E_{L+1}|P_0\rangle+E_0|P_{L+1}\rangle=M_j|P_L\rangle+M_d|P_{L+1}\rangle.
\end{equation}

Putting those $L+1$ first equations together, we get
\begin{equation}\label{IV-3-E3}
|P_{L+1}\rangle= M_j\Bigl(\frac{1}{E_0-M_d}M_j\Bigr)^L|P_0\rangle\sim|P_0\rangle+\dots
\end{equation}
and
\begin{equation}
 E_{L+1}=\langle P_0|P_{L+1}\rangle=\frac{\alpha}{1-\alpha}.
\end{equation}

Putting this back into $E(\mu)$, we get:
\begin{equation}\label{IV-3-EmuLD}\boxed{
E(\mu)\sim -\alpha+{\rm e}^{\mu}\frac{\alpha}{1-\alpha}
}\end{equation}
and
\begin{equation}\label{IV-3-gjLD}\boxed{\boxed{
g(j)=\alpha+j\log(j)-j\bigl(\log(\alpha/(1-\alpha))+1\bigr).
}}\end{equation}

We may note that taking the limit $\mu\!\rightarrow\!-\infty$ in (\ref{IV-1-EmuTASEP}) gives the same result:
\begin{equation}\label{IV-3-EmuTASEP}
E(\mu)=\frac{a}{a+1} \frac{ {\rm e}^\mu -1 }{{\rm e}^\mu + a}\sim-\frac{1}{1+a}+\frac{1}{a}{\rm e}^\mu=-\alpha+{\rm e}^{\mu}\frac{\alpha}{1-\alpha}
\end{equation}
which indicates that this expression for $E(\mu)$ remains valid for all $\mu<0$. We will be more specific in section \ref{V}.

Finally, note that in this case, the second largest eigenvalue is $-\beta$ for $\varepsilon\rightarrow 0$ (and corresponds to a completely full system), so that the gap between the first two eigenvalues of $M_\mu$ is finite and equal, at leading order, to $\Delta E=(\beta-\alpha)$. 

The corresponding results for $\beta<\alpha$ (the `full' phase) can be obtained by exchanging $\alpha$ with $\beta$.

\subsubsection{Coexistence line}
\label{IV-1-b}

We now consider the slightly more complex case where $\alpha=\beta<1$. This time, there are two states with equal eigenvalues for $\mu=-\infty$:
\begin{equation}\label{IV-3-P0c}
|P_0\rangle=|0\rangle=|00\dots00\rangle~~{\rm with}~~E_0=-\alpha
\end{equation}
and
\begin{equation}\label{IV-3-P0c2}
|\tilde{P}_0\rangle=|1\rangle=|11\dots11\rangle~~{\rm with}~~\tilde{E}_0=-\alpha.
\end{equation}

As in the previous case, the first corrections to those eigenvalues are the rates with which we can go from these configurations back to themselves, but since they are degenerate, we must also consider how we can go from one to the other. As before, it takes the same $L+1$ steps to go from $|0\rangle$ to itself, or from $|1\rangle$ to itself, so that the first correction to both $E$ and $\tilde{E}$ is ${\rm e}^{\mu}\frac{\alpha}{1-\alpha}$. At this stage, those states are still degenerate. To lift the degeneracy, we have to consider the shortest way to go from $|0\rangle$ to $|1\rangle$, or the opposite. This means completely filling or emptying the system, and can be done in $L(L+1)/2$ steps. This tells us that the difference between the two highest eigenvalues is of order $\varepsilon^{L(L+1)/2}={\rm e}^{\frac{L}{2}\mu}$. For symmetry reasons, the main eigenvector is then $\frac{1}{2}(|0\rangle+|1\rangle)$, and the second one is $\frac{1}{2}(|0\rangle-|1\rangle)$.

In conclusion, we have, as in the previous case, 
\begin{equation}\label{IV-3-EmuLDHD}\boxed{
E(\mu)\sim -\alpha+{\rm e}^{\mu}\frac{\alpha}{1-\alpha}
}\end{equation}
and
\begin{equation}\label{IV-3-gjLDHD}\boxed{\boxed{
g(j)=\alpha+j\log(j)-j\bigl(\log(\alpha/(1-\alpha))+1\bigr)
}}\end{equation}
but this time, the gap behaves as $\Delta E\sim{\rm e}^{\frac{L}{2}\mu}$.

~~

To put these considerations in a more systematic format, we may use the so-called `resolvent formalism' \cite{Fredholm1903}. We present it here for two reasons: it will be useful to us in the next section, and it allows, in the case of a perturbation around degenerate states, to rigorously define an effective interaction matrix between those states. This gives us, in essence, a reduced dynamics for the system in the subset of phase space which contains only the dominant configurations.

This formalism can be stated as follows: for a general matrix $M$ with eigenvalues $E_i$ and eigenvectors $ |P_i\rangle$ and $\langle P_i|$, we may write
\begin{equation}\label{IV-3-proj}
\oint_{C}\frac{dz}{i 2\pi}\frac{1}{z-M}=\sum\limits_{E_i\in C} |P_i\rangle\langle P_i|
\end{equation}
where the sum is over the eigenvalues of $M$ which lie inside of the contour $C$. Moreover, we have
\begin{equation}\label{IV-3-Eproj}
\oint_{C}\frac{dz}{i 2\pi}\frac{z}{z-M}=\sum\limits_{E_i\in C} E_i|P_i\rangle\langle P_i|.
\end{equation}

Going back to the matter at hand, which is $M_\mu$ with $\alpha=\beta$, a good way to isolate the two dominant eigenvectors is to consider that same contour integral, with a contour close enough to $-\alpha$ so that the two highest eigenvalues are inside it, but not any of the others. This allows us to define an effective matrix $M_{eff}$ such that:
\begin{equation}\label{IV-3-Eproj2}
M_{eff}=-\alpha+\oint_{C}\frac{dz}{i 2\pi}\frac{z}{z-\alpha-M_d-\varepsilon M_j}
\end{equation}
where $C$ is a small circle centred at $0$.

We can now expand this expression in terms of $\varepsilon$:
\begin{equation}
M_{eff}=-\alpha+\oint_{C}\frac{dz}{i 2\pi} \sum\limits_{k=0}^{\infty}\frac{z}{z-(M_d+\alpha)}\Bigl(M_j \frac{1}{z-(M_d+\alpha)} \Bigr)^k \varepsilon ^k
\end{equation}
which is a sum over paths of length $k$, with transitions given by $M_j$ and a `potential' given by $(z-M_d-\alpha)^{-1}$. We see that the only terms which contribute to the integral (i.e. that give first order poles which yield non-zero residues) are those for which $M_d$ is taken at $-\alpha$ exactly twice, which is to say the paths that go through $|0\rangle$ or $|1\rangle$ twice.

It is now fairly straightforward to find the amplitudes of $M_{eff}$ between $|0\rangle$ and $|1\rangle$: we only have to project that expression between those states, and since $M_d=-\alpha$ in both of those, we only have to consider all the paths going from one of those states to another without going through them at any other point. Since we are doing a perturbative expansion in $\varepsilon$, we only need the term with the lowest number of steps. Between $|0\rangle$ and itself, or $|1\rangle$ and itself, there is only one path of the lowest length, which is $L+1$, and the amplitude for that path is $\frac{\alpha}{1-\alpha}$. Between $|0\rangle$ and $|1\rangle$, or the opposite, the shortest length is $L(L+1)/2$. There are many suitable paths for that transition, and the total amplitude is a quantity $X$ which we do not need explicitly, since that factor does not appear in the dominant term in $E(\mu)$ (it would be of order ${\rm e}^{\frac{L}{2}\mu}$).

All in all, we have an effective matrix given by:
\begin{equation}
M_{eff}=\begin{bmatrix} -\alpha +{\rm e}^{\mu}\frac{\alpha}{1-\alpha} &X {\rm e}^{\frac{L}{2}\mu}\\ X {\rm e}^{\frac{L}{2}\mu} &-\alpha +{\rm e}^{\mu}\frac{\alpha}{1-\alpha}  \end{bmatrix}
\end{equation}
which is easily diagonalised, and we can retrieve the results we found earlier.

\subsubsection{Equal rates point}
\label{IV-1-c}

For the last case, where all the jumping rates are equal ($\alpha=\beta=1$), we find $L+1$ states with an eigenvalue equal to $-1$ for $\mu=-\infty$. Those states are given by $|k\rangle=|\{1\}_k \{0\}_{L-k}\rangle$, i.e. configurations made of a block of $1$'s followed by a block of $0$'s. Those are called `anti-shocks', being symmetric to the usual shocks which have a low density region followed by a high density one.

Using the resolvent formalism, we find:
\begin{align}
\langle k| M_{eff} |k\rangle&\sim-1+\varepsilon^{L+1},\\
\langle k+1| M_{eff} |k\rangle&\sim\varepsilon^{k+1},\\
\langle k-1| M_{eff} |k\rangle&\sim\varepsilon^{L-k+1},
\end{align}
as well as terms of the type
\begin{align}
\langle k+2| M_{eff} |k\rangle&\sim X\varepsilon^{2k+3},\\
\langle k-2| M_{eff} |k\rangle&\sim Y\varepsilon^{2L-2k+3},\\
\langle k+3| M_{eff} |k\rangle&\sim Z\varepsilon^{3k+6},
\end{align}
and so on. We can check those last terms to be of sub-leading order in $E(\mu)$, and we will neglect them right away.

We are left with
\begin{equation}
M_{eff}=-1+\varepsilon^{L+1}+\sum\limits_{k=1}^{L}\varepsilon^{k}|k\rangle\langle k-1|+\varepsilon^{L-k+1}|k-1\rangle\langle k|.
\end{equation}
We can transform it through a matrix similarity to have all the non-diagonal coefficients be equal to $\varepsilon^{(L+1)/2}$, which yields:
\begin{equation}\label{IV-3-MeffHuckel}
\tilde M_{eff}=-1+\varepsilon^{L+1}+\varepsilon^{(L+1)/2}\sum\limits_{k=1}^{L}|k\rangle\langle k-1|+|k-1\rangle\langle k|.
\end{equation}

This is a well known tridiagonal matrix, used for instance to model the electronic interactions in conjugated dienes through the H\"{u}ckel method \cite{Coulson1978}. It is easily diagonalised (which is left as an exercise to the reader). Its eigenvalues are
\begin{equation}\label{IV-3-Ek}
E^{(k)}=-1+2\varepsilon^{(L+1)/2} \cos(k \pi/(L+2))
\end{equation}
for $k\in[\![1,L+1]\!]$. The highest one is
\begin{equation}\label{IV-3-Ek2}
E^{(1)}=-1+2\varepsilon^{(L+1)/2}\cos(\pi/(L+2))
\end{equation}
and the gap to the second one is
\begin{equation}
\Delta E=2\varepsilon^{(L+1)/2}\Bigl(\cos(\pi/(L+2))-\cos(2 \pi/(L+2))\Bigr)\sim  \frac{3\pi^2}{L^2}{\rm e}^{\mu/2}.
\end{equation}

Ultimately, we find that
\begin{equation}\label{IV-3-EER-}\boxed{
E(\mu)\sim-1+2{\rm e}^{\mu/2}
}\end{equation}
and
\begin{equation}\label{IV-3-gjLDc}\boxed{\boxed{
g(j)=1+2j\log(j)-2j.
}}\end{equation}
Moreover, knowing that the eigenvector associated to that first eigenvalue is distributed according to a sine function, which is to say that the probability of $|k\rangle$ is:
\begin{equation}
{\rm P}(k)\sim\sin\biggl(\frac{\pi k}{L+1}\biggr)^2
\end{equation}
we find that the mean density $\rho_n$ at site $n$ is of the form:
\begin{equation}\boxed{
\rho_n=1-\frac{n}{L+1}+\frac{1}{2\pi}\sin\biggl(\frac{2\pi n}{L+1}\biggr).
}\end{equation}
Note that this probability, being given by the product of the coefficients of the right and left dominant eigenvectors on state $|k\rangle$, corresponds to the probability of observing $|k\rangle$ conditioned on a low current in the original process $M_{eff}$, even though it is obtained from $\tilde M_{eff}$, because these matrices are similar.

\subsection{High current limit}
\label{IV-2}

We now consider the limit where $\mu\rightarrow\infty$. In this case, the diagonal part of $M_\mu$ is negligible, and the non-diagonal part is dominant. Since this non-diagonal part depends on $\mu$, we do not need to conserve the sub-dominant part in order to transform $E(\mu)$ to $g(j)$. To make certain upcoming calculations easier, we will shift our choice of $\mu_i$'s slightly, to
\begin{align}
\mu_0&=\frac{\mu}{L+1}+\frac{1}{L+1}\log(2\alpha\beta)-\log(\sqrt{2}\alpha),\\
\mu_i&=\frac{\mu}{L+1}+\frac{1}{L+1}\log(2\alpha\beta),\\
\mu_L&=\frac{\mu}{L+1}+\frac{1}{L+1}\log(2\alpha\beta)-\log(\sqrt{2}\beta),
\end{align}
so that we may write
\begin{equation}\label{IV-3-Mhc}
M_\mu\sim(2\alpha\beta{\rm e}^{\mu})^{\frac{1}{L+1}} M_j
\end{equation}
with
\begin{equation}\label{IV-3-M+hc}
M_j=\frac{1}{\sqrt{2}} S_1^+ +\sum\limits_{n=1}^{L-1}S_n^- S_{n+1}^+ +\frac{1}{\sqrt{2}}S_L^-
\end{equation}
where $S_n^\pm$ are the operators for the creation or annihilation of a particle at site $n$. Note that the dependence in $\mu$, but also $\alpha$ and $\beta$, is only in a global pre-factor in $M_\mu$, and hence in $E(\mu)$. This tells us that we should not expect any phase transitions in this limit (except perhaps at $\alpha=0$ or $\beta=0$): the largest eigenvalue of $M_\mu$ will simply be that pre-factor multiplied by a constant:
\begin{equation}
E(\mu)\propto(2\alpha\beta{\rm e}^{\mu})^{\frac{1}{L+1}}
\end{equation}
 This tells us already most of what we want to know  about the behaviour of the large deviations of the current for extremely large positive fluctuations, but we can learn much more, as it turns out that $M_j$ is exactly diagonalisable.

~

We may recognise $M_j$ to be the upper half of the Hamiltonian of an open XX spin chain \cite{Bilstein1999}. Moreover, it happens to commute with its transpose. We know, from the Perron-Frobenius theorem, that the highest eigenvalue of that matrix is real and non-degenerate. It is therefore also the highest eigenvalue of its transpose, with the same eigenvectors (because they commute). This allows us to define $H=\frac{1}{2}(M_j+{}^t\!M_j)$, which has the same highest eigenvalue and the same eigenvectors as $M_j$. $H$ is given by:
\begin{equation}\boxed{
H=\frac{1}{\sqrt{8}} S_1^x +\frac{1}{2}\sum\limits_{n=1}^{L-1}(S_n^- S_{n+1}^++S_n^+ S_{n+1}^-) +\frac{1}{\sqrt{8}}S_L^x
}\end{equation}
which is the Hamiltonian for the open XX chain with spin $1/2$ and extra boundary terms $S_1^x$ and $S_L^x$ (with $S^x=S^++S^-$). This spin chain was studied for general boundary conditions in \cite{Bilstein1999}. We will present here a simpler version of their calculations, only applicable to our situation but much less intricate than the general solution.

The main problem in dealing with $H$ is that it is not entirely quadratic. The first step in diagonalising it is to remedy this by considering two extra sites, one at $0$ and one at $L+1$, which we couple with our system by defining an new Hamiltonian:
\begin{equation}
\tilde{H}=\frac{1}{\sqrt{8}}S_0^x S_1^x +\frac{1}{2}\sum\limits_{n=1}^{L-1}(S_n^- S_{n+1}^++S_n^+ S_{n+1}^-) +\frac{1}{\sqrt{8}}S_L^xS_{L+1}^x.
\end{equation}

Since $[\tilde{H},S_0^x]=[\tilde{H},S_{L+1}^x]=0$, this modified Hamiltonian has four sectors, corresponding to the eigenspaces of $S_0^x$ and $S_{L+1}^x$. Since each of those has two eigenvalues $1$ and $-1$, we can recover $H$ by projecting $\tilde{H}$ onto the eigenspaces of $S_0^x$ and $S_{L+1}^x$ where both eigenvalues are $1$:
\begin{equation}
H=\frac{1}{4}\bigl(\!\langle 0_0|+\langle 1_0|\bigr)\!\otimes\!\bigl(\!\langle 0_{L+1}|+\langle 1_{L+1}|\bigr)\tilde{H}\bigl(|0_0\rangle+|1_0\rangle\!\bigr)\!\otimes\!\bigl(|0_{L+1}\rangle+|1_{L+1}\rangle\!\bigr).
\end{equation}

We are now left with diagonalising $\tilde{H}$. The rest of the calculation is a rather standard approach to quadratic spin chains.

~~

We first perform a Jordan-Wigner transformation on the operators $S_n^\pm$ :
\begin{align}
c_n=\Biggl(\prod\limits_{m=0}^{n-1}(-1)^{n_m}\Biggr)S_n^-~~~~&,~~~~c_n^\dag=\Biggl(\prod\limits_{m=0}^{n-1}(-1)^{n_m}\Biggr)S_n^+,\\
S_n^-=\Biggl(\prod\limits_{m=0}^{n-1}(-1)^{c_m^\dag c_m}\Biggr)c_n~~~~&,~~~~S_n^+=\Biggl(\prod\limits_{m=0}^{n-1}(-1)^{c_m^\dag c_m}\Biggr)c_n^\dag,
\end{align}
where $n_m$ is the number of particles on site $m$, with values in $\{0,1\}$. This yields fermionic operators:
\begin{align}
\{c_n^\dag,c_m\}&=\delta_{n,m},\\
\{c_n^\dag,c_m^\dag\}&=0,\\
\{c_n,c_m\}&=0.
\end{align}

The elements of $\tilde{H}$ become:
\begin{align}
S_n^+S_{n+1}^-&=c_n^\dag c_{n+1},\\
S_n^-S_{n+1}^+&=c_{n+1}^\dag c_n,\\
S_0^x S_1^x&=(c_0^\dag-c_0)(c_1^\dag+c_1),\\
S_L^xS_{L+1}^x&=(c_L^\dag-c_L)(c_{L+1}^\dag+c_{L+1}),
\end{align}
so that
\begin{equation}
\tilde{H}=\frac{1}{\sqrt{8}}(c_0^\dag-c_0)(c_1^\dag+c_1) +\frac{1}{2}\sum\limits_{n=1}^{L-1}(c_n^\dag c_{n+1}+c_{n+1}^\dag c_n) +\frac{1}{\sqrt{8}}(c_L^\dag-c_L)(c_{L+1}^\dag+c_{L+1}).
\end{equation}

We now perform a Bogoliubov transformation \cite{Kittel1987} on $\tilde{H}$, writing it as
\begin{equation}
\tilde{H}={\cal E}_0+\sum\limits_{k}{\cal E}_k d_k^\dag d_k
\end{equation}
with all the ${\cal E}_k>0$, and the $d_k$'s to be determined.

We want the $d_k$'s to be fermionic, so that $[\tilde{H},d_k^\dag]={\cal E}_k d_k^\dag$, which is the equation we will now try to solve. We have two trivial solutions with energy $0$ (called zero-modes): $(c_0^\dag+c_0)$ and $(c_{L+1}^\dag-c_{L+1})$. For the other solutions, we write:
\begin{equation}
d_k^\dag=\frac{X^{(k)}}{\sqrt{2}}(c_0^\dag-c_0)+\sum\limits_{n=1}^{L}A_i^{(k)}c_n^\dag+B_n^{(k)}c_n+\frac{Y^{(k)}}{\sqrt{2}}(c_{L+1}^\dag+c_{L+1})
\end{equation}
and $[\tilde{H},d_k^\dag]={\cal E}_k d_k^\dag$ becomes:
\begin{align}
A_{n+1}^{(k)}+A_{n-1}^{(k)}&=2{\cal E}_k A_n^{(k)}~~~~{\rm for}~~ n\in\llbracket2,L-1\rrbracket,\\
-B_{n+1}^{(k)}-B_{n-1}^{(k)}&=2{\cal E}_k B_n^{(k)}~~~~{\rm for}~~ n\in\llbracket2,L-1\rrbracket,\\
X^{(k)}+A_2^{(k)}&=2{\cal E}_k A_1^{(k)},\\
X^{(k)}-B_2^{(k)}&=2{\cal E}_k B_1^{(k)},\\
A_{1}^{(k)}+B_{1}^{(k)}&=2{\cal E}_k X^{(k)},\\
Y^{(k)}+A_{L-1}^{(k)}&=2{\cal E}_k A_L^{(k)},\\
-Y^{(k)}-B_{L-1}^{(k)}&=2{\cal E}_k B_L^{(k)},\\
A_{L}^{(k)}-B_{L}^{(k)}&=2{\cal E}_k Y^{(k)}.
\end{align}
All those equations can be written in a more compact form by defining:
\begin{align}
A_{L+1}^{(k)}&=Y^{(k)},\\
A_{L+1+n}^{(k)}&=(-1)^{n}B_{L+1-n}^{(k)},\\
A_{0}^{(k)}&=X^{(k)},
\end{align}
for which they become:
\begin{align}
A_{n+1}^{(k)}+A_{n-1}^{(k)}&=2{\cal E}_k A_n^{(k)}~~~~{\rm for}~~ n\in\llbracket1,2L\rrbracket \label{IV-3-Anbulk},\\
A_{2L}^{(k)}+(-1)^{L}A_{0}^{(k)}&=2{\cal E}_k A_{2L+1}^{(k)},\\
(-1)^{L}A_{2L+1}^{(k)}+A_{1}^{(k)}&=2{\cal E}_k A_{0}^{(k)}.
\end{align}

These are the same equations which we would have found for a periodic $XX$ spin chain with $2L+2$ sites, the only (but important) difference being that $d_k^\dag$ mixes $c_k$'s and $c_k^\dag$'s, so that the total spin is not conserved.

We look for plane wave solutions of the form $A_n=r^n$, with $2{\cal E}=r+\frac{1}{r}$. This automatically solves eq.(\ref{IV-3-Anbulk}). The other two equations become:
\begin{align}
r^{2L}+(-1)^{L}&=r^{2L}+r^{2L+2},\\
(-1)^{L}r^{2L+1}+r&=r+\frac{1}{r}
\end{align}
and both simplify into
\begin{equation}
r^{2L+2}=(-1)^{L}.
\end{equation}

We have $2L+2$ solutions to this equation, given by $r=\omega_k={\rm e}^{\frac{i\pi(L-2k+2)}{2L+2}}$ for $ k\in\llbracket1,2L+2\rrbracket$, so that $A_n^{(k)}=\omega_k^n$, and the energies are given by:
\begin{equation}\boxed{
{\cal E}_k=\cos\biggl(\frac{(L-2k+2)\pi}{2L+2}\biggr)=\sin\biggl(\frac{(2k-1)\pi}{2L+2}\biggr)~~~~~~~~{\rm for}~~ k\in\llbracket1,2L+2\rrbracket.
}\end{equation}

We can then write the $d_k^\dag$'s as:
\begin{equation}
d_k^\dag=\frac{1}{\sqrt{2L+2}}\Biggl(\frac{1}{\sqrt{2}}(c_0^\dag-c_0)+\sum\limits_{n=1}^{L}\omega_k^n c_n^\dag-(-\omega_k)^{-n}c_n+\frac{\omega_k^{L+1}}{\sqrt{2}}(c_{L+1}^\dag+c_{L+1})\Biggr)
\end{equation}
and the inverse relations as:
\begin{align}
\frac{1}{\sqrt{2}}(c_0^\dag-c_0)&=\frac{1}{\sqrt{2L+2}}\sum\limits_{k=1}^{2L+2}d_k^\dag,\\
c_n^\dag&=\frac{1}{\sqrt{2L+2}}\sum\limits_{k=1}^{2L+2} \omega_k^{-n} d_k^\dag,\\
c_n&=\frac{1}{\sqrt{2L+2}}\sum\limits_{k=1}^{2L+2}- (-\omega_k)^{n} d_k^\dag,\\
\frac{1}{\sqrt{2}}(c_{L+1}^\dag+c_{L+1})&=\frac{1}{\sqrt{2L+2}}\sum\limits_{k=1}^{2L+2} \omega_k^{-L-1} d_k^\dag\label{CL+1}.
\end{align}

Note that the $d_k^\dag$'s are fermions, but, because there are $2L+2$ of them, and only $L+2$ of the $c_k^\dag$'s, they are not all independent: we have $\omega_{L+1+k}=-\omega_k$, so that $d_{L+1+k}^\dag=-d_k$.

~~

We now need to determine the constant term ${\cal E}_0$ in $\tilde{H}$. Considering only the scalar terms in $\sum\limits_{k}{\cal E}_k d_k^\dag d_k$, we find:
\begin{align}
&\sum\limits_{k=1}^{L+1}{\cal E}_k\frac{1}{2L+2}\Biggl(\frac{1}{2}(c_0^\dag-c_0)(c_0-c_0^\dag)+\sum\limits_{n=1}^{L}c_n^\dag c_n+c_n c_n^\dag+\frac{1}{2}(c_{L+1}^\dag+c_{L+1})(c_{L+1}+c_{L+1}^\dag)\Biggr)\nonumber\\
&~~=\frac{1}{2}\sum\limits_{k=1}^{L+1}{\cal E}_k.
\end{align}

Since there is no scalar part in $\tilde{H}$, we must therefore have:
\begin{equation}
{\cal E}_0=-\frac{1}{2}\sum\limits_{k=1}^{L+1}{\cal E}_k.
\end{equation}

The highest eigenvalue can then be obtained by considering the state with all the energy levels occupied:
\begin{equation}\boxed{
E={\cal E}_0+\sum\limits_{k=1}^{L+1}{\cal E}_k=\frac{1}{2}\sin\Bigl(\frac{\pi}{2L+2}\Bigr)^{-1}\sim \frac{L}{\pi}.
}\end{equation}
The corresponding eigenstate is defined by
\begin{equation}\label{IV-3-psi}\boxed{
|\psi\rangle=\prod\limits_{k=1}^{L+1}d_k^\dag|\Omega\rangle
}\end{equation}
which is such that $d_k^\dag|\psi\rangle=0$ for $k\in[\![1,L+1]\!]$. The vector $|\Omega\rangle$ is arbitrary (provided that it is not in the kernel of any of the $d_k^\dag$'s that we apply to it).

Remembering the global factor which we took out of $M_\mu$ at the beginning of this section, we finally get:
\begin{equation}\boxed{
E(\mu)\sim\frac{L}{\pi}{\rm e}^{\mu/L}
}\end{equation}
and
\begin{equation}\label{IV-3-gMMC}\boxed{\boxed{
g(j)\sim L\bigl( j\log(j)-j(1-\log(\pi))\bigl)
}}\end{equation}
which is proportional to $L$, consistently with eq.(\ref{IV-2-gLMC4}) which was obtained for positive fluctuations of the current in the MC phase.

~~

We now have what we wanted, but we can have a look at the eigenvector (\ref{IV-3-psi}) as well.

The first and easiest calculation that we can do here is that of the two-point correlations in $|\psi\rangle$. The connected correlation between the occupancies of sites $n$ and $m$ is given by:
\begin{align}
C_{nm}&=\langle \psi|c^\dag_n c_n c^\dag_m c_m|\psi\rangle-\langle \psi|c^\dag_n c_n |\psi\rangle \langle \psi|c^\dag_m c_m|\psi\rangle\nonumber\\
&=-\langle \psi|c^\dag_n c^\dag_m|\psi\rangle \langle \psi|c_n c_m|\psi\rangle+\langle \psi|c^\dag_n c_m |\psi\rangle \langle \psi|c_n c^\dag_m |\psi\rangle
\end{align}
(using Wick's theorem). We find that:
\begin{align}
\langle \psi|c^\dag_n c_m |\psi\rangle&=\frac{1}{2L+2}\sum\limits_{k=1}^{L+1} \omega_k^{m-n},  \label{IV-cdncm}\\
\langle \psi|c_n c^\dag_m |\psi\rangle&=\frac{1}{2L+2}\sum\limits_{k=1}^{L+1} (-\omega_k)^{n-m},   \label{IV-cncdm}\\
\langle \psi|c^\dag_n c^\dag_m|\psi\rangle&=-\frac{1}{2L+2}\sum\limits_{k=1}^{L+1} \omega_k^{-n}(-\omega_{k})^{-m},   \label{IV-cdncdm}\\
\langle \psi|c_n c_m|\psi\rangle&=-\frac{1}{2L+2}\sum\limits_{k=1}^{L+1} (-\omega_{k})^{n}\omega_k^{m}, \label{IV-cncm}
\end{align}
If $n$ and $m$ have same parity, each of those terms sum to $0$ (unless $n=m$ in the first two sums, but here we consider two different sites). If not, we get:
\begin{align}
\langle \psi|c^\dag_n c_m |\psi\rangle&=\frac{1}{L+1}\frac{{\rm e}^{i\pi(L)(m-n)/(2L+2)}}{1-{\rm e}^{-i\pi(m-n)/(L+1)}} , \\
\langle \psi|c^\dag_n c^\dag_m|\psi\rangle&=-\frac{1}{L+1}(-1)^m\frac{{\rm e}^{i\pi(L)(-m-n)/(2L+2)}}{1-{\rm e}^{i\pi(m+n)/(L+1)}} , \\
\langle \psi|c_n c_m|\psi\rangle&= -\frac{1}{L+1}(-1)^n\frac{{\rm e}^{i\pi(L)(m+n)/(2L+2)}}{1-{\rm e}^{-i\pi(m+n)/(L+1)}},
\end{align}
so that
\begin{equation}\boxed{
C_{mn}=\frac{1}{4(L+1)^2\sin^2\Bigl(\frac{\pi(m+n)}{(2L+2)}\Bigr)}-\frac{1}{4(L+1)^2\sin^2\Bigl(\frac{\pi(m-n)}{(2L+2)}\Bigr)}.
}\end{equation}

The correlations are therefore exactly $0$ for sites which are an even number of bonds apart (as is the case for a half-filled periodic chain \cite{Popkov2010}), and behave as
\begin{equation}\label{IV-3-corrMC}
C_{mn}\sim -\frac{1}{\pi^2 (m-n)^2}
\end{equation}
otherwise, if the two sites are far away enough from the boundaries. Note that those correlations do not vanish with the size of the system, in contrast with the steady state of the ASEP at $\mu=0$, where they behave as $L^{-1}$ in the maximal current phase and vanished exponentially in the high and low density phases \cite{Derrida1993a}.

~~

We can now examine the (un-normalised) probability of any given configuration. This can be expressed as
\begin{equation}\label{Phnpm}
P(\{h_n,p_n\})=\langle \psi|\prod\limits_{n=0}^{N}c_{h_n}^{} c_{h_n}^{\dag}\frac{(c_{L+1}^{\dag}+c_{L+1})^2}{2}\prod\limits_{n=N+1}^{L}c_{p_n}^{\dag} c_{p_n}^{}|\psi\rangle
\end{equation}
which is to say that we select only the configuration that has holes at positions $\{h_n\}$ and particles at positions $\{p_n\}$ in $|\psi\rangle$, and project it onto its Hermitian conjugate. Note that the term $\frac{(c_{L+1}^{\dag}+c_{L+1})^2}{2}$ makes no difference (since it is equal to a constant factor $\frac{1}{2}$), but is there to have effectively $L+1$ sites instead of $L$, which will be useful shortly. From the expression of this term in (\ref{CL+1}), we see that it corresponds to having a hole (or, in fact, a particle) at site $L+1$. Note also that the terms in (\ref{Phnpm}) can be reordered as long as any pair $\{c_n,c^\dag_n\}$ is kept in the same order, so that we can regroup all the $c$'s from the first product to the left, for instance.

We can now use Wick's theorem \cite{Stephan2010} on this expression, and write it as the Pfaffian of an anti-symmetric matrix ${\cal A}$ whose upper triangle consists of all the mean values $\langle \psi|ab|\psi\rangle$ where $a$ and $b$ are two terms from the product in (\ref{Phnpm}), taken in the same order. We can write it as a block matrix:
\begin{equation}
{\cal A}=\begin{bmatrix}
A_1 & \langle \psi|c_{h_n}c_{h_m}^\dag |\psi\rangle & \langle \psi|c_{h_n} c_{p_m}^\dag |\psi\rangle & \langle \psi|c_{h_n} c_{p_m}|\psi\rangle \\
-\langle \psi|c_{h_m}c_{h_n}^\dag |\psi\rangle & A_2 & \langle \psi|c_{h_n}^\dag c_{p_m}^\dag  |\psi\rangle &  \langle \psi|c_{h_n}^\dag c_{p_m} |\psi\rangle\\
-\langle \psi|c_{h_m} c_{p_n}^\dag  |\psi\rangle & -\langle \psi|c_{h_m}^\dag c_{p_n}^\dag |\psi\rangle & A_3 & \langle \psi|c_{p_n}^\dag c_{p_m} |\psi\rangle\\
-\langle \psi|c_{h_m} c_{p_n} |\psi\rangle & -\langle \psi|c_{h_m}^\dag c_{p_n}  |\psi\rangle & -\langle \psi|c_{p_m}^\dag c_{p_n} |\psi\rangle & A_4
\end{bmatrix}
\end{equation}
where $A_1|_{n,m}= \langle \psi|c_{h_n}c_{h_m} |\psi\rangle$ if $n<m$ and $A_1|_{n,m}=- \langle \psi|c_{h_m}c_{h_n} |\psi\rangle$ if $n>m$. The same goes for $A_2$ with $c_{h_n}^\dag$, $A_3$ with $c_{p_n}$ and $A_4$ with $c_{p_n}^\dag$.

Looking at the expression given in (\ref{IV-cdncm}) to (\ref{IV-cncm}), we see that $- \langle \psi|c_m c_n |\psi\rangle=\langle \psi|c_n c_m |\psi\rangle$, $- \langle \psi|c_m^\dag c_n^\dag |\psi\rangle=\langle \psi|c_n^\dag c_m^\dag |\psi\rangle$ and $- \langle \psi|c_m c_n^\dag |\psi\rangle=\langle \psi|c_n^\dag c_m |\psi\rangle-\delta_{n,m}$.  We also note that all those block matrices can be factorised in a simple way: if we define
\begin{align}
X^-_{n,k}=-(-w_k)^{h_n},~~~~~~&~~~~~~X^+_{n,k}=w_k^{-h_n},\\
Y^-_{n,k}=-(-w_k)^{p_n},~~~~~~&~~~~~~Y^+_{n,k}=w_k^{-p_n},
\end{align}
then ${\cal A}$ can be rewritten as
\begin{equation}
{\cal A}=\begin{bmatrix}
X^-(X^+)^\dag & X^-(X^-)^\dag & X^-(Y^-)^\dag & X^-(Y^+)^\dag\\
X^+(X^+)^\dag -1_h& X^+(X^-)^\dag & X^+(Y^-)^\dag& X^+(Y^+)^\dag \\
Y^+(X^+)^\dag & Y^+(X^-)^\dag & Y^+(Y^-)^\dag& Y^+(Y^+)^\dag \\
Y^-(X^+)^\dag & Y^-(X^-)^\dag & Y^-(Y^-)^\dag-1_p& Y^-(Y^+)^\dag 
\end{bmatrix}
\end{equation}
where $1_h$ and $1_p$ are identity matrices whose respective sizes are the numbers of holes and particles (one of which is occupying site $L+1$, so that the sum of the two numbers is $L+1$).

In order to calculate $P(\{h_n,p_n\})={\rm Pf}\bigl[{\cal A} \bigr]$, we need to first consider its square $P(\{h_n,p_n\})^2={\rm Det}\bigl[{\cal A} \bigr]$. After exchanging a few lines and columns in that determinant, we can write it in a factorised form:
\begin{equation}
{\rm Det}\bigl[{\cal A} \bigr]={\rm Det}\left[\begin{array}{ c| c} \substack{X^+\\Y^-}&-1_{L+1}\\ \hline \substack{X^-\\Y^+}&0\end{array}\right]\centerdot\left[\begin{array}{ c| c} (X^+)^\dag ~(Y^-)^\dag&(X^-)^\dag ~(Y^+)^\dag\\ \hline 1_{L+1}&0\end{array}\right].
\end{equation}
Each block in this last expression is of a square matrix of size $L+1$ (which is where the fact that we included site $L+1$ becomes useful). That determinant then reduces to:
\begin{equation}
P(\{h_n,p_n\})^2={\rm Det}\bigl[\substack{X^-\\Y^+} \bigr]{\rm Det}\bigl[(X^-)^\dag ~(Y^+)^\dag\bigr]=\big|{\rm Det}\bigl[\substack{X^-\\Y^+} \bigr]\big|^2.
\end{equation}

After a few final simplifications, we find
\begin{equation}\boxed{
P(\{h_n,p_n\})={\rm Det}\bigl[\omega^{h_n k}, \omega^{-p_n k} \bigr]
}\end{equation}
where $\omega={\rm e}^{\frac{i\pi}{L+1}}$. We recognise this to be Vandermonde determinant which gives, for any configuration:
\begin{equation}\boxed{\boxed{
P(\{n_n\})=\prod\limits_{n_n=n_m}[\sin(r_m - r_n)]\prod\limits_{n_n\neq n_m}[\sin(r_m + r_n)]
}}\end{equation}
where $n_n$ is the occupancy of site $n$, and $r_n=n \pi/(2L+2)$. Note that all these probabilities are still un-normalised.

\begin{figure}[ht]
\begin{center}
 \includegraphics[width=0.5\textwidth]{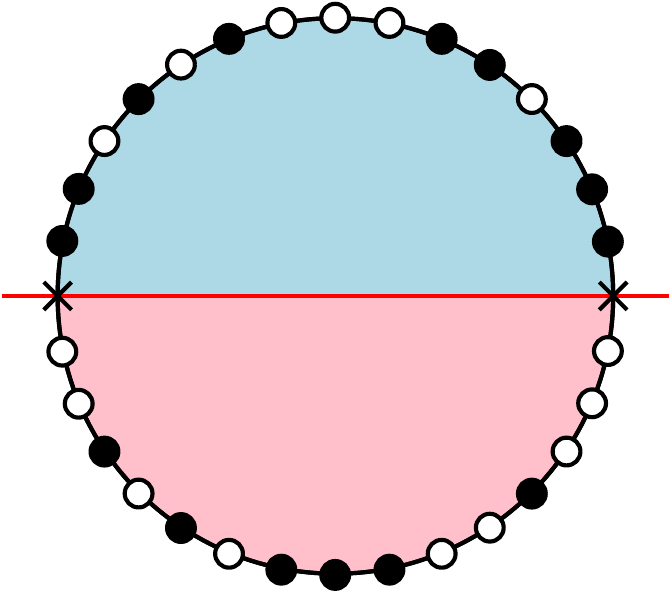}
  \caption{Dyson-Gaudin gas equivalent for the configuration $(110101000110111)$ for the open ASEP conditioned on a large current. The lower part of the system is deduced from the upper part by an axial anti-symmetry.}
\label{fig-DGgas}
 \end{center}
 \end{figure}

This distribution is exactly that of a Dyson-Gaudin gas \cite{Gaudin1973}, which is a discrete version of the Coulomb gas, on a periodic lattice of size $2L+2$, with two defect sites (at $0$ and $L+1$) that have no occupancy, and a reflection anti-symmetry between one side of the system and the other (fig.-\ref{fig-DGgas}). The first (upper) part of the gas is given by the configuration we are considering, and the second (lower) is deduced by anti-symmetry. The interaction potential between two particles at positions $r_n$ and $r_m$ is then given by:
\begin{equation}
V(r_n,r_m)=-\log\bigl(\sin(r_m - r_n)\bigr).
\end{equation}

It was shown in \cite{Popkov2010} that the large current limit of the steady state of the periodic ASEP of size $L$ also converges to a Dyson-Gaudin gas (the standard periodic case, without defects or symmetry).

We should note that the trick consisting in taking the sum of $M_j$ and its transpose to reconstruct an XX spin chain is not in fact necessary \cite{Schutz}. All the calculations we saw can be performed, in a slightly different way, on $M_j$ directly, which has the added advantage that the imaginary part of the other eigenvalues is not lost.

\subsection{Generic dynamical phase transition}
\label{IV-3}

As we stated in the introduction to the present section, none of the methods we used to analyse the $\mu\rightarrow\pm\infty$ limits appear to rely on the ASEP being integrable. It could be, however, that they do require it in a subtle way. This is, in fact, not the case, as it turns out that those two limits can be treated in the exact same way, and yield very similar results, for a relatively broad class of models which are in general not integrable \cite{Lazarescua}.

~

Let us consider a generalisation of the open TASEP, with site-dependent jump rates $p_i$, as well as an arbitrary (but finite) extra interaction potential $V({\cal C})$ which may depend on the whole configuration ${\cal C}$, so that the rate for a particle to jump from site $i$, turning configuration ${\cal C}$ into ${\cal C}'$, is of the form $p_i{\rm e}^{(V({\cal C}')-V({\cal C}))/2}$. The methods we used for the TASEP can be applied in precisely the same way to that new model.

In the low current limit, we found for the TASEP that the first order in perturbation of the largest eigenvalue did not depend on the size of the system, in every case. This comes from the fact that, whether the dominant eigenstate at order $0$ is degenerate or not, the fastest way for the system to leave one of these states and come back to it is to have one particle jump through the whole system once, which corresponds to one quantum of current. This remains true for the generalised model we are considering now, so that {\it the large deviation function of the current does not depend on the size of the system for low currents}.

In the large current limit, the similarity is even stronger, as was noticed in \cite{Schutz2015}. The non-diagonal part of the deformed Markov matrix which we are keeping in that limit is entirely equivalent to that of the simple TASEP, up to a matrix similarity, so that its whole spectrum is the same (up to a multiplicative constant, in fact). The eigenvectors are also related to those of the TASEP, and the product of the right and left dominant eigenvectors, which give the stationary density conditioned on a large current, is identical. More importantly, {\it the large deviation function of the current is linear in the size of the system for large currents}.

One will be able to find a detailed account of these statements in \cite{Lazarescua}.

~

In the case of the open TASEP, these two limits are compatible with the results obtained for small fluctuations of the current around the MC phase: the large deviation function is independent of the size of the system for negative fluctuations, and proportional to it for positive fluctuations, so that a dynamical phase transition takes place at the average current. In the next section, we will see that this comes from a transition in the appropriate coarse-graining scheme in the large size limit. For currents which are below $\frac{1}{4}$, which is the average current in the MC phase, the system can be describes through a hydrodynamic (i.e. mean field) description, where the large size behaviour of the system depends only on the local density. In that scenario, as we shall see shortly, the optimal way to produce a fluctuation of the current is to have a localised fluctuation of the density. The localised nature of that fluctuation means that its cost, in terms of probability, will not depend on the size of the system.

However, as we saw in section \ref{II-2}, the largest current which can be produced in the mean-field approach is $\frac{1}{4}$. In order to have the current fluctuate above that value, one has to introduce correlations, spread out through the entire system, which can be seen in the large current limit. Since these correlations are needed everywhere in the system, the probability cost is extensive, hence the factor $L$ in the large deviation function of the current. Moreover, because of those correlations, the local density fluctuates much less that in the hydrodynamic case, so that these states are sometimes called `hyperuniform' \cite{Jack2015}.

~

One would hope that these observations remain true in the case of generalised rates, and that the behaviour of the large deviation function of the current in these two limits is a good indication of the presence of a dynamical phase transition somewhere in between. It is not known, for the moment, whether there are any universal features to that transition. It would be particularly interesting to examine the case where the jumping rates are disordered.

\newpage

\section{Hydrodynamic description and dynamical phase diagram}
\label{V}

In the previous sections, we obtained, by various exact calculation methods, the behaviour of the large deviation function of the current of the open TASEP in three different limits: around the average current, and for very high and very low currents. In this final section, we will see how the gaps between those limits can be bridged through a hydrodynamic description of the system, based on the macroscopic fluctuation theory (MFT, \cite{1742-5468-2007-07-P07023}). Although this method is much less rigorous than the exact calculations performed earlier (because it relies on a poorly controlled exchange of limits), we will see that it gives much more information on the fluctuations of the currents, in the cases where it applies, and gives us access to the complete dynamical phase diagram of the open ASEP. Five phases will be obtained: a low and a high density phase, which are the continuation of those same phases for the average current, and which meet the empty and full phases in the low current limit ; a shock phase, which stems from the shock line, for positive fluctuations of the current ; an anti-shock phase, which is obtained by decreasing the current from the maximal current phase of the average current ; and a maximal current phase, which is obtained for $j>\frac{1}{4}$. In the first four of these phases, the large deviation function of the current is independent of the size of the system, and the results obtained with this method agree with what we obtained earlier in the appropriate limits. In the last phase, the hydrodynamic description breaks down, as the system goes through the dynamical phase transition which we discussed at the end of the previous section.

We will start by presenting the principle of the method, which was first used in \cite{Bodineau2006} on the open ASEP, but works equally well for more general one-dimensional driven lattice gases. We will then use it to describe all the dynamical phases of the ASEP, obtaining the large deviation function of the current, as well as the form of the typical profiles conditioned on that current, save for the maximal current phase where the method breaks down. This will allow us to draw the dynamical phase diagram of the system. Finally, we will comment on evidence showing that this method yields not only the dominant state and eigenvalue of the deformed Markov matrix, but also a large family of higher eigenstates.

\subsection{Macroscopic Fluctuation Theory: from WASEP to ASEP}
\label{V-1}

The method which we will be using in this section, and which can be found in \cite{Bodineau2006}, is based on the macroscopic fluctuation theory \cite{1742-5468-2007-07-P07023}, or MFT for short. The principle is the following: for a noisy Langevin equation with a Gaussian noise, such as a noisy Burgers equation, the probability of a certain history is given by the Gaussian weight of the noise which generates it. For instance, for a driven one-dimensional interacting particle system, with a conductivity $\sigma(\rho)$ and a diffusion coefficient $D(\rho)$, subject to a field $F$, the local density verifies the equation
\begin{equation}\label{LangevinGas}
{\rm d}_t \rho=-\nabla j~~~~{\rm with}~~~~j=F\sigma(\rho)-D(\rho)\nabla\rho+\sqrt{\sigma(\rho)}\xi,
\end{equation}
where $\xi$ is a Gaussian noise with mean $0$ and variance $1$, and it is implied that $\rho$ and $j$ depend on $x$ and $t$. The probability of a certain history $\{j(x,t),\rho(x,t)\}$ is thus given by
\begin{equation}
-\log\Bigl({\rm P}\bigl[\{j(x,t),\rho(x,t)\}\bigr]\Bigr)\sim\int_0^t d\tau\int_0^1\frac{\bigl[j-F\sigma(\rho)+D(\rho)\nabla\rho\bigr]^2}{2\sigma(\rho)}dx
\end{equation}
with the constraint that ${\rm d}_t \rho=-\nabla j$. This gives explicitly the joint large deviation function of the current and density. In order to obtain that of the current alone, one then has to minimise that quadratic action with respect to the density (as stated by the contraction principle ; c.f. section \ref{I-1}). This minimisation will give us not only the large deviation function of the current, but also the optimal density profile to produce that current, which is the value of $\rho$ realising the minimum.

However, this cannot be directly applied to the ASEP: looking back at to eq.(\ref{II-1-Jx}), which is the deterministic part of the Langevin equation, we see that $D=\frac{1+q}{2L}$ vanishes at large sizes, or, equivalently (through a rescaling), that the field $F\sim L(1-q)$ diverges, leaving us with an inviscid Burgers equation.

We will use this approach nonetheless, by starting from the weakly asymmetric simple exclusion process (or WASEP), for which $F$ is finite (which is to say $(1-q)\sim L^{-1}$), or in fact the general system described by eq.(\ref{LangevinGas}). We will then minimise the MFT action with respect to the density, after which we will take $F$ to be of order $L$ again.

~~

We therefore start with the joint large deviation function of the current and density, given by
\begin{equation}\label{IV-2-gjSigmaDt}
g(j,\rho)=\lim\limits_{t\rightarrow+\infty}\frac{1}{t}\int_0^t d\tau\int_0^1\frac{\bigl[j-F\sigma(\rho)+D(\rho)\nabla\rho\bigr]^2}{2\sigma(\rho)}dx
\end{equation}
with ${\rm d}_t \rho=-\nabla j$. We intend to contract this to the large deviation function of only the space integrated current. We will assume that the best density to produce a given constant average current, in the long time limit, is time-independent. The constraint thus becomes ${\rm d}_t \rho=0=-\nabla j$, which is to say that $j$ does not depend on $x$ or $t$ any more. The time integral can then be taken out of the large deviation function, and we get
\begin{equation}\label{IV-2-gjSigmaD}
g(j,\rho)=\int_0^1\frac{\bigl[j-F\sigma(\rho)+D(\rho)\nabla\rho\bigr]^2}{2\sigma(\rho)}dx
\end{equation}
which is the action we will minimise over $\rho$.

The first step in that direction is to expand the square
\begin{equation}
g(j,\rho)=\int_0^1\frac{\bigl[j-F\sigma(\rho)\bigr]^2+\bigl[D(\rho)\nabla\rho\bigr]^2}{2\sigma(\rho)}dx+\int_0^1\frac{\bigl[j-F\sigma(\rho)\bigr]}{\sigma(\rho)}D(\rho)\nabla\rho ~dx
\end{equation}
and notice that the cross product between the gradient and the rest produces a constant term
\begin{equation}
\int_0^1\frac{\bigl[j-F\sigma(\rho)\bigr]}{\sigma(\rho)}D(\rho)\nabla\rho ~dx=\int_{\rho_a}^{\rho_b}\frac{\bigl[j-F\sigma(\rho)\bigr]}{\sigma(\rho)}D(\rho)d\rho
\end{equation}
which does not need to be minimised. We are left with having to cancel the functional derivative of the first part alone.

Let us write
\begin{equation}
A(\rho)=\frac{\bigl[j-F\sigma(\rho)\bigr]^2}{2\sigma(\rho)}~~~~{\rm and}~~~~B(\rho)=\frac{\bigl[D(\rho)\bigr]^2}{2\sigma(\rho)}.
\end{equation}
We want to minimise $\int_0^1dx~A(\rho)+B(\rho)(\nabla\rho)^2$. The minimising profile satisfies the Euler-Lagrange equation:
\begin{equation}
A'(\rho)-B'(\rho)(\nabla\rho)^2-2B(\rho)\Delta\rho=0
\end{equation}
which, multiplied by $\nabla\rho$, gives
\begin{equation}
\nabla\bigl[A(\rho)-B(\rho)(\nabla\rho)^2\bigr]=0.
\end{equation}
The minimising profile is therefore such that
\begin{equation}
A(\rho)-B(\rho)(\nabla\rho)^2=K
\end{equation}
where $K$ is an integration constant. That constant can be found to be $0$ for $F\rightarrow\infty$ \cite{Bodineau2006}. Replacing $A(\rho)$ and $B(\rho)$ by their expressions, and taking a square root, we obtain the equation defining the optimal profile:
\begin{equation}\label{optRho}\boxed{
j-F\sigma(\rho)\pm D(\rho)\nabla\rho=0.
}\end{equation}

This equation is the same as the mean field equation, except for the sign of the gradient. To construct this profile, for a given current, one may therefore use the same construction as we saw in section \ref{II-2}, but allow the gradient to have either sign. We will come back to this in the case of the ASEP.

We also obtain, from this minimisation, the large deviation of the current, by injecting eq.(\ref{optRho}) into $g(j,\rho)$. The portions of the optimal profile where $\nabla\rho$ has the same sign as in the mean field equation do not contribute to the integral, as they make the integrand vanish, so that:
\begin{equation}
g(j)=\int\frac{\bigl[j-F\sigma(\rho)+D(\rho)\nabla\rho\bigr]^2}{2\sigma(\rho)}dx=\int\frac{4\bigl[j-F\sigma(\rho)\bigr]^2}{2\sigma(\rho)}dx=\int\frac{2\bigl[j-F\sigma(\rho)\bigr]D(\rho)\nabla\rho}{\sigma(\rho)}dx
\end{equation}
where the integral is over all the portions of space where $j-F\sigma(\rho)=D(\rho)\nabla\rho$. We end up with a combination of the values of the primitive of $(j-F\sigma(\rho))/\sigma(\rho)$ at the points delimiting those portions of space.

~

We now consider the case of the TASEP, where $F=L$, $\sigma(\rho)=\rho(1-\rho)$ and $D=\frac{1}{2}$, and we rescale time by a factor $L$, so that the optimal profile equation for a current $j$ is given by
\begin{equation}\label{IV-4-Jx}
j=\rho(1-\rho)\mp\frac{1}{2L}\nabla\rho.
\end{equation}

\begin{figure}[ht]
\begin{center}
 \includegraphics[width=0.8\textwidth]{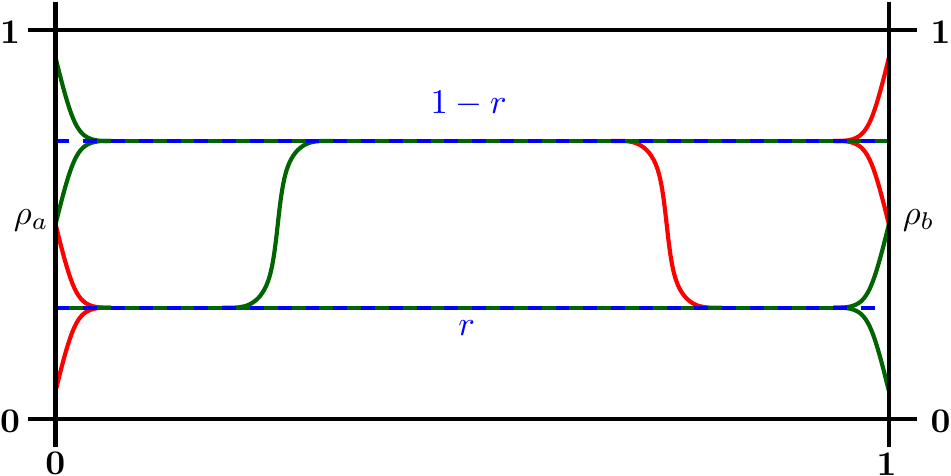}
  \caption{Various optimal profiles for a fixed $r$. The parts in green satisfy the mean field equation, and do not contribute to $g(j)$, whereas the portions in red do.}
\label{fig-ProfileCost}
 \end{center}
 \end{figure}

For a given current $j$, and two boundary conditions $\rho_a$ and $\rho_b$, we can repeat the construction of density profiles we saw in section \ref{II-2}, without the constraint on the sign of the variations of $\rho$. Note that this constraint enforced that for given boundary conditions, only one value of the current was possible. In the present case, the current is a free variable. The allowed building blocks for the profiles are represented on fig.\ref{fig-ProfileCost} ; the portions in red are the only ones contributing to $g(j)$.

The large deviation function of the current for that profile is then given by an integral over the red portions of the profile:
\begin{equation}
g(j)=\int\frac{\bigl[j-\rho(1-\rho)\bigr]}{\rho(1-\rho)}\nabla\rho~ dx=\sum\limits_{i}\biggl[j\log\Bigl(\frac{\rho}{1-\rho}\Bigr)-\rho \biggr]_{\rho_i^-}^{\rho_i^+}
\end{equation}
where $\rho_i^-$ and $\rho_i^+$ are the densities at the boundary of each red section. Note that in every case, one of the boundaries is $r$ or $1-r$. We shall write each of these terms as
\begin{equation}\label{IV-2-bound}\boxed{
f(j;r1,r2)=\int_{r_1}^{r_2}\frac{\bigl[j-\rho(1-\rho)\bigr]}{\rho(1-\rho)}d\rho=j\log\Bigl(\frac{1-r_1}{r_1}\frac{r_2}{1-r_2}\Bigr)+r_1-r_2
}\end{equation}
where $j$ has to be equal to either $r_1(1-r_1)$ or $r_2(1-r_2)$. This is the same as the function $F^{\rm res}$ from \cite{Bodineau2006}.

~

Through this procedure, there are many profiles we can build for a given set $\{j,\rho_a,\rho_b\}$, since any number of shocks and anti-shocks can be added between $r$ and $1-r$. The true optimal profile is the one which minimises the large deviation function, which is to say the one with the fewest red portions. If at least one of $r$ and $1-r$ is between $\rho_a$ and $\rho_b$, then the optimal profile is monotonic. Otherwise, two candidates must be compared: one where $\rho$ goes from $\rho_a$ to $r$, and then to $\rho_b$, and one where $\rho$ goes from $\rho_a$ to $1-r$, and then to $\rho_b$.

The other extremising profiles are conjectured to correspond to excited states of the system, conditioned on the average current. We will say more about these in section \ref{V-3}.

~~

In the next section, we will list all the possible configurations for $j=r(1-r)$, $\rho_a$ and $\rho_b$ that lead to different forms of $g(j)$ (i.e. with different combinations of the function $f$), and determine, in each of those phases, the expressions of $g(j)$, $E(\mu)$, $j(\mu)$, and the boundaries of the phase. We will also compare the asymptotic behaviours of those results with everything we found in the previous sections, to confirm their validity. We will then summarise all we know about the maximal current phase, which is not accessible by this method (and which is defined by $j>\frac{1}{4}$). Finally, we will put all this together in order to draw the phase diagram of the open ASEP with respect to $\rho_a$, $\rho_b$ and $\mu$.

In all cases, we will be noting $r$ the density for which $j=r(1-r)$ which is below $\frac{1}{2}$. Also note that we will do all the calculations for the TASEP, knowing that the same for the ASEP can be obtained merely by multiplying $E(\mu)$ by $(1-q)$. We also recall that we have defined two other boundary parameters $a=\frac{1-\rho_a}{\rho_a}$ and $b=\frac{\rho_b}{1-\rho_b}$, which we will use in certain formulae to make them more compact. Finally, we will, for the same reason, sometimes use, instead of $\mu$, the variable $u$ defined by:
\begin{equation}
u=\frac{1}{1+{\rm e}^{\mu}}.
\end{equation}
The non-perturbed case is given by $u=\frac{1}{2}$, $u=0$ corresponds to an infinite current, and $u=1$ to zero current.

\subsection{Dynamical phase diagram of the open ASEP}
\label{V-2}

Each phase is described in turn.

\subsubsection{High/low density phases}
\label{V-2-a}

We start with the low density phase, from which we can deduce the high density phase through $\rho_a\leftrightarrow 1-\rho_b$.

\begin{figure}[ht]
\begin{center}
 \includegraphics[width=0.5\textwidth]{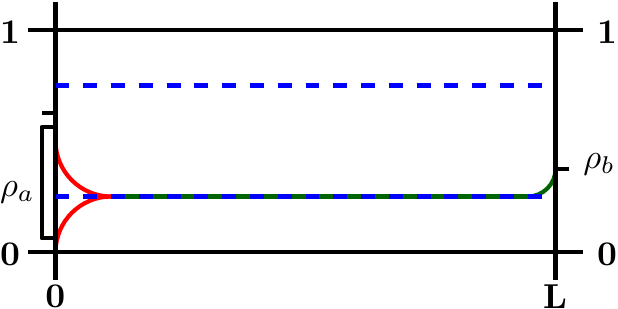}
  \caption{Optimal profiles for a fixed $\rho_c$ in the low density phase. Only the portion in red contribute to $g(j)$.}
\label{fig-ProfileLD}
 \end{center}
 \end{figure}

This phase is defined by $\rho_a<1-\rho_b$, $\rho_a<1-r$ and $\rho_b<1-r$. The optimal profile is almost always on $\rho=r$, with possible boundary layers at both ends, with only the one at the left boundary contributing to $g(j)$ (fig.-\ref{fig-ProfileLD}).

The large deviation function of the current is in this case:
\begin{equation}
g(j)=f(j;\rho_a,r)=j\log\Bigl(\frac{1-\rho_a}{\rho_a}\frac{r}{1-r}\Bigr)+\rho_a-r,
\end{equation}
which agrees with what we found in eq.(\ref{IV-2-gjASEP}) from exact calculations. This was first obtained in \cite{Bodineau2006}.
The generating function of the cumulants of the current is
\begin{equation}\label{IV-4-ELD}
E(\mu)=\frac{a}{a+1} \frac{ {\rm e}^\mu -1 }{{\rm e}^\mu + a}= \frac{{\rm e}^\mu }{{\rm e}^\mu + a}-\frac{1}{1+a},
\end{equation}
which was also obtained in \cite{PhysRevLett.107.010602} through a numerical approach to the Bethe equations, and the current is, in terms of $\mu$:
\begin{equation}
 j(\mu)=\frac{a~{\rm e}^{\mu}}{({\rm e}^{\mu}+a)^2}=\rho_a(1-\rho_a)\frac{u(1-u)}{(\rho_a+u-2u\rho_a)^2}.
\end{equation}
The boundaries of the phase are given by:
\begin{align}
\rho_a&<1-\rho_b\\
u&>\frac{\rho_a^2}{1-2\rho_a+2\rho_a^2}~~~~~~~~~~~{\rm with}~~\rho_a>\frac{1}{2},\\
u&>\frac{\rho_a \rho_b}{1-\rho_b-\rho_a+2\rho_a\rho_b}~~~~{\rm with}~~\rho_b>\frac{1}{2},\\
u&>\rho_a~~~~~~~~~~~~~~~~~~~~~~~~~~~{\rm with}~~\rho_a<\frac{1}{2}~~,~~\rho_b<\frac{1}{2},
\end{align}
where this last condition corresponds to $j<\frac{1}{4}$, which is the boundary with the MC phase.

~~~~

According to this, the LD phase goes all the way up to $u=1$. This expression of $E(\mu)$ is consistent with what we found for $\mu\rightarrow -\infty$, i.e. $u\rightarrow 1$, i.e. eq.(\ref{IV-3-gjLD}).

We may also note that, on the line $\rho_a=\frac{1}{2}$, which corresponds to the LD-MC transition line for $\mu=0$, we find:
\begin{equation}
E(\mu)=\frac{1}{2} \frac{{\rm e}^\mu -1}{{\rm e}^\mu+1}
\end{equation}
which is consistent with the expression found in \cite{derrida1999universal} for the half-filled periodic TASEP (we recall that the open system with $\rho_a=\frac{1}{2}$ and $\rho_b<1/2$ is equivalent to a half-filled periodic system of twice the size). The $\mu\rightarrow 0^-$ limit gives:
\begin{equation}
E(\mu)\sim \frac{\mu}{4}-\frac{\mu^3}{48}
\end{equation}
which is the same as eq.(\ref{IV-2-gMC5}).

\subsubsection{Shock phase}
\label{V-2-b}

\begin{figure}[ht]
\begin{center}
 \includegraphics[width=0.5\textwidth]{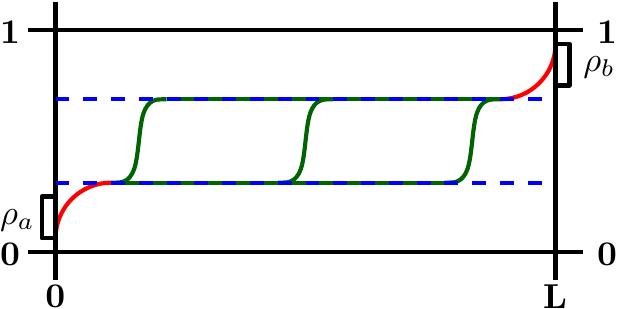}
  \caption{A few optimal profiles for a fixed $\rho_c$ in the shock phase. Only the portions in red contribute to $g(j)$.}
\label{fig-ProfileSP}
 \end{center}
 \end{figure}

We consider the case where $\rho_a<r$ and $\rho_b>1-r$. Here, there is a number of optimal profiles of order $L$. Each of them has a boundary layer around each boundary, both of them contributing to $g(j)$, and two constant regions, where $\rho=r$ near the left boundary and $\rho=1-r$ near the right boundary, separated by a shock that can be placed anywhere in the system (fig.-\ref{fig-ProfileSP}).

The large deviation function of the current is given by:
\begin{equation}
g(j)=f(j;\rho_a,r)+f(j;1-r,\rho_b)=j\log\biggl(\frac{(1-\rho_a)\rho_b}{\rho_a(1-\rho_b)}\frac{r^2}{(1-r)^2}\biggr)+\rho_a-\rho_b+1-2r.
\end{equation}
The generating function of the cumulants of the current is
\begin{equation}\label{IV-4-ELD2}
E(\mu)=\frac{2{\rm e}^{\mu/2} }{{\rm e}^{\mu/2} + \sqrt{ab}}-\frac{1}{1+a}-\frac{1}{1+b}
\end{equation}
and the current is:
\begin{equation}
 j(\mu)=\frac{\sqrt{ab}~{\rm e}^{\mu/2} }{({\rm e}^{\mu/2} + \sqrt{ab})^2}=\sqrt{\frac{(1-\rho_a)\rho_b}{\rho_a(1-\rho_b)}\frac{(1-u)}{u}}\Biggl(\sqrt{\frac{(1-\rho_a)\rho_b}{\rho_a(1-\rho_b)}}+\sqrt{\frac{(1-u)}{u}}\Biggr)^{-2}.
\end{equation}
The boundaries of the phase are given by:
\begin{align}
u&<\frac{\rho_a \rho_b}{1-\rho_b-\rho_a+2\rho_a\rho_b}~~~~{\rm with}~~\rho_a<\frac{1}{2}~~,~~\rho_b>\frac{1}{2},\\
u&<\frac{(1-\rho_a) (1-\rho_b)}{1-\rho_b-\rho_a+2\rho_a\rho_b}~~~~{\rm with}~~\rho_a<\frac{1}{2}~~,~~\rho_b>\frac{1}{2},\\
u&>\frac{\rho_a(1-\rho_b)}{\rho_b+\rho_a-2\rho_a\rho_b}~~~~~~~~~~{\rm with}~~\rho_a<\frac{1}{2}~~,~~\rho_b>\frac{1}{2},
\end{align}
where the last condition corresponds to $j<\frac{1}{4}$. We may note that the volume which is defined by these boundaries is symmetric under any permutation of $\rho_a$, $1-\rho_b$ and $u$.

~~

The shock phase concerns one of the asymptotic results we found before. For $\mu\rightarrow 0^+$, which imposes $\rho_a=1-\rho_b$, we find:
\begin{equation}
E(\mu)\sim\frac{a}{(1+a)^2}\mu+\frac{a(a-1)}{4(a+1)^3}\mu^2
\end{equation}
which is what we found in eq.(\ref{IV-2-ESL+}).

\subsubsection{Anti-shock phase}
\label{V-2-c}

\begin{figure}[ht]
\begin{center}
 \includegraphics[width=0.5\textwidth]{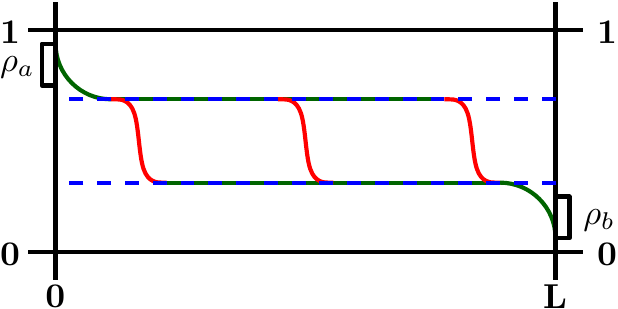}
  \caption{A few optimal profiles for a fixed $\rho_c$ in the Anti-shock phase. Only the portions in red contribute to $g(j)$.}
\label{fig-ProfileASP}
 \end{center}
 \end{figure}

The last phase we can access through the MFT is for $\rho_a>(1-r)$ and $\rho_b<r$. In this case, there also is a number of optimal profiles of order $L$: the first go down from $\rho_a$ to $(1-r)$, then down from $(1-r)$ to $r$ through an anti-shock that can be placed anywhere, and that contributes to $g(j)$, and finally down from $r$ to $\rho_b$ (fig.-\ref{fig-ProfileASP}).

The large deviation function of the current is given by:
\begin{equation}
g(j)=f(j;1-r,r)=2j\log\Bigl(\frac{r}{1-r}\Bigr)+1-2r.
\end{equation}
The generating function of the cumulants of the current is
\begin{equation}\label{IV-4-ELD3}
E(\mu)=\frac{2{\rm e}^{\mu/2} }{{\rm e}^{\mu/2} + 1}-1=\tanh(\mu/4)
\end{equation}
and the current is:
\begin{equation}
 j(\mu)=\frac{1-\tanh(\mu/4)}{4}=\frac{-2(u-u^2)+\sqrt{u-u^2}}{1-4(u-u^2)}.
\end{equation}
The boundaries of the phase are given by:
\begin{align}
u&<\frac{\rho_a^2}{1-2\rho_a+2\rho_a^2}~~~~{\rm with}~~\rho_a>\frac{1}{2}~~,~~\rho_a>1-\rho_b,\\
u&<\frac{(1-\rho_b)^2}{1-2\rho_b+2\rho_b^2}~~~~{\rm with}~~\rho_b<\frac{1}{2}~~,~~\rho_a<1-\rho_b,\\
u&>\frac{1}{2},
\end{align}
where the last condition corresponds to $j<\frac{1}{4}$.

We note that this phase corresponds to one of the examples that can be found in \cite{Bodineau2006}. The expression for $E(\mu)$ also comes up as a side note in \cite{PhysRevLett.107.010602}.

~~

The limit $\mu\rightarrow 0^-$ gives:
\begin{equation}
E(\mu)\sim\frac{\mu}{4}-\frac{\mu^3}{192}
\end{equation}
which is consistent with eq.(\ref{IV-2-gLMC7}). The limit $\mu\rightarrow -\infty$, which implies $\rho_a=1-\rho_b=1$, gives:
\begin{equation}
E(\mu)\sim-1+2{\rm e}^{\mu/2} 
\end{equation}
which is the same as equation (\ref{IV-3-EER-}), and this is the last asymptotic limit that we had to check.

 \begin{figure}[hp]
\begin{center}
 \includegraphics[width=0.98\textwidth]{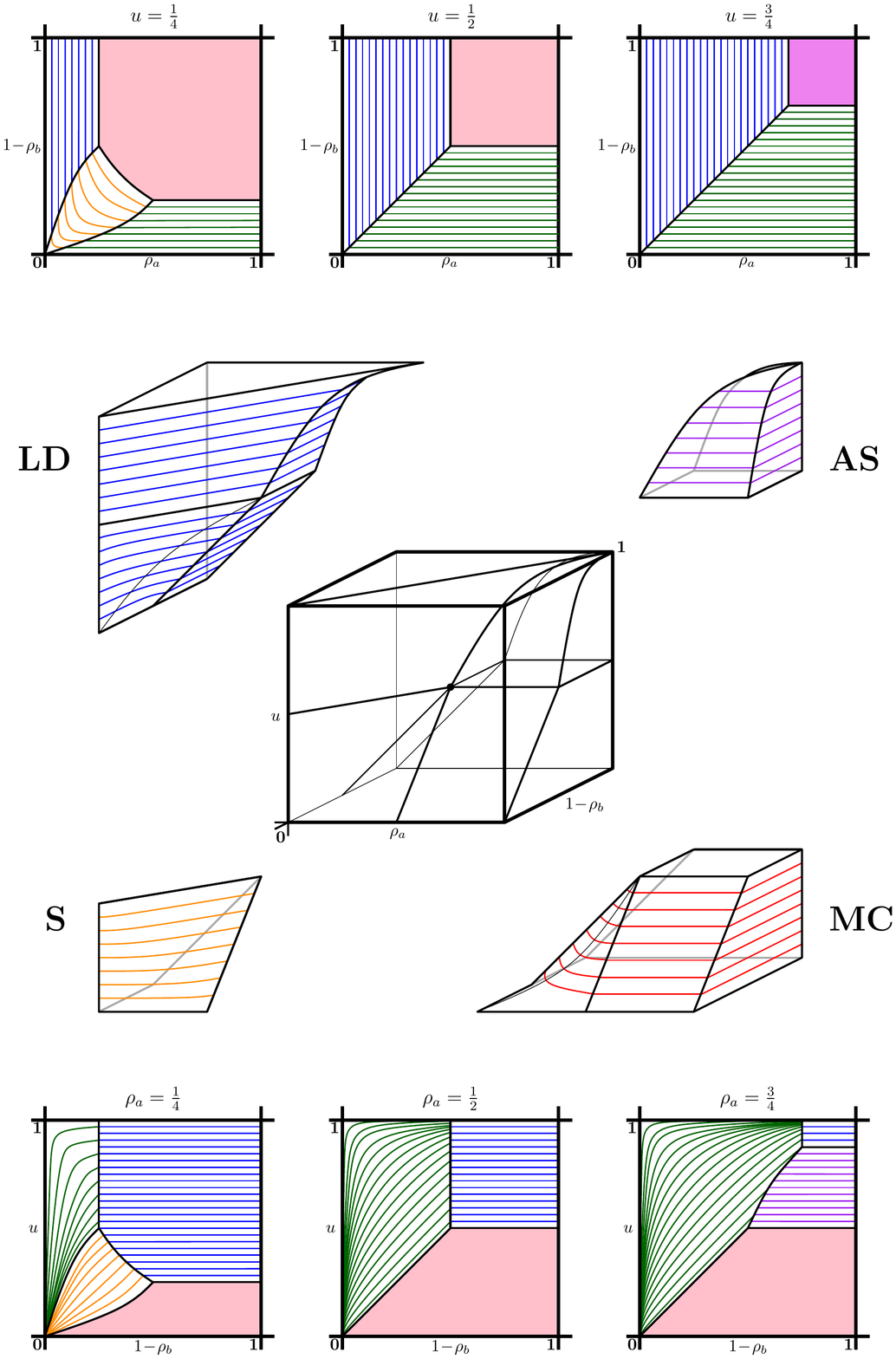}
  \caption{Phase diagram of the open ASEP in the s-ensemble. Top: diagrams at fixed $u$. Centre: complete diagram with phase boundaries and exploded view. Bottom: diagrams at fixed $\rho_a$.}
\label{fig-3DPhaseDiag}
 \end{center}
 \end{figure}

\subsubsection{Maximal current phase}
\label{V-2-d}

There is one phase left for us to examine, to a much lesser extent than all the the others because the MFT breaks down in this case: the maximal current phase. Once we take out the phases we have already considered, we are left with a volume, in the three-dimensional phase space with variables $\rho_a$, $\rho_b$ and $u$, defined by:
\begin{align}
u&<\frac{1}{2}~~~~~~~~~~~~~~~~~~~~~~{\rm with}~~\rho_a>\frac{1}{2}~~,~~\rho_b<\frac{1}{2},\\
u&<\rho_a~~~~~~~~~~~~~~~~~~~~~{\rm with}~~\rho_a>\frac{1}{2}~~,~~\rho_b>\frac{1}{2},\\
u&<1-\rho_b~~~~~~~~~~~~~~~~{\rm with}~~\rho_a<\frac{1}{2}~~,~~\rho_b<\frac{1}{2},\\
u&<\frac{\rho_a(1-\rho_b)}{\rho_b+\rho_a-2\rho_a\rho_b}~~~~{\rm with}~~\rho_a<\frac{1}{2}~~,~~\rho_b>\frac{1}{2}.
\end{align}

We know that, asymptotically:
\begin{equation}
g(j)\sim(j-J)^{5/2}\frac{32\sqrt{3}L}{5\pi(1-q)^{3/2}}
\end{equation}
for $\mu\rightarrow 0^+$ with $\rho_a>\frac{1}{2}$ and $\rho_b<\frac{1}{2}$ (i.e. right next to the MC phase for the steady state), which we found in eq.(\ref{IV-2-gMC4}), and that:
\begin{equation}
g(j)\sim L j\log(j)-Lj(1-\log(\pi))
\end{equation}
for $\mu\rightarrow\infty$, as we saw in eq.(\ref{IV-3-gMMC}). The best estimate of $g(j)$ from the MFT is obtained for $\rho=\frac{1}{2}$ in the whole system, which yields $g(j)\sim L(j-J)^2$, showing that this hydrodynamic description does indeed break down in the MC phase.

Since this last result is valid independently of the boundary parameters, we know that all the plane $u=0$ belongs to the same phase. We have, however, no way to be certain that {\it all} of the volume we have described above is just one phase. That being said, we know that the entire region corresponds to a mean current higher than $\frac{1}{4}$. There is no way for the system to produce such a current through a hydrodynamic profile, for which the maximal possible current is $\frac{1}{4}$ if $\rho=\frac{1}{2}$, so that, in order to increase the current, the system must produce correlations, which is why the MFT breaks down. Those correlations must be negative for neighbouring sites (if the particles are next to holes, they will jump more easily and produce more current), which is consistent with what we found in the large current limit in eq.(\ref{IV-3-corrMC}). What's more, those correlations must be created everywhere in the system, because a single non-correlated zone would cause a blockage and bring the current back down to $\frac{1}{4}$. We can also argue that the mean density should be around $\frac{1}{2}$, because it is easier to get to a large current starting from $\frac{1}{4}$ than from anything lower. In all these remarks, the boundaries play little part: independently of them, the system must be around $\rho=\frac{1}{2}$, and anti-correlated at every point. We therefore don't expect any sub-phases in this region.

\subsubsection{Phase diagram}
\label{V-2-e}

Now that we have considered all the possible combinations of $\rho_a$, $\rho_b$ and $u$, we can draw the phase diagram of the ASEP in the s-ensemble (fig.-\ref{fig-3DPhaseDiag}). Each phase is represented using a different colour: blue for the low density phase, green for the high density phase, orange for the shock phase, purple for the anti-shock phase, and red/pink for the maximal current phase. The full diagram can be seen in the centre of the figure, with black lines marking the corners of the phases, and an exploded view of the LD, MC, shock (S) and anti-shock (AS) phases is also shown, with coloured lines representing slices for regularly spaced values of $u$. The HD phase can be deduced from the LD phase through the symmetry $\rho_a\leftrightarrow 1-\rho_b$. The top and bottom parts of the figure contain slices of the diagram for specific values of $u$ (top) and $\rho_a$ (bottom), with a few iso-current lines drawn in all phases except the MC phase. Note that those iso-current lines do not represent evenly spaced values of the current ($j$ varies, in fact, more slowly as one approaches the MC phase).

\subsection{Hydrodynamic excited states}
\label{V-3}

As we noted earlier, the extremisation of the MFT action yields many density profiles, which are local minima of that action, and of which we only considered the optimal ones to build the dynamical phase diagram of the current. One may naturally wonder if the other extremising profiles have any physical significance. We conjecture that they do in fact correspond to the low excitations (i.e. relaxation modes) of the model. In particular, profiles which correspond to a vanishing $\mu$, i.e. $\frac{{\rm d}}{{\rm d}j} g(j)=0$, if they exist, are relaxation modes of the unperturbed dynamics, and the corresponding values of the Legendre transform of $g(j)$ are the eigenvalues of these relaxation modes, which is to say the inverses of the relaxation times. In this section, we give some evidence towards that conjecture. More details on this conjecture will appear in \cite{Lazarescu}.

~

\begin{figure}[h]
\begin{center}
 \includegraphics[width=\textwidth]{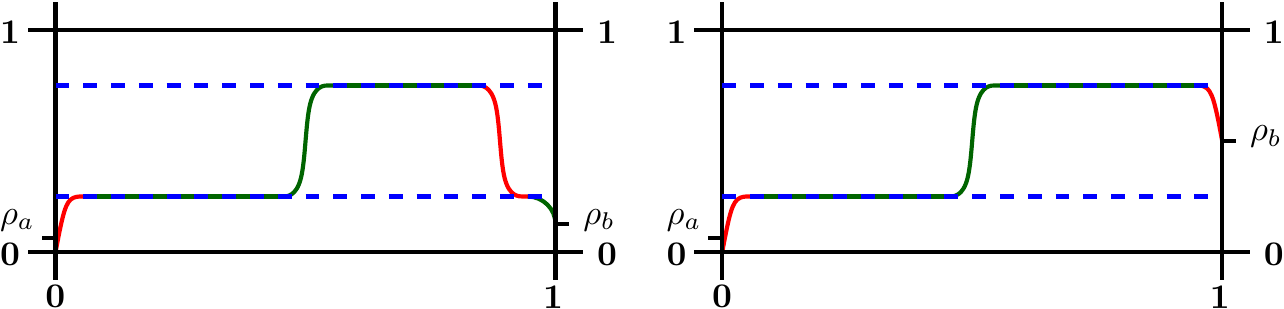}
  \caption{Second best extremising profiles in the LD phase, for $\rho_b$ smaller or larger than $r$. The profile on the left still exists for $\rho_b>r$, but becomes the third best.}
\label{fig-ProfExc}
 \end{center}
 \end{figure}

Consider the system in its low density phase, with some fixed values of $j=r(1-r)$ and $\rho_a$, and $\rho_b<1-r$. Looking for the second best profile obtained from the aforementioned extremisation, we notice that it depends on the position of $\rho_b$ with respect to $r$: if $\rho_b<r$, it is obtained by adding a shock and an anti-shock to the optimal profile, whereas if $\rho_b>r$, one may go directly down to $\rho_b$ after the shock, so that the profile with an anti-shock becomes the third best (fig.-\ref{fig-ProfExc}).

~

In the case where $\rho_b<r$, the large deviation function of the current for the second best profile is given by
\begin{equation}
g(j)=j\log\biggl(\frac{1-\rho_a}{\rho_a}\Bigl(\frac{r}{1-r}\Bigr)^3\biggr)+1+\rho_a-3r.
\end{equation}
The deformation parameter $\mu$ vanishes for
\begin{equation}
r=\Bigl[1+\bigl(\frac{1-\rho_a}{\rho_a}\bigr)^{1/3}\Bigr]^{-1}
\end{equation}
and we get a value for the Legendre transform of $g(j)$ equal to
\begin{equation}
E^\star=-1-\rho_a+3\rho_c.
\end{equation}

~

If $\rho_b>r$, that profile still exists, but another one appears, which is more probable (fig.-\ref{fig-ProfExc} right). The large deviation function of the current for that profile is
\begin{equation}
g(j)=j\log\biggl(\frac{1-\rho_a}{\rho_a}\frac{\rho_b}{1-\rho_b}\bigl(\frac{\rho_c}{1-\rho_c}\bigr)^2\biggr)+1+\rho_a-\rho_b-2\rho_c.
\end{equation}
The deformation parameter $\mu$ vanishes for
\begin{equation}
r=\Bigl[1+\bigl(\frac{1-\rho_a}{\rho_a}\frac{\rho_b}{1-\rho_b}\bigr)^{1/2}\Bigr]^{-1}
\end{equation}
and we get a value for the Legendre transform of $g(j)$ equal to
\begin{equation}
E^\star=-1-\rho_a+\rho_b+2\rho_c.
\end{equation}

 \begin{figure}[h]
\begin{center}
 \includegraphics[width=0.5\textwidth]{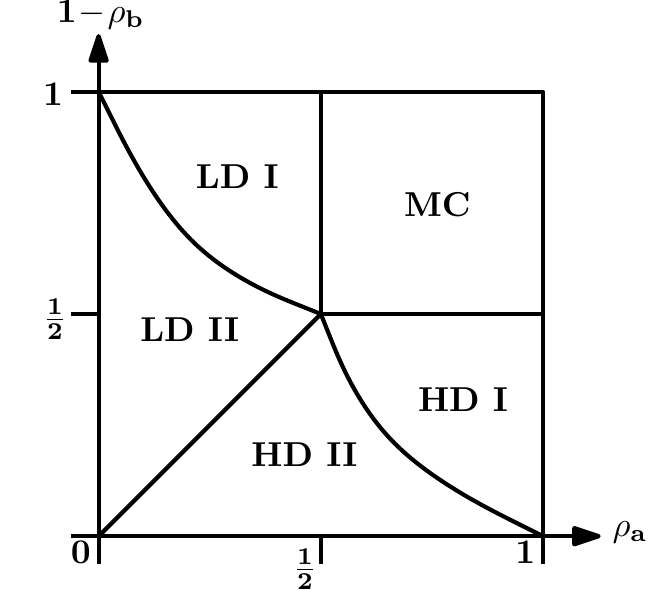}
  \caption{Phase diagram of the first excited state of the open ASEP ; an extra transition appears in the LD and HD phases, corresponding to a change of behaviour of the associated density profile.}
\label{fig-PhaseDiagExc}
 \end{center}
 \end{figure}

The line on which the second profile appears is that for which
\begin{equation}
\rho_b=r=\Bigl[1+\bigl(\frac{1-\rho_a}{\rho_a}\bigr)^{1/3}\Bigr]^{-1},
\end{equation}
which is to say, using the alternative boundary parameters $a$ and $b$:
\begin{equation}\boxed{
a~ b^3=1.
}\end{equation}
This line is therefore a transition line in the phase diagram of the second best density profile of the open ASEP (fig.-\ref{fig-PhaseDiagExc}). The same calculations can be done for the HD phase by exchanging $a$ and $b$.

The value obtained for $E^\star$, as well as the location of this transition line, agree perfectly with results obtained in \cite{1742-5468-2006-12-P12011} for the eigenvalue of the first excited state of the open ASEP, through a numerical resolution of the Bethe equations, and which were later verified in \cite{Proeme2011}. This leads us to conjecture that all the non-optimal extremal profiles obtained from the MFT are excited states of the (biased, unless $\mu=0$) open ASEP, with the value of the Legendre transform of the large deviation function giving the corresponding eigenvalue. One should note that each shock and anti-shock in the profile adds an order $L$ to the degeneracy of that eigenvalue (because it does not depend on their position), and that finite size corrections will lift that degeneracy, in which case the true eigenstates will be specific superpositions of these profiles. More evidence for that statement will appear in \cite{Lazarescu}.

\newpage

\section{Conclusion and outlook}

We have seen, in this review, how to obtain the large deviation function of the average current of particles in the steady state of one-dimensional bulk-driven lattice gases, and in particular of the asymmetric simple exclusion process with open boundaries.

We first reviewed the mathematical tools necessary to pose and treat the problem at hand. We defined large deviation functions, which are a natural generalisation of free energies, and saw that they relate to generating functions of cumulants through a Legendre transform. We then considered the special case of time-additive observables in time-continuous Markov processes, and saw how the problem of obtaining the generating function of those observables in the long time limit reduces to that of computing the largest eigenvalue of a deformed Markov matrix.

We then introduced the reader to the open ASEP, starting with a rapid overview of the existing literature related to the model and its variants. We also presented two important results pertaining to its steady state: first, the phase diagram of the average current, as well as the typical density profiles, using a mean-field approach ; then, the famous matrix Ansatz, giving the exact probability distribution of the steady state using a matrix product formulation, and which allows to obtain that same phase diagram through an exact calculation.

Being interested in the fluctuations of the current rather than its average, we posed the problem of obtaining its large deviation function, which is equivalent to that of calculating the largest eigenvalue of a deformed Markov matrix. That matrix being integrable, we saw how to use the coordinate Bethe Ansatz (in the periodic case) and the Q-operator method (in the open case) in order to obtain an exact expression of the generating function of the cumulants of the current, written as an implicit pair of series in an intermediate parameter. Treating those series in the large size limit, we obtained the asymptotic behaviour of the large deviation function of the current for small fluctuations, in each of the phases of the average current, and found a few dynamical phase transitions. We noticed in particular that for fluctuations of the current smaller than $\frac{1}{4}$, which is the maximal current obtained from a hydrodynamic (mean field) description of the system, the large deviation function does not depend on the size of the system, whereas if the current is larger than $\frac{1}{4}$, it is proportional to the size of the system.

Having obtained the behaviour of the large deviation function of the current in the limit of small fluctuations, we then looked at the limit of extreme fluctuations, either positive or negative, where it turns out that the deformed Markov matrix can be diagonalised directly, without invoking the integrability of the system. In the large negative fluctuation limit, where the current goes to $0$, the deformed Markov matrix is a perturbation of a diagonal matrix, and we find that the large deviation function of the current is independent of the size of the system. In the large positive fluctuation limit, where the current goes to infinity, the system is equivalent to an open XX spin chain, and the large deviation function of the current is proportional to the size of the system. We also note that these properties do not depend on many details of the system, and would be equally valid with added interactions between the particle or site-wise disorder of the jumping rates. The difference in scaling with respect to the size of the system suggests that a generic dynamical phase transition exists at an intermediate current for all those systems.

Finally, we considered a less exact but more powerful approach, based on the macroscopic fluctuation theory, to obtain the large deviation function of any finite fluctuation of the current, using a hydrodynamic description of the system in the large size limit and a non-rigorous exchange of limits. This allowed us to build the complete dynamical phase diagram of the current for the open ASEP. In four of the five phases we obtained, which are the ones corresponding to a current smaller that $\frac{1}{4}$, we found large deviation functions independent of the size of the system, and perfect agreement with the exact results obtained before. In the last phase, we found no such agreement, and we surmised that the MFT breaks down due to the necessary presence of local correlations in the system in that phase. In the phases where the MFT does apply, we gave some evidence for a conjecture stating that the non-optimal extremisers of the MFT action are the slow relaxation modes of the system.

~

Many challenges related to the contents of this review remain to be tackled, and we will conclude by mentioning a few of them.

First of all, the Q-operator method summarised in section \ref{III-2-b} has only yielded the dominant eigenvalue of the deformed Markov matrix so far, but it allows in principle access to the whole spectrum. Obtaining it, through that method or any of the other ones mentioned in section \ref{III-2}, would in particular allow to verify that the non-optimal profiles mentioned in section \ref{V-3} are really related to the low excitations of the model.

On the subject of those local minima of the MFT action, it would be interesting to understand for which models they might appear and whether they always correspond to excited states. For one-dimensional lattice gases, a finite drive in the bulk seems essential: those local minima appear for the ASEP because of the appearance of shocks and anti-shocks which can be combined in many ways, which is a consequence of the diffusion part of the Langevin equation being inversely proportional to the size of the system. It is natural to wonder what might be the case in other types of systems, such as higher-dimensional ones, for instance.

Finally, it would certainly be interesting to understand exactly how generic the dynamical phase transition we discussed in section \ref{IV-3} really is, and whether it has any universal features. In the case of the open ASEP, that transition arises because of the fact that the current which can be obtained from a hydrodynamic description is bounded. If one tries to impose a current higher that $\frac{1}{4}$ (which is the upper bound), which is not forbidden in the underlying microscopic model, the hydrodynamic description breaks down, and another macroscopic description of the model is needed. It would stand to reason that such a transition would appear for any model where a hydrodynamic limit introduces bounds for an observable. The question of whether those transitions have any universal features is for the moment entirely open.

~

\textbf{Acknowledgements:} The author would like to thank K. Mallick, V. Pasquier, F. van Wijland, J. Tailleur, M. Evans, R. Blythe, G. Schutz, D. Karevski, T. Sadhu, C. Maes, W. de Roeck, C. Nardini and C. Perez Espigares for useful and interesting discussions.

This work was financed by the Interuniversity Attraction Pole - Phase VII/18 (Dynamics,
Geometry and Statistical Physics) at KU Leuven.

\newpage

\bibliographystyle{mybibstyle}

\bibliography{Biblio}{}

\begin{thebibliography}{100}

\bibitem{Lazarescu2013}
A.~Lazarescu.
\newblock \emph{{Exact Large Deviations of the Current in the Asymmetric Simple
  Exclusion Process with Open Boundaries (PhD thesis)}}.
\newblock Ph.D. thesis (2013).

\bibitem{Touchette20091}
H.~Touchette.
\newblock \emph{{The large deviation approach to statistical mechanics}}.
\newblock Physics Reports \textbf{478(1-3)}, 1--69 (2009).

\bibitem{Touchette2011}
H.~Touchette and R.~J. Harris.
\newblock \emph{{Large Deviation Approach to Nonequilibrium Systems}}.
\newblock Nonequilibrium Statistical Physics of Small Systems: Fluctuation
  Relations and Beyond pp. 335--360 (2013).

\bibitem{Touchette2010}
H.~Touchette, R.~J. Harris and J.~Tailleur.
\newblock \emph{{First-order phase transitions from poles in asymptotic
  representations of partition functions}}.
\newblock Physical Review E - Statistical, Nonlinear, and Soft Matter Physics
  \textbf{81(3)}, 030101 (2010).

\bibitem{Donsker2010}
M.~D. Donsker and S.~R.~S. Varadhan.
\newblock \emph{{Asymptotic evaluation of certain Markov process expectations
  for large time, III}}.
\newblock Communications on Pure and Applied Mathematics \textbf{29(4)},
  389--461 (1976).

\bibitem{Donsker1975}
M.~D. Donsker and S.~R.~S. Varadhan.
\newblock \emph{{Asymptotic evaluation of certain Markov process expectations
  for large time, III}}.
\newblock Communications on Pure and Applied Mathematics \textbf{29(4)},
  389--461 (1976).

\bibitem{Donsker1976}
M.~D. Donsker and S.~R.~S. Varadhan.
\newblock \emph{{Asymptotic evaluation of certain Markov process expectations
  for large time, III}}.
\newblock Communications on Pure and Applied Mathematics \textbf{29(4)},
  389--461 (1976).

\bibitem{Donsker1983}
M.~D. Donsker and S.~R.~S. Varadhan.
\newblock \emph{{Asymptotic evaluation of certain Markov process expectations
  for large time, III}}.
\newblock Communications on Pure and Applied Mathematics \textbf{29(4)},
  389--461 (1976).

\bibitem{deuschel1989large}
F.~den Hollander.
\newblock \emph{{Large deviations}}, vol.~14 (Academic press, 2000).
\newblock ISBN 0-8218-1989-5.

\bibitem{Nemoto2014a}
T.~Nemoto and S.~I. Sasa.
\newblock \emph{{Computation of large deviation statistics via iterative
  measurement-and-feedback procedure}}.
\newblock Physical Review Letters \textbf{112(9)} (2014).

\bibitem{jack2010large}
R.~L. Jack and P.~Sollich.
\newblock \emph{{Large deviations and ensembles of trajectories in stochastic
  models}}.
\newblock Progress of Theoretical Physics Supplement \textbf{184(Supplement
  1)}, 14 (2009).

\bibitem{Chetrite2014}
R.~Chetrite and H.~Touchette.
\newblock \emph{{Nonequilibrium Markov processes conditioned on large
  deviations}}.
\newblock Annales Henri Poincar\'{e}  (2014).

\bibitem{Chetrite2015}
R.~Chetrite and H.~Touchette.
\newblock \emph{{Variational and optimal control representations of conditioned
  and driven processes}}.
\newblock arXiv:1506.05291  (2015).

\bibitem{Lebowitz99agallavotti-cohen}
J.~L. Lebowitz and H.~Spohn.
\newblock \emph{{A Gallavotti-Cohen Type Symmetry in the Large Deviation
  Functional for Stochastic Dynamics}}.
\newblock Journal of Statistical Physics \textbf{95}, 333--365 (1998).

\bibitem{Kurchan1998}
J.~Kurchan.
\newblock \emph{{Fluctuation theorem for stochastic dynamics}}.
\newblock Journal of Physics A: Mathematical and General \textbf{31(16)}, 15
  (1997).

\bibitem{PhysRevLett.74.2694}
G.~Gallavotti and E.~G.~D. Cohen.
\newblock \emph{{Dynamical ensembles in nonequilibrium statistical mechanics}}.
\newblock Physical Review Letters \textbf{74(14)}, 2694--2697 (1995).

\bibitem{PhysRevLett.71.2401}
D.~J. Evans, E.~G.~D. Cohen and G.~P. Morriss.
\newblock \emph{{Probability of second law violations in shearing steady
  states}}.
\newblock Physical Review Letters \textbf{71(15)}, 2401--2404 (1993).

\bibitem{Evans1994}
D.~J. Evans and D.~J. Searles.
\newblock \emph{{Equilibrium microstates which generate second law violating
  steady states}}.
\newblock Physical Review E \textbf{50(2)}, 1645--1648 (1994).

\bibitem{MacDonald1968}
C.~T. MacDonald, J.~H. Gibbs and a.~C. Pipkin.
\newblock \emph{{Kinetics of biopolymerization on nucleic acid templates.}}
\newblock Biopolymers \textbf{6(1)}, 1--5 (1968).

\bibitem{MacDonald1969}
C.~T. MacDonald and J.~H. Gibbs.
\newblock \emph{{Concerning the kinetics of polypeptide synthesis on
  polyribosomes}}.
\newblock Biopolymers \textbf{7(5)}, 707--725 (1969).

\bibitem{1742-5468-2008-06-P06009}
D.~a. Adams, B.~Schmittmann and R.~K.~P. Zia.
\newblock \emph{{Far-from-equilibrium transport with constrained resources}}.
\newblock Journal of Statistical Mechanics: Theory and Experiment
  \textbf{2008(06)}, 14 (2008).

\bibitem{Greulich2012}
P.~Greulich, L.~Ciandrini, R.~J. Allen and M.~C. Romano.
\newblock \emph{{Mixed population of competing totally asymmetric simple
  exclusion processes with a shared reservoir of particles}}.
\newblock Physical Review E - Statistical, Nonlinear, and Soft Matter Physics
  \textbf{85(1)}, 011142 (2012).

\bibitem{Greulich2008}
P.~Greulich and A.~Schadschneider.
\newblock \emph{{Single-Bottleneck Approximation for Driven Lattice Gases with
  Disorder and Open Boundary Conditions}}.
\newblock Journal of Statistical Mechanics: Theory and Experiment
  \textbf{2008(04)}, P04009 (2007).

\bibitem{Ciandrini2010}
L.~Ciandrini, I.~Stansfield and M.~C. Romano.
\newblock \emph{{Role of the particle's stepping cycle in an asymmetric
  exclusion process: A model of mRNA translation}}.
\newblock Physical Review E - Statistical, Nonlinear, and Soft Matter Physics
  \textbf{81(5)}, 051904 (2010).

\bibitem{Reese2011}
L.~Reese, A.~Melbinger and E.~Frey.
\newblock \emph{{Crowding of molecular motors determines microtubule
  depolymerization}}.
\newblock Biophysical Journal \textbf{101(9)}, 2190--2200 (2011).

\bibitem{0034-4885-74-11-116601}
T.~Chou, K.~Mallick and R.~K.~P. Zia.
\newblock \emph{{Paradigmatic Model To Biological Transport}}.
\newblock Reports on Progress in Physics \textbf{74(11)}, 116601 (2011).

\bibitem{sandow1994partially}
S.~Sandow.
\newblock \emph{{Partially asymmetric exclusion process with open boundaries}}.
\newblock Physical Review E \textbf{50(4)}, 2660--2667 (1994).

\bibitem{Faddeev1996}
L.~D. Faddeev.
\newblock \emph{{How Algebraic Bethe Ansatz works for integrable model}}.
\newblock Les-Houches lectures p.~59 (1996).

\bibitem{Baxter1982}
R.~J. Baxter.
\newblock \emph{{Exactly solved models in statistical mechanics}} (Academic
  Press, 2007).
\newblock ISBN 0486462714.

\bibitem{Prolhac2008}
S.~Prolhac.
\newblock \emph{{Fluctuations and skewness of the current in the partially
  asymmetric exclusion process}}.
\newblock Journal of Physics A: Mathematical and Theoretical \textbf{41(36)},
  21 (2008).

\bibitem{Prolhac2009}
S.~Prolhac.
\newblock \emph{{A combinatorial solution for the current fluctuations in the
  exclusion process}}.
\newblock arXiv preprint arXiv:0904.2356 \textbf{(2)}, 7 (2009).

\bibitem{prolhac2010tree}
S.~Prolhac.
\newblock \emph{{Tree structures for the current fluctuations in the exclusion
  process}}.
\newblock Journal of Physics A: Mathematical and Theoretical \textbf{43(10)},
  43 (2009).

\bibitem{Prolhac2008a}
S.~Prolhac and K.~Mallick.
\newblock \emph{{Current Fluctuations in the exclusion process and Bethe
  Ansatz}}.
\newblock Journal of Physics A: Mathematical and Theoretical \textbf{41(17)},
  17 (2008).

\bibitem{Crampe2010}
N.~Cramp\'{e}, E.~Ragoucy and D.~Simon.
\newblock \emph{{Eigenvectors of open XXZ and ASEP models for a class of
  non-diagonal boundary conditions}}.
\newblock Journal of Statistical Mechanics: Theory and Experiment
  \textbf{2010(11)}, P11038 (2010).

\bibitem{crampe2011matrix}
N.~Crampe, E.~Ragoucy and D.~Simon.
\newblock \emph{{Matrix Coordinate Bethe Ansatz: Applications to XXZ and ASEP
  models}}.
\newblock Journal of Physics A: Mathematical and Theoretical \textbf{44(40)},
  18 (2011).

\bibitem{1742-5468-2006-12-P12011}
J.~de~Gier and F.~H.~L. Essler.
\newblock \emph{{Exact Spectral Gaps of the Asymmetric Exclusion Process with
  Open Boundaries}}.
\newblock Journal of Statistical Mechanics: Theory and Experiment
  \textbf{2006(12)}, 42 (2006).

\bibitem{de2005bethe}
J.~de~Gier and F.~H.~L. Essler.
\newblock \emph{{Bethe ansatz solution of the asymmetric exclusion process with
  open boundaries}}.
\newblock Physical Review Letters \textbf{95(24)}, 240601 (2005).

\bibitem{simon2009construction}
D.~Simon.
\newblock \emph{{Construction of a Coordinate Bethe Ansatz for the asymmetric
  simple exclusion process with open boundaries}}.
\newblock Journal of Statistical Mechanics: Theory and Experiment
  \textbf{2009(07)}, P07017 (2009).

\bibitem{derrida1993exact}
B.~Derrida, M.~R. Evans, V.~Hakim and V.~Pasquier.
\newblock \emph{{Exact solution of a 1D asymmetric exclusion model using a
  matrix formulation}}.
\newblock Journal of Physics A: Mathematical and General \textbf{26(7)},
  1493--1517 (1999).

\bibitem{kardar1986dynamic}
M.~Kardar, G.~Parisi and Y.~C. Zhang.
\newblock \emph{{Dynamic scaling of growing interfaces}}.
\newblock Physical Review Letters \textbf{56(9)}, 889--892 (1986).

\bibitem{johansson2000shape}
K.~Johansson.
\newblock \emph{{Shape Fluctuations and Random Matrices}}.
\newblock Communications in mathematical physics \textbf{209(2)}, 51 (1999).

\bibitem{tracy:095204}
C.~A. Tracy and H.~Widom.
\newblock \emph{{Total current fluctuations in the asymmetric simple exclusion
  process}}.
\newblock Journal of Mathematical Physics \textbf{50(9)}, 095204 (2009).

\bibitem{PhysRevLett.104.230602}
T.~Sasamoto and H.~Spohn.
\newblock \emph{{One-Dimensional kardar-Parisi-zhang equation: An exact
  solution and its universality}}.
\newblock Physical Review Letters \textbf{104(23)}, 230602 (2010).

\bibitem{Prahofer2002}
M.~Praehofer and H.~Spohn.
\newblock \emph{{Current fluctuations for the totally asymmetric simple
  exclusion process}}.
\newblock In and out of equilibrium \textbf{51}, 19 (2001).

\bibitem{BenArous2011}
G.~B. Arous and I.~Corwin.
\newblock \emph{{Current fluctuations for TASEP: A proof of the
  Pr\"{a}hofer-Spohn conjecture}}.
\newblock Annals of Probability \textbf{39(1)}, 104--138 (2011).

\bibitem{ferrari2010interacting}
P.~L. Ferrari.
\newblock \emph{{From interacting particle systems to random matrices}}.
\newblock Journal of Statistical Mechanics: Theory and Experiment
  \textbf{2010(10)}, 18 (2010).

\bibitem{sasamoto2007fluctuations}
T.~Sasamoto.
\newblock \emph{{Fluctuations of the one-dimensional asymmetric exclusion
  process using random matrix techniques}}.
\newblock Journal of Statistical Mechanics: Theory and Experiment
  \textbf{2007(07)}, 41 (2007).

\bibitem{Prolhac2011}
S.~Prolhac and H.~Spohn.
\newblock \emph{{The One-dimensional KPZ Equation and the Airy Process}}.
\newblock Journal of Statistical Mechanics: Theory and Experiment
  \textbf{2011(03)}, 15 (2011).

\bibitem{amir2011probability}
G.~Amir, I.~Corwin and J.~Quastel.
\newblock \emph{{Probability Distribution of the Free Energy of the Continuum
  Directed Random Polymer in 1+1 dimensions}}.
\newblock Communications on Pure and Applied Mathematics \textbf{64(4)}, 68
  (2010).

\bibitem{Calabrese2010}
P.~Calabrese, P.~L. Doussal and A.~Rosso.
\newblock \emph{{Free-energy distribution of the directed polymer at high
  temperature}}.
\newblock EPL (Europhysics Letters) \textbf{90(2)}, 6 (2010).

\bibitem{Imamura2011}
T.~Imamura, T.~Sasamoto and H.~Spohn.
\newblock \emph{{KPZ, ASEP and Delta-Bose Gas}}.
\newblock Journal of Physics: Conference Series \textbf{297}, 012016 (2011).

\bibitem{Spohn2012}
H.~Spohn.
\newblock \emph{{Stochastic integrability and the KPZ equation}}.
\newblock IAMP News Bulletin pp. 1--6 (2012).

\bibitem{Bertini1997}
L.~Bertini and G.~Giacomin.
\newblock \emph{{Stochastic Burgers and KPZ Equations from Particle Systems}}
  (1997).

\bibitem{Lecomte2007a}
V.~Lecomte, U.~C. T\"{a}uber and F.~van Wijland.
\newblock \emph{{Current distribution in systems with anomalous diffusion:
  renormalization group approach}}.
\newblock Journal of Physics A: Mathematical and Theoretical \textbf{40},
  1447--1465 (2007).

\bibitem{PhysRevLett.104.230601}
K.~A. Takeuchi and M.~Sano.
\newblock \emph{{Universal fluctuations of growing interfaces: Evidence in
  turbulent liquid crystals}}.
\newblock Physical Review Letters \textbf{104(23)}, 230601 (2010).

\bibitem{corwin2012kardar}
I.~Corwin.
\newblock \emph{{The Kardar-Parisi-Zhang equation and universality class}}.
\newblock Random Matrices: Theory and Applications \textbf{1(01)}, 57 (2011).

\bibitem{halpin1995kinetic}
T.~Halpin-Healy and Y.-C. Zhang.
\newblock \emph{{Kinetic roughening phenomena, stochastic growth, directed
  polymers and all that. Aspects of multidisciplinary statistical mechanics}}.
\newblock Physics Reports \textbf{254(4-6)}, 215--414 (1995).

\bibitem{kriecherbauer2010pedestrian}
T.~Kriecherbauer and J.~Krug.
\newblock \emph{{A pedestrian's view on interacting particle systems, KPZ
  universality, and random matrices}}.
\newblock Journal of Physics A: Mathematical and Theoretical \textbf{43(40)},
  52 (2008).

\bibitem{Quastel2010}
J.~Quastel.
\newblock \emph{{Weakly Asymmetric Exclusion and KPZ}}.
\newblock \emph{Proceedings of the International Congress of Mathematicians
  2010 (ICM 2010) - Vol. I: Plenary Lectures and Ceremonies, Vols. II-IV:
  Invited Lectures}, pp. 2310--2324 (Hindustan Book Agency, India, 2010).
\newblock ISBN 9789814324304.

\bibitem{Karzig2010}
T.~Karzig and F.~{Von Oppen}.
\newblock \emph{{Signatures of critical full counting statistics in a
  quantum-dot chain}}.
\newblock Physical Review B - Condensed Matter and Materials Physics
  \textbf{81(4)}, 045317 (2010).

\bibitem{Batchelor2001}
M.~T. Batchelor, J.~de~Gier and B.~Nienhuis.
\newblock \emph{{The quantum symmetric XXZ chain at Delta=-1/2, alternating
  sign matrices and plane partitions}}.
\newblock Journal of Physics A: Mathematical and General \textbf{34(19)}, 7
  (2001).

\bibitem{DeGier2002}
J.~de~Gier, M.~T. Batchelor, B.~Nienhuis and S.~Mitra.
\newblock \emph{{The XXZ spin chain at Delta = - 1/2: Bethe roots, symmetric
  functions, and determinants}}.
\newblock Journal of Mathematical Physics \textbf{43(8)}, 4135--4146 (2002).

\bibitem{Blythe2009}
R.~A. Blythe, W.~Janke, D.~A. Johnston and R.~Kenna.
\newblock \emph{{Continued Fractions and the Partially Asymmetric Exclusion
  Process}}.
\newblock Journal of Physics A: Mathematical and Theoretical \textbf{42(32)},
  325002 (2009).

\bibitem{Derrida2004a}
B.~Derrida, C.~Enaud and J.~L. Lebowitz.
\newblock \emph{{The asymmetric Exclusion Process and Brownian Excursions}}.
\newblock Journal of Statistical Physics \textbf{115(1/2)}, 23 (2003).

\bibitem{Majumdar2005}
S.~N. Majumdar and A.~Comtet.
\newblock \emph{{Airy distribution function: From the area under a brownian
  excursion to the maximal height of fluctuating interfaces}}.
\newblock Journal of Statistical Physics \textbf{119(3-4)}, 777--826 (2005).

\bibitem{Majumdar2004}
S.~N. Majumdar and A.~Comtet.
\newblock \emph{{Exact maximal height distribution of fluctuating interfaces}}.
\newblock Physical Review Letters \textbf{92(22)}, 225501--1 (2004).

\bibitem{corteel2011tableaux}
S.~Corteel and L.~K. Williams.
\newblock \emph{{Tableaux combinatorics for the asymmetric exclusion process}}.
\newblock Advances in Applied Mathematics \textbf{39(3)}, 293--310 (2007).

\bibitem{sasamoto1999one}
T.~Sasamoto.
\newblock \emph{{One-dimensional partially asymmetric simple exclusion process
  with open boundaries: orthogonal polynomials approach}}.
\newblock Journal of Physics A: Mathematical and General \textbf{32(41)},
  7109--7131 (1999).

\bibitem{Uchiyama2004}
M.~Uchiyama, T.~Sasamoto and M.~Wadati.
\newblock \emph{{Asymmetric Simple Exclusion Process with Open Boundaries and
  Askey-Wilson Polynomials}}.
\newblock Journal of Physics A: Mathematical and General \textbf{37(18)}, 21
  (2003).

\bibitem{gerard2009canopy}
X.~G. Viennot.
\newblock \emph{{Canopy of binary trees, Catalan tableaux and the asymmetric
  exclusion process}}  (2009).

\bibitem{Schütz20011}
G.~Schutz.
\newblock \emph{{Exactly Solvable Models for Many-Body Systems Far from
  Equilibrium}}.
\newblock vol.~19 of \emph{Phase Transitions and Critical Phenomena}, pp.
  3--251 (Academic Press, 2001).
\newblock ISBN 0122203194.

\bibitem{Derrida199865}
B.~Derrida.
\newblock \emph{{An exactly soluble non-equilibrium system: The asymmetric
  simple exclusion process}}.
\newblock Physics Reports \textbf{301(1-3)}, 65--83 (1998).

\bibitem{1742-5468-2007-07-P07023}
B.~Derrida.
\newblock \emph{{Non equilibrium steady states: fluctuations and large
  deviations of the density and of the current}}.
\newblock Journal of Statistical Mechanics: Theory and Experiment
  \textbf{2007(07)}, 35 (2007).

\bibitem{Schmittmann19953}
B.~Schmittmann and R.~K.~P. Zia.
\newblock \emph{{Statistical mechanics of driven diffusive systems}}.
\newblock B.~Schmittmann and R.~K.~P. Zia (Eds.), \emph{Phase Transitions and
  Critical Phenomena}, vol.~17 of \emph{Phase Transitions and Critical
  Phenomena}, pp. 3--214 (Academic Press, 1995).
\newblock ISBN 9780122203176.

\bibitem{appert2008universal}
C.~Appert-Rolland, B.~Derrida, V.~Lecomte and F.~{Van Wijland}.
\newblock \emph{{Universal cumulants of the current in diffusive systems on a
  ring}}.
\newblock Physical Review E - Statistical, Nonlinear, and Soft Matter Physics
  \textbf{78(2)}, 21122 (2008).

\bibitem{PhysRevLett.92.180601}
T.~Bodineau and B.~Derrida.
\newblock \emph{{Current fluctuations in nonequilibrium diffusive systems: An
  additivity principle}}.
\newblock Physical Review Letters \textbf{92(18)}, 180601--1 (2004).

\bibitem{Bertini2002}
L.~Bertini, a.~{De Sole}, D.~Gabrielli, G.~Jona-Lasinio and C.~Landim.
\newblock \emph{{Macroscopic fluctuation theory for stationary non-equilibrium
  states}}.
\newblock Journal of Statistical Physics \textbf{107(3-4)}, 635--675 (2002).

\bibitem{Bertini2007}
L.~Bertini, a.~{De Sole}, D.~Gabrielli, G.~Jona-Lasinio and C.~Landim.
\newblock \emph{{Stochastic interacting particle systems out of equilibrium}}.
\newblock Journal of Statistical Mechanics: Theory and Experiment
  \textbf{2007(07)}, 36 (2007).

\bibitem{bertini2005current}
L.~Bertini, a.~D. Sole, D.~Gabrielli, G.~Jona-Lasinio and C.~Landim.
\newblock \emph{{Current Fluctuations in Stochastic Lattice Gases}}.
\newblock Physical Review Letters \textbf{94(January)}, 030601 (2005).

\bibitem{Bertini2006}
L.~Bertini, A.~{De Sole}, D.~Gabrielli, G.~Jona-Lasinio and C.~Landim.
\newblock \emph{{Large deviation approach to non equilibrium processes in
  stochastic lattice gases}}.
\newblock Bulletin of the Brazilian Mathematical Society \textbf{37(4)},
  611--643 (2006).

\bibitem{spohn1991large}
H.~Spohn.
\newblock \emph{{Large scale dynamics of interacting particles}}.
\newblock Texts and monographs in physics (Springer-Verlag, 1991).
\newblock ISBN 9783642843730.

\bibitem{Enaud2004}
B.~Derrida and C.~Enaud.
\newblock \emph{{Large deviation functional of the weakly asymmetric exclusion
  process}}.
\newblock Journal of Statistical Physics \textbf{114(3/4)}, 537--562 (2003).

\bibitem{Bodineau2006}
T.~Bodineau and B.~Derrida.
\newblock \emph{{Current large deviations for asymmetric exclusion processes
  with open boundaries}}.
\newblock Journal of Statistical Physics \textbf{123(2)}, 277--300 (2006).

\bibitem{Derrida2004}
B.~Derrida, B.~Doucot and P.~E. Roche.
\newblock \emph{{Current fluctuations in the one dimensional Symmetric
  Exclusion Process with open boundaries}}.
\newblock Journal of Statistical Physics \textbf{115(3/4)}, 717--748 (2003).

\bibitem{Tailleur2007}
J.~Tailleur, J.~Kurchan and V.~Lecomte.
\newblock \emph{{Mapping nonequilibrium onto equilibrium: The macroscopic
  fluctuations of simple transport models}}.
\newblock Physical Review Letters \textbf{99(15)}, 150602 (2007).

\bibitem{Tailleur2008}
J.~Tailleur, J.~Kurchan and V.~Lecomte.
\newblock \emph{{Mapping out of equilibrium into equilibrium in one-dimensional
  transport models}}.
\newblock Journal of Physics A: Mathematical and Theoretical \textbf{41(50)},
  505001 (2008).

\bibitem{Lecomte2010}
V.~Lecomte, A.~Imparato and F.~{Van Wijland}.
\newblock \emph{{Current fluctuations in systems with diffusive dynamics, in
  and out of equilibrium}}.
\newblock Progress of Theoretical Physics Supplement \textbf{184}, 276--289
  (2009).

\bibitem{Prolhac2009a}
S.~Prolhac and K.~Mallick.
\newblock \emph{{Cumulants of the current in the weakly asymmetric exclusion
  process}}.
\newblock Journal of Physics A: Mathematical and Theoretical \textbf{42(17)},
  24 (2009).

\bibitem{PhysRevE.72.066110}
T.~Bodineau and B.~Derrida.
\newblock \emph{{Distribution of current in nonequilibrium diffusive systems
  and phase transitions}}.
\newblock Physical Review E - Statistical, Nonlinear, and Soft Matter Physics
  \textbf{72(6)}, 66110 (2005).

\bibitem{Simon2011}
D.~Simon.
\newblock \emph{{Bethe Ansatz for the Weakly Asymmetric Simple Exclusion
  Process and Phase Transition in the Current Distribution}}.
\newblock Journal of Statistical Physics \textbf{142(5)}, 931--951 (2011).

\bibitem{Belitsky2013}
V.~Belitsky and G.~M. Sch\"{u}tz.
\newblock \emph{{Microscopic Structure of Shocks and Antishocks in the ASEP
  Conditioned on Low Current}}.
\newblock Journal of Statistical Physics \textbf{152}, 93--111 (2013).

\bibitem{Lecomte2012}
V.~Lecomte, J.~P. Garrahan and F.~van Wijland.
\newblock \emph{{Inactive dynamical phase of a symmetric exclusion process on a
  ring}}.
\newblock Journal of Physics A: Mathematical and Theoretical \textbf{45(17)},
  175001 (2012).

\bibitem{Jack2015}
R.~L. Jack, I.~R. Thompson and P.~Sollich.
\newblock \emph{{Hyperuniformity and Phase Separation in Biased Ensembles of
  Trajectories for Diffusive Systems}}.
\newblock Physical Review Letters \textbf{114(1)}, 060601 (2015).

\bibitem{Krapivsky}
P.~L. Krapivsky, K.~Mallick and T.~Sadhu.
\newblock \emph{{Dynamical properties of single-file diffusion}}.
\newblock arXiv:1505.01287  (2015).

\bibitem{Mallick}
P.~L. Krapivsky, K.~Mallick and T.~Sadhu.
\newblock \emph{{Tagged particle in single-file diffusion}}.
\newblock arXiv:1506.00865 .

\bibitem{derrida1998exact}
B.~Derrida and J.~L. Lebowitz.
\newblock \emph{{Exact Large Deviation Function in the Asymmetric Exclusion
  Process}}.
\newblock Physical Review Letters \textbf{80(2)}, 8 (1998).

\bibitem{derrida1999universal}
B.~Derrida and C.~Appert.
\newblock \emph{{Universal large-deviation function of the
  Kardar--Parisi--Zhang equation in one dimension}}.
\newblock Journal of Statistical Physics \textbf{94(1-2)}, 1--30 (1999).

\bibitem{Derrida1997}
B.~Derrida and K.~Mallick.
\newblock \emph{{Exact diffusion constant for the one-dimensional partially
  asymmetric exclusion model}}.
\newblock Journal of Physics A: Mathematical and General \textbf{30(4)},
  1031--1046 (1999).

\bibitem{Golinelli2004}
O.~Golinelli and K.~Mallick.
\newblock \emph{{Bethe Ansatz calculation of the spectral gap of the asymmetric
  exclusion process}}.
\newblock Journal of Physics A: Mathematical and General \textbf{37(10)},
  3321--3331 (2003).

\bibitem{Golinelli2005}
O.~Golinelli and K.~Mallick.
\newblock \emph{{Spectral gap of the totally asymmetric exclusion process at
  arbitrary filling}}.
\newblock Journal of Physics A: Mathematical and General \textbf{38(7)},
  1419--1425 (2004).

\bibitem{Prolhac2013}
S.~Prolhac.
\newblock \emph{{Spectrum of the totally asymmetric simple exclusion process on
  a periodic lattice -- first excited states}}.
\newblock arXiv preprint arXiv:1404.1315 \textbf{46(41)}, 30 (2014).

\bibitem{Prolhac2014a}
S.~Prolhac.
\newblock \emph{{Spectrum of the totally asymmetric simple exclusion process on
  a periodic lattice -- first excited states}}.
\newblock J. Phys. A: Math. Theor. \textbf{47}, 375001 (2014).

\bibitem{Prolhac2015}
S.~Prolhac.
\newblock \emph{{Current fluctuations for totally asymmetric exclusion on the
  relaxation scale}}.
\newblock Journal of Physics A: Mathematical and Theoretical \textbf{48},
  06FT02 (2015).

\bibitem{Popkov2010}
V.~Popkov, G.~M. Sch\"{u}tz and D.~Simon.
\newblock \emph{{Asymmetric simple exclusion process on a ring conditioned on
  enhanced flux}}.
\newblock Journal of Statistical Mechanics: Theory and Experiment
  \textbf{2010(10)}, P10007 (2010).

\bibitem{derrida1992exact}
B.~Derrida, M.~R. Evans and K.~Mallick.
\newblock \emph{{Exact diffusion constant of a one-dimensional asymmetric
  exclusion model with open boundaries}}.
\newblock Journal of Statistical Physics \textbf{79(5-6)}, 833--874 (1995).

\bibitem{krug1991boundary}
J.~Krug.
\newblock \emph{{Boundary-induced phase transitions in driven diffusive
  systems}}.
\newblock Physical Review Letters \textbf{67(14)}, 1882--1885 (1991).

\bibitem{Schutz1993}
G.~Sch\"{u}tz and E.~Domany.
\newblock \emph{{Phase transitions in an exactly soluble one-dimensional
  exclusion process}}.
\newblock Journal of Statistical Physics \textbf{72(1-2)}, 277--296 (1993).

\bibitem{Sasamoto1999}
T.~Sasamoto.
\newblock \emph{{Density profile of the one-dimensional partially asymmetric
  simple exclusion process with open boundaries}}.
\newblock Journal of the Physical Society of Japan \textbf{69(4)}, 1055--1067
  (2000).

\bibitem{Derrida1993a}
B.~Derrida and M.~R. Evans.
\newblock \emph{{Exact correlation functions in an asymmetric exclusion model
  with open boundaries}}.
\newblock Journal de Physique I \textbf{3(2)}, 311--322 (1993).

\bibitem{uchiyama2005correlation}
M.~Uchiyama and M.~Wadati.
\newblock \emph{{Correlation Function of Asymmetric Simple Exclusion Process
  with Open Boundaries}}.
\newblock Journal of Nonlinear Mathematical Physics \textbf{12(sup1)}, 14
  (2004).

\bibitem{1751-8121-40-46-R01}
R.~A. Blythe and M.~R. Evans.
\newblock \emph{{Nonequilibrium steady states of matrix-product form: a
  solver's guide}}.
\newblock Journal of Physics A: Mathematical and Theoretical \textbf{40(46)},
  R333 (2007).

\bibitem{Crampe}
N.~Crampe, K.~Mallick, E.~Ragoucy and M.~Vanicat.
\newblock \emph{{Open two-species exclusion processes with integrable
  boundaries}}.
\newblock Journal of Physics A: Mathematical and Theoretical \textbf{48},
  175002 (2015).

\bibitem{Queiroz2012}
S.~L.~A. {De Queiroz}.
\newblock \emph{{Current-activity versus local-current fluctuations in a driven
  flow with exclusion}}.
\newblock Physical Review E - Statistical, Nonlinear, and Soft Matter Physics
  \textbf{86(4)}, 1--9 (2012).

\bibitem{depken2005exact}
M.~Depken and R.~Stinchcombe.
\newblock \emph{{Exact probability function for bulk density and current in the
  asymmetric exclusion process}}.
\newblock Physical Review E - Statistical, Nonlinear, and Soft Matter Physics
  \textbf{71(3)}, 36120 (2005).

\bibitem{derrida2002exa}
B.~Derrida, J.~L. Lebowitz and E.~R. Speer.
\newblock \emph{{Exact free energy functional for a driven diffusive open
  stationary nonequilibrium system.}}
\newblock Physical review letters \textbf{89(3)}, 030601 (2002).

\bibitem{DeGier2011}
J.~de~Gier, C.~Finn and M.~Sorrell.
\newblock \emph{{Relaxation rate of the reverse biased asymmetric exclusion
  process}}.
\newblock J. Phys. A: Math. Theor. \textbf{44}, 405002 (2011).

\bibitem{Blythe2000}
R.~A. Blythe, M.~R. Evans, F.~Colaiori and F.~H.~L. Essler.
\newblock \emph{{Exact solution of a partially asymmetric exclusion model using
  a deformed oscillator algebra}}.
\newblock Journal of Physics A: Mathematical and General \textbf{33(12)}, 27
  (1999).

\bibitem{Dudzinski2000}
M.~Dudzinski and G.~M. Sch\"{u}tz.
\newblock \emph{{Relaxation spectrum of the asymmetric exclusion process with
  open boundaries}}.
\newblock Journal of Physics A: Mathematical and General \textbf{33(47)},
  8351--8363 (2000).

\bibitem{Kolomeisky1998}
A.~B. Kolomeisky, G.~M. Sch\"{u}tz, E.~B. Kolomeisky and J.~P. Straley.
\newblock \emph{{Phase diagram of one-dimensional driven lattice gases with
  open boundaries}}.
\newblock Journal of Physics A: Mathematical and General \textbf{31(33)},
  6911--6919 (1999).

\bibitem{Santen2002}
L.~Santen and C.~Appert.
\newblock \emph{{The asymmetric exclusion process revisited: Fluctuations and
  dynamics in the domain wall picture}}.
\newblock Journal of Statistical Physics \textbf{106(1-2)}, 187--199 (2002).

\bibitem{Varadhan1996}
S.~Varadhan.
\newblock \emph{{Ito's Stochastic Calculus and Probability Theory}}.
\newblock N.~Ikeda, S.~Watanabe, M.~Fukushima and H.~Kunita (Eds.),
  \emph{It\^{o}’s Stochastic Calculus and Probability Theory} (Springer
  Japan, Tokyo, 1996).
\newblock ISBN 9784431685340.

\bibitem{Bahadoran}
C.~Bahadoran.
\newblock \emph{{A quasi-potential for conservation laws with boundary
  conditions}}.
\newblock arXiv:1010.3624  (2010).

\bibitem{Proeme2011}
A.~Proeme, R.~A. Blythe and M.~R. Evans.
\newblock \emph{{Dynamical Transition in the Open-boundary Totally Asymmetric
  Exclusion Process}}.
\newblock Journal of Physics A: Mathematical and Theoretical \textbf{44(3)}, 27
  (2010).

\bibitem{PhysRevLett.107.010602}
J.~de~Gier and F.~H.~L. Essler.
\newblock \emph{{Large deviation function for the current in the open
  asymmetric simple exclusion process}}.
\newblock Physical Review Letters \textbf{107(1)}, 10602 (2011).

\bibitem{derrida1995exact}
B.~Derrida, M.~R. Evans and K.~Mallick.
\newblock \emph{{Exact diffusion constant of a one-dimensional asymmetric
  exclusion model with open boundaries}}.
\newblock Journal of Statistical Physics \textbf{79(5-6)}, 833--874 (1995).

\bibitem{Lazarescu2014}
A.~Lazarescu and V.~Pasquier.
\newblock \emph{{Bethe Ansatz and Q-operator for the open ASEP}}.
\newblock Journal of Physics A: Mathematical and Theoretical \textbf{47(1)}, 46
  (2014).

\bibitem{gorissen2012exact}
M.~Gorissen, A.~Lazarescu, K.~Mallick and C.~Vanderzande.
\newblock \emph{{Exact current statistics of the asymmetric simple exclusion
  process with open boundaries}}.
\newblock Physical Review Letters \textbf{109(17)}, 170601 (2012).

\bibitem{Evans2009}
M.~R. Evans, P.~a. Ferrari and K.~Mallick.
\newblock \emph{{Matrix representation of the stationary measure for the
  multispecies TASEP}}.
\newblock Journal of Statistical Physics \textbf{135(2)}, 217--239 (2009).

\bibitem{prolhac2009matrix}
S.~Prolhac, M.~R. Evans and K.~Mallick.
\newblock \emph{{Matrix product solution of the multispecies partially
  asymmetric exclusion process}}.
\newblock Journal of Physics A: Mathematical and Theoretical \textbf{42(16)},
  27 (2008).

\bibitem{Boutillier2002}
C.~Boutillier, P.~Fran ois, K.~Mallick and S.~Mallick.
\newblock \emph{{A matrix ansatz for the diffusion of an impurity in the
  asymmetric exclusion process}}.
\newblock Journal of Physics A: Mathematical and General \textbf{35(46)},
  9703--9730 (2002).

\bibitem{derrida1999bethe}
B.~Derrida and M.~R. Evans.
\newblock \emph{{Bethe Ansatz Solution for a Defect Particle in the Asymmetric
  Exclusion Process}}.
\newblock Journal of Physics A: Mathematical and General \textbf{32(26)}, 23
  (1999).

\bibitem{Sasamoto2000}
T.~Sasamoto.
\newblock \emph{{One-Dimensional Partially Asymmetric Simple Exclusion Process
  on a Ring with a Defect Particle}}.
\newblock Physical Review E \textbf{61(5)}, 23 (1999).

\bibitem{Derrida1993}
B.~Derrida, S.~a. Janowsky, J.~L. Lebowitz and E.~R. Speer.
\newblock \emph{{Exact solution of the totally asymmetric simple exclusion
  process: Shock profiles}}.
\newblock Journal of Statistical Physics \textbf{73(5-6)}, 813--842 (1993).

\bibitem{Mallick1996}
K.~Mallick.
\newblock \emph{{Shocks in the asymmetry exclusion model with an impurity}}.
\newblock Journal of Physics A: Mathematical and General \textbf{29(17)},
  5375--5386 (1999).

\bibitem{evans1999exact}
M.~R. Evans, N.~Rajewsky and E.~R. Speer.
\newblock \emph{{Exact solution of a cellular automaton for traffic}}.
\newblock Journal of Statistical Physics \textbf{95(1-2)}, 54 (1998).

\bibitem{rajewsky1998asymmetric}
N.~Rajewsky, L.~Santen, a.~Schadschneider and M.~Schreckenberg.
\newblock \emph{{The asymmetric exclusion process: Comparison of update
  procedures}}.
\newblock Journal of Statistical Physics \textbf{92(1-2)}, 47 (1997).

\bibitem{Evans2011}
M.~R. Evans, Y.~Kafri, K.~E.~P. Sugden and J.~Tailleur.
\newblock \emph{{Phase diagram of two-lane driven diffusive systems}}.
\newblock Journal of Statistical Mechanics: Theory and Experiment
  \textbf{2011(06)}, P06009 (2011).

\bibitem{Evans2003}
M.~R. Evans, R.~Juh\'{a}sz and L.~Santen.
\newblock \emph{{Shock formation in an exclusion process with creation and
  annihilation.}}
\newblock Physical review. E, Statistical, nonlinear, and soft matter physics
  \textbf{68(2 Pt 2)}, 026117 (2003).

\bibitem{Harris2004}
R.~J. Harris and R.~B. Stinchcombe.
\newblock \emph{{Disordered asymmetric simple exclusion process: Mean-field
  treatment}}.
\newblock Physical Review E - Statistical, Nonlinear, and Soft Matter Physics
  \textbf{70(1 2)}, 016108 (2004).

\bibitem{Stinchcombe2011}
R.~B. Stinchcombe and S.~L.~a. {De Queiroz}.
\newblock \emph{{Smoothly varying hopping rates in driven flow with
  exclusion}}.
\newblock Physical Review E - Statistical, Nonlinear, and Soft Matter Physics
  \textbf{83(6)}, 061113 (2011).

\bibitem{Janowsky1994}
S.~A. Janowsky and J.~L. Lebowitz.
\newblock \emph{{Exact results for the asymmetric simple exclusion process with
  a blockage}}.
\newblock Journal of Statistical Physics \textbf{77(1-2)}, 35--51 (1994).

\bibitem{Janowsky1992}
S.~A. Janowsky and J.~L. Lebowitz.
\newblock \emph{{Finite-size effects and shock fluctuations in the asymmetric
  simple-exclusion process}}.
\newblock Physical Review A \textbf{45(2)}, 618--625 (1992).

\bibitem{Popkov1999}
V.~Popkov and G.~M. Schuetz.
\newblock \emph{{Steady-state selection in driven diffusive systems with open
  boundaries}}.
\newblock EPL (Europhysics Letters) \textbf{1(1)}, 6 (2000).

\bibitem{PhysRevB.28.1655}
S.~Katz, J.~L. Lebowitz and H.~Spohn.
\newblock \emph{{Phase transitions in stationary nonequilibrium states of model
  lattice systems}}.
\newblock Physical Review B \textbf{28(3)}, 1655--1658 (1983).

\bibitem{Carlos2015}
C.~P\'{e}rez-Espigares, F.~Redig and C.~Giardin.
\newblock \emph{{The spatial fluctuation theorem}}.
\newblock arXiv:1502.03364  (2015).

\bibitem{Hieida1998}
Y.~Hieida.
\newblock \emph{{Application of the Density Matrix Renormalization Group Method
  to a Non-Equilibrium Problem}}.
\newblock J. Phys. Soc. Jpn \textbf{67(369)}, 8 (1998).

\bibitem{Gorissen2009}
M.~Gorissen, J.~Hooyberghs and C.~Vanderzande.
\newblock \emph{{Density-matrix renormalization-group study of current and
  activity fluctuations near nonequilibrium phase transitions}}.
\newblock Physical Review E - Statistical, Nonlinear, and Soft Matter Physics
  \textbf{79(2)}, 020101 (2009).

\bibitem{1751-8121-44-11-115005}
M.~Gorissen and C.~Vanderzande.
\newblock \emph{{Finite size scaling of current fluctuations in the totally
  asymmetric exclusion process}}.
\newblock Journal of Physics A: Mathematical and Theoretical \textbf{44(11)},
  18 (2010).

\bibitem{Bunin2012}
G.~Bunin, Y.~Kafri and D.~Podolsky.
\newblock \emph{{Non-differentiable large-deviation functionals in
  boundary-driven diffusive systems}}.
\newblock Journal of Statistical Mechanics: Theory and Experiment
  \textbf{2012(10)}, L10001 (2012).

\bibitem{Bunin2012a}
G.~Bunin, Y.~Kafri and D.~Podolsky.
\newblock \emph{{Large deviations in boundary-driven systems: Numerical
  evaluation and effective large-scale behavior}}.
\newblock EPL (Europhysics Letters) \textbf{99(2)}, 20002 (2012).

\bibitem{Bunin2013}
G.~Bunin, Y.~Kafri and D.~Podolsky.
\newblock \emph{{Cusp Singularities in Boundary-Driven Diffusive Systems}}.
\newblock Journal of Statistical Physics \textbf{152(1)}, 112--135 (2013).

\bibitem{giardina2011simulating}
C.~Giardina, J.~Kurchan, V.~Lecomte and J.~Tailleur.
\newblock \emph{{Simulating Rare Events in Dynamical Processes}}.
\newblock Journal of Statistical Physics \textbf{145(4)}, 787--811 (2011).

\bibitem{giardina2006direct}
C.~Giardin\`{a}, J.~Kurchan and L.~Peliti.
\newblock \emph{{Direct evaluation of large-deviation functions}}.
\newblock Physical Review Letters \textbf{96(12)}, 120603 (2006).

\bibitem{Lecomte2007}
V.~Lecomte and J.~Tailleur.
\newblock \emph{{A numerical approach to large deviations in continuous-time}}.
\newblock Journal of Statistical Mechanics: Theory and Experiment
  \textbf{2007(03)}, P03004--P03004 (2006).

\bibitem{Tailleur2009}
J.~Tailleur and V.~Lecomte.
\newblock \emph{{Simulation of large deviation functions using population
  dynamics}}.
\newblock \emph{AIP Conference Proceedings}, vol. 1091, pp. 212--219 (AIP,
  2009).

\bibitem{Evans1998}
M.~R. Evans, Y.~Kafri, H.~M. Koduvely and D.~Mukamel.
\newblock \emph{{Phase Separation and Coarsening in One-Dimensional Driven
  Diffusive Systems: Local Dynaimcs Leading to Long-Range Hamiltonians}}.
\newblock Phys. Rev. E \textbf{58(3)}, 2764 (1998).

\bibitem{chaichian1996introduction}
P.~Podle\'{s} and E.~M\"{u}ller.
\newblock \emph{{Introduction to Quantum Groups}}, vol.~10 (World Scientific
  Publishing Company Incorporated, 1998).

\bibitem{Gaudin1983}
M.~Gaudin.
\newblock \emph{{La fonction d'onde de Bethe}} (Masson, 1983).
\newblock ISBN 2225796076.

\bibitem{Cao2013a}
J.~Cao, W.~L. Yang, K.~Shi and Y.~Wang.
\newblock \emph{{Off-diagonal Bethe ansatz solution of the XXX spin chain with
  arbitrary boundary conditions}}.
\newblock Nuclear Physics B \textbf{875(1)}, 152--165 (2013).

\bibitem{Wen2015}
F.-K. Wen, Z.-Y. Yang, S.~Cui, J.-P. Cao and W.-L. Yang.
\newblock \emph{{Spectrum of the Open Asymmetric Simple Exclusion Process with
  Arbitrary Boundary Parameters}}.
\newblock Chinese Physics Letters \textbf{32}, 050503 (2015).

\bibitem{SamuelBelliard2014}
S.~Belliard and R.~A. Pimenta.
\newblock \emph{{Modified algebraic Bethe ansatz for XXZ chain on the segment -
  II - general cases}}.
\newblock Nuclear Physics B \textbf{894}, 1--22 (2014).

\bibitem{Baseilhac2006}
P.~Baseilhac.
\newblock \emph{{The q-deformed analogue of the Onsager algebra: Beyond the
  Bethe ansatz approach}}.
\newblock Nuclear Physics B \textbf{754(1)}, 309--328 (2006).

\bibitem{Baseilhac2012}
P.~Baseilhac and S.~Belliard.
\newblock \emph{{The half-infinite XXZ chain in Onsager's approach}}.
\newblock Nuclear Physics B \textbf{873(3)}, 550--584 (2013).

\bibitem{Faldella2014}
S.~Faldella, N.~Kitanine and G.~Niccoli.
\newblock \emph{{Complete spectrum and scalar products for the open spin-1/2
  XXZ quantum chains with non-diagonal boundary terms}}.
\newblock J. Stat. Mech. \textbf{2014}, P01011 (2014).

\bibitem{Crampe2014}
N.~Crampe, E.~Ragoucy and M.~Vanicat.
\newblock \emph{{Integrable approach to simple exclusion processes with
  boundaries. Review and progress}}.
\newblock Journal of Statistical Mechanics: Theory and Experiment
  \textbf{2014(11)}, P11032 (2014).

\bibitem{Bethe1931}
H.~Bethe.
\newblock \emph{{Zur Theorie der Metalle - I. Eigenwerte und Eigenfunktionen
  der linearen Atomkette}}.
\newblock Zeitschrift fur Physik \textbf{71(3-4)}, 205--226 (1931).

\bibitem{Gantmacher2000}
F.~R. Gantmacher.
\newblock \emph{{Matrix Theory, Vol. 2}} (American Mathematical Society, 2000).

\bibitem{Korff2005}
C.~Korff.
\newblock \emph{{Auxiliary matrices on both sides of the equator}}.
\newblock Journal of Physics A: Mathematical and General \textbf{38(1)}, 22
  (2004).

\bibitem{Sedgewick2009}
F.~Flajolet and R.~Sedgewick.
\newblock \emph{{Analytic combinatorics}} (Cambridge University Press, 2009).
\newblock ISBN 9780521898065.

\bibitem{Corless1996}
R.~M. Corless, G.~H. Gonnet, D.~E.~G. Hare, D.~J. Jeffrey and D.~E. Knuth.
\newblock \emph{{On the LambertW function}}.
\newblock Advances in Computational Mathematics \textbf{5(1)}, 329--359 (1996).

\bibitem{Derrida1996}
B.~Derrida and R.~Zeitak.
\newblock \emph{{Distribution of domain sizes in the zero temperature Glauber
  dynamics of the 1 D Potts model}}.
\newblock Physical Review E \textbf{54(3)}, 24 (1996).

\bibitem{Fredholm1903}
I.~Fredholm.
\newblock \emph{{Sur une classe d'\'{e}quations fonctionnelles}}.
\newblock Acta Mathematica \textbf{27(1)}, 365--390 (1903).

\bibitem{Coulson1978}
C.~A. Coulson, B.~O'Leary and R.~B. Mallion.
\newblock \emph{{Huckel theory for organic chemists}} (Academic Press (London
  and New York), 1978).
\newblock ISBN 0121932508.

\bibitem{Bilstein1999}
U.~Bilstein and B.~Wehefritz.
\newblock \emph{{The XX--model with boundaries. Part I: Diagonalization of the
  finite chain}}.
\newblock Journal of Physics A: Mathematical and General \textbf{32(2)}, 56
  (1998).

\bibitem{Kittel1987}
C.~Kittel.
\newblock \emph{{Quantum Theory of Solids, 2nd Revised Edition}} (Wiley, 1987).
\newblock ISBN 978-0-471-62412-7.

\bibitem{Stephan2010}
J.~M. St\'{e}phan, G.~Misguich and V.~Pasquier.
\newblock \emph{{R\'{e}nyi entropy of a line in two-dimensional Ising models}}.
\newblock Physical Review B - Condensed Matter and Materials Physics
  \textbf{82(12)}, 1--8 (2010).

\bibitem{Gaudin1973}
M.~Gaudin.
\newblock \emph{{Gaz coulombien discret \`{a} une dimension}}.
\newblock Journal de Physique \textbf{34(7)}, 511--522 (1973).

\bibitem{Schutz}
G.~M. Sch\"{u}tz.
\newblock \emph{{Private communication}} .

\bibitem{Lazarescua}
A.~Lazarescu.
\newblock \emph{{Generic Dynamical Phase Transition in One-Dimensional Lattice
  Gases With Exclusion}}.
\newblock (in preparation) .

\bibitem{Schutz2015}
G.~M. Sch\"{u}tz.
\newblock \emph{{Conditioned Stochastic Particle Systems and Integrable Quantum
  Spin Systems}}.
\newblock \emph{From Particle Systems to Partial Differential Equations II},
  vol. 129, pp. 371--393 (2015).
\newblock ISBN 978-3-319-16636-0.

\bibitem{Lazarescu}
A.~Lazarescu.
\newblock \emph{{Hydrodynamic Spectrum of Bulk-Driven One-Dimensional Lattice
  Gases}}.
\newblock (in preparation) .

\end{thebibliography}

\end{document}